\documentclass[12pt, review, sort&compress]{elsarticle}
\usepackage{graphicx}
\usepackage{xcolor}
\usepackage[margin=2.5cm]{geometry}
\usepackage[utf8]{inputenc}
\usepackage{amsmath}
\usepackage{siunitx}
\usepackage[version=4]{mhchem}

\newcommand{\AlTaO}[0]{\ce{Al_{x}Ta_{y}O_{z}}}
\newcommand{\AlO}[0]{\ce{Al2O3}}
\newcommand{\TaO}[0]{\ce{Ta2O5}}

\newcommand{\change}[1]{{#1}}

\newcommand{\Ttwo}[0]{\ce{Al_{32.4}Ta_{9.9}O_{57.7}}}
\newcommand{\Tthree}[0]{\ce{Al_{29.3}Ta_{9.3}O_{61.4}}}

\newcommand{\Tfive}[0]{\ce{Al_{19.9}Ta_{11.1}O_{69.0}}}

\newcommand{\DFTone}[0]{\ce{Al_{33.3}Ta_{8.4}O_{58.3}}}
\newcommand{\DFTtwo}[0]{\ce{Al_{32.0}Ta_{8.0}O_{60.0}}}
\newcommand{\DFTthree}[0]{\ce{Al_{30.8}Ta_{7.7}O_{61.5}}}
\newcommand{\DFTfour}[0]{\ce{Al_{29.6}Ta_{7.4}O_{63.0}}}

\DeclareSIUnit{\atomicmassunit}{\text{u}}
\DeclareSIUnit\angstrom{\text {Å}}
\DeclareSIUnit{\atpercent}{\text{at.}\%}
\DeclareSIUnit{\sccm}{\text{sccm}}
\DeclareSIUnit{\elementarycharge}{e}

\title{Effect of oxygen content on optical, structural, and dielectric properties of \AlTaO{} thin films}

\begin{document}

\begin{frontmatter}


\begin{graphicalabstract}
\centering
\includegraphics[width=\textwidth]{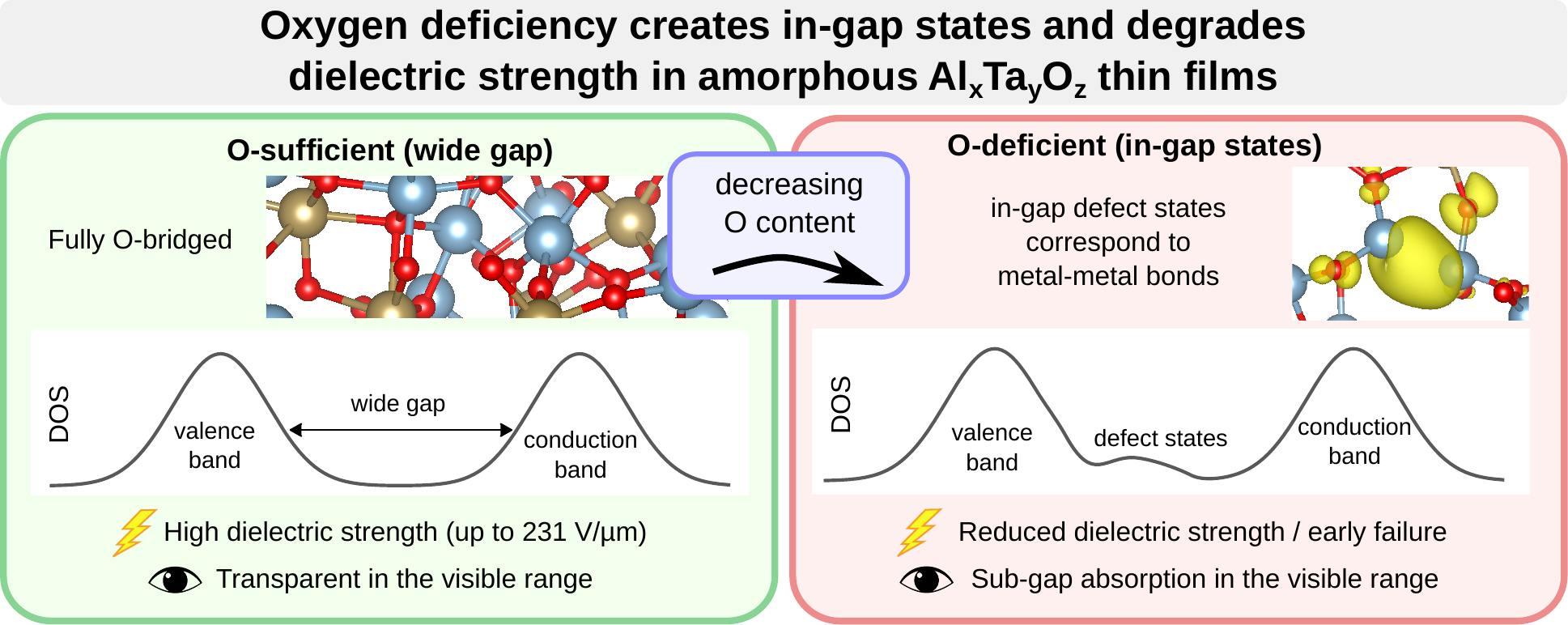}
\end{graphicalabstract}

\begin{highlights}
\change{
\item Oxygen content tunes Al–Ta–O breakdown from none to 231 V/µm.
\item DFT shows oxygen deficiency creates metal–metal motifs and in-gap states.
\item Valence-band XPS detects in-gap states in the most O-deficient films.
\item Absorption in visible range correlates with in-gap states in O-deficient films.
\item Breakdown degradation links to bonding topology in amorphous films, not crystallinity.
}
\end{highlights}

\author[inst1]{Pavel Ondračka}
\author[inst2]{Richard Drevet}
\author[inst1]{Daniel Franta}
\author[inst1]{Jan Dvořák}
\author[inst1]{Ivan Ohlídal}
\author[inst1]{Petr Vašina}

\address[inst1]{Department of Plasma Physics and Technology, Masaryk University, Kotlářská 2, Brno 61137, Czechia}
\address[inst2]{Institut de Thermique, Mécanique et Matériaux (ITheMM), EA 7548, Université de Reims Champagne-Ardenne (URCA), Bât.6, Moulin de la Housse, BP 1039, 51687 Reims Cedex 2, France}

\begin{abstract}
 This study reports on the optical, structural, and dielectric properties of aluminum tantalum oxide (\AlTaO{}) thin films deposited at low temperature on silicon and steel substrates by pulsed direct current reactive magnetron sputtering of a target containing 80 at.\% aluminum and 20 at.\% tantalum in Ar/O$_2$ atmosphere.
 Oxygen flow rates ranging from \qtyrange{5.0}{20}{\sccm} corresponded to O content changes in \qtyrange{57.7}{69.6}{\atpercent} and resulted in large differences in dielectric behavior, from films with no measurable dielectric strength to a dielectric strength of \qty{231}{\volt\per\micro\metre}, respectively.
 {\it Ab initio} calculations were employed to explain the large property changes, and we show that a decrease in the dielectric strength can be linked to the formation of metal-metal bonds in the material, when the O content is less than what would correspond to a stoichiometric \TaO{} and \AlO{} mixture.
 The electronic states corresponding to the metal--metal bonds are located in the band gap close to the top of the valence band, leading to an effective band gap reduction, which is directly supported by X-ray photoelectron spectroscopy valence band measurements and by a broad optical absorption in the visible region.

\end{abstract}

\end{frontmatter}

\section{Introduction}
Functional oxide thin films are key technological solutions to improve the surface properties of materials. They provide oxidation protection against many aggressive environments, and their chemical stability makes them suitable for long-term applications~\cite{Arunadevi2022, Chavali2019, Yang2019}. Their physicochemical properties are relevant for industrial applications requiring electrical or thermal insulation, resistance to mechanical stress, and optical transparency~\cite{Palumbo2019, AnjuBalaraman2022, Devaray2022}.

The most common binary oxides used in the industry for decades are \ce{SiO2}, \ce{TiO2}, \AlO{}, \TaO{}, \ce{ZnO}, \ce{HfO2}, \ce{Gd2O3}, \ce{ZrO2}, \ce{SnO2}, \ce{V2O5}, and \ce{Nb2O5}~\cite{Robertson2015, Robertson2004, McPherson2003, Verwey1990}. Among them, \AlO{} and \TaO{} thin films are particularly interesting for a wide range of technological devices. \AlO{} thin films can be easily produced at low cost by several deposition methods, and they are characterized by high band gap values in the range of \qtyrange{7}{9}{\eV} as a function of their crystalline state (amorphous, $\alpha$-\AlO{}, $\gamma$-\AlO{}, $\delta$-\AlO{}, $\theta$-\AlO{}, $\eta$-\AlO{}, or $\kappa$-\AlO{})~\cite{Pawowski1988}. They are highly transparent to visible light, and they are commonly used for electrical and thermal insulation of the surface of various materials~\cite{Musil2010, Eklund2009, Drevet2023}. \TaO{} thin films are also widely used for their thermal and chemical stability. They are described in the literature as having a high dielectric constant, high refractive index, and a band gap value of \qty{4.4}{\eV}~\cite{Chaneliere1998, AlmeidaAlves2018, Ezhilvalavan1999, Chandra2008}. These properties make \AlO{} and \TaO{} thin films relevant for dielectric applications in microelectronics (sensors, CMOS transistors, solar cells, energy storage capacitors, flash memory, piezoelectric actuators, etc.) and for optical applications (astronomical mirrors, car headlamps, camera lenses, optical sensors, etc.). However, further improving oxide thin-film performance remains a major scientific challenge for the development of new technological equipment~\cite{Dong2023}.

An innovative research direction consists of exploring new compositions of ternary oxide materials~\cite{Sivaperuman2024, Drevet2024, Chen2015, Qu2020}. They are reported to exhibit enhanced properties compared with their binary oxide counterparts. At present, much is known about the industrial performances and applications of ternary oxide thin films such as \ce{SrTiO3}, \ce{BaTiO3}, \ce{Al_{x}Si_{y}O_{z}}, and \ce{Al_{x}Ti_{y}O_{z}} among others~\cite{Gupta2017, Rebouta2015, ALRjoub2020, Xiao2018}. Recent experimental results have shown the impact of the concentration in aluminum, tantalum, and oxygen on the dielectric properties of \AlTaO{} thin films~\cite{Drevet2024}. \change{
However, the role of chemical bonding and oxygen-deficiency–induced defect states remains comparatively less explored. In particular, establishing clear links between bonding motifs, electronic structure, and dielectric breakdown remains important for optimizing these films.
}

The relationship between structure and the optical and dielectric properties of \AlTaO{} is investigated here using a combined theoretical and experimental approach. Thin \AlTaO{} films were deposited at low temperatures by reactive magnetron sputtering from an Al–Ta target under varying oxygen flow rates, and the resulting structure–property relationships were interpreted using density functional theory (DFT) calculations 
\change{
together with X-ray photoelectron spectroscopy (XPS) and optical spectroscopy. By correlating oxygen content, electronic structure, and dielectric breakdown behavior, we identify oxygen-deficiency–induced bonding motifs that give rise to in-gap states and degrade dielectric strength.}

\section{Materials and Methods}
\subsection{Oxide thin films deposition}
\AlTaO{} thin films were deposited by pulsed direct current reactive magnetron sputtering using the HVM Flexilab Y610 thin film deposition system (HVM Plasma Ltd.) equipped with three sputtering heads arranged in a confocal and top-down sputtering configuration. Double-sided polished Si (100) and steel substrates were ultrasonically cleaned for \qty{3}{\minute} in acetone and \qty{3}{\minute} in isopropyl alcohol. The deposition chamber was evacuated by pumping for two hours until reaching a background pressure value of below \qty{1}{\milli\Pa}. Argon ion-cleaning was used for \qty{15}{\minute} before deposition, using a \qty{13.56}{\mega\hertz} RF self-bias of \qty{200}{\volt} to remove surface contamination and oxidation from the substrates. The rotation of the sample holder was set to 5 rotations per minute (rpm). For the deposition, argon was continuously injected at a flow of \qty{60}{\sccm} (standard cubic centimeters per minute). Various oxygen flows ranging from \qtyrange{5}{20}{\sccm} were studied. The corresponding pressure ranged from \qtyrange{1.0}{1.2}{\Pa}. The temperature of the substrates was maintained at \qty{150}{\celsius} during the deposition. The sputtered 2-inch target was made of a mixture of \qty{80}{\atpercent} aluminum and \qty{20}{\atpercent} tantalum (Testbourne Ltd.). This target was powered by a Pinnacle Plus+ power supply (Advanced Energy) at \qty{250}{\watt}. The direct current was pulsed at a frequency of \qty{350}{\kilo\hertz} and a pulse reversal time of \qty{0.6}{\micro\second}. The target-to-substrate distance was \qty{87}{\mm}.

\subsection{Structure and composition}
The morphology and the chemical composition of the obtained thin films were studied from cross-sections of the samples with a Tescan MIRA3 scanning electron microscope (SEM) equipped with an Oxford Instruments X-MAXN 50 energy dispersive X-ray (EDX) spectrometer. The measurements were obtained by using an acceleration voltage of \qty{5}{\kV}.
The individual coatings (as well as the {\it ab initio} simulations) are referenced by their chemical compositions throughout the manuscript, in the form of \AlTaO{}, where $x$, $y$ and $z$ are the atomic concentrations (at.\%) of Al, Ta and O respectively. 
The structure of the oxide thin films was characterized by X-ray diffraction (XRD) with a Rigaku SmartLab Type F diffractometer using monochromatic Cu-K$\alpha$ radiation ($\lambda = \qty{0.15406}{\nm}$). The obtained XRD patterns were compared to the Powder Diffraction Files (PDF) provided by the International Centre for Diffraction Data (ICDD). The files of \TaO{} (\# 00-025-0922), Ta (\# 00-025-1280) and \AlO{} (\# 00-034-0493) were used in this study.

\subsection{Dielectric strength measurements}
The dielectric strengths of the thin films were measured on layers deposited on steel substrates using a pin-plate electrode configuration. The pin was in contact with the top of the dielectric thin film. The top and the bottom of the coated sample were connected to a DC power supply used to increase the voltage between the two electrodes progressively. Since oxide thin films are electrical insulators, there was no current flowing through the sample until the dielectric breakdown occurred. The corresponding voltage value was divided by the total thickness of the film to obtain the dielectric strength. The thicknesses of the films were determined from cross-section SEM images. The voltage measurements were repeated at ten different positions for each sample to obtain mean values and standard deviations.

\subsection{Optical characterization}
Optical characterization was based on a method of processing a set of heterogeneous data from several optical instruments in a wide spectral range. Experimental data were collected using the Woollam RC2 variable angle spectroscopic UV-VIS-NIR ellipsometer with two rotating compensators, Bruker Vertex 70v and 80v FTIR spectrophotometers,  Perkin Elmer Lambda 1050+ double-beam UV-VIS-NIR spectrophotometer, and McPherson VUVAS 1000 VUV spectrophotometer. The total spectral range covered was \qtyrange{0.003}{10.8}{\eV}.
Detailed descriptions of the experimental setups for these instruments can be found, for example, in the Ref.~\cite{Franta2025-Ta2O5}.
All ellipsometric and spectrophotometric measurements were fitted simultaneously using the Universal Dispersion Model~\cite{Franta2018}. This approach simplifies the construction of complex dispersion models that meet fundamental criteria for physical correctness, such as time-reversal symmetry, Kramers–Kronig relations, and adherence to sum rules.
In this model, the dielectric response is expressed as a sum of contributions representing
individual absorption processes. It allows not only the determination of optical constants but also features parameters with direct physical meaning, e.g., the optical band gap. Contributions from interband transitions and phonon contributions and absorption from defect states within the band gap were considered. Surface roughness of the samples was modeled with the Rayleigh-Rice theory.
\change{
For the optical constants used in the analysis below, we report the response of the top (inhomogeneous) \AlTaO{} sublayer, i.e., the near-surface gradient layer in the two-sublayer model.}
More details about the structural and dispersion model used for the fitting as well as fits of all optical measurements and the most important fit parameters can be found in the Supplementary Information (Section\,SI3).

\subsection{X-ray photoelectron spectroscopy (XPS)}
XPS valence band spectra were measured at a KRATOS Axis Supra X-ray Photoelectron Spectrometer equipped with an Al\,K$_\alpha$ source at \qty{1486.6}{\eV} and a hemispherical analyzer. Due to large differences between the dielectric strength of different samples, all samples were measured electrically isolated from the sample stage and charge neutralization was performed with a low energy electron source~\cite{Baer2020}.
The binding energy scale was later shifted using the signal of adventitious carbon at \qty{284.8}{\eV} as a reference. While we are aware of the shortcomings of this method, highlighted for example in~\cite{Greczynski2020a,Greczynski2020b}, no other option was available and we note that in the following analysis no conclusions are taken based on the precise binding energy values.
The spectra of valence band, O\,1s, Ta\,4f, Al\,2p, and C\,1s regions were obtained at \qty{20}{\eV} pass energy, \qty{0.05}{\eV} step and in 5 alternating sweeps. The total pressure during the measurements was lower than \qty{5}{\micro\pascal}.

\subsection{{\it Ab initio} modeling}
Amorphous \AlTaO{} models with sizes around 150 atoms were prepared by means of density functional theory~\cite{Hohenberg1964, Kohn1965} (DFT) with the molecular dynamics (MD) melt and quench scheme.
40 Al, 10 Ta and variable number of 70, 75, 80, or 85 O atoms, resulting in compositions \DFTone, \DFTtwo, \DFTthree, and \DFTfour, respectively, were randomly placed inside a cubic simulation cell with size corresponding to the initial mass density of \qty{4.3}{\gram\per\cm\cubed}, based on expected experimental values, and equilibrated at \qty{3000}{\K} to provide a randomized starting structure. The cells were subsequently cooled to \qty{300}{\K} over a \qty{2}{\ps} MD run and the atomic positions and cell sizes were later fully relaxed at \qty{0}{\K}. The temperature during the MD runs was controlled by a Nosé-Hoover thermostat~\cite{Nose1984} with an NVT ensemble. 
The MD runs and cell relaxations were performed at the generalized gradient approximation (GGA) level with the PBE exchange-correlation functional~\cite{Perdew1996}, while the electronic structure was calculated later with the PBE0 hybrid functional~\cite{Adamo1999}. The hybrid functional was chosen to overcome the shortcomings of the standard DFT GGA functionals in order to better describe the localized states in the band gap as well as to obtain a more accurate estimate of the band gap value. Further technical details of the DFT calculations are available in the supplementary information (SI), Section~SI1. The {\it ab initio} calculations are available under the Creative Commons license in the NOMAD Archive~\cite{NOMADdata, Scheidgen2023}.

\section{Results and discussion}

\subsection{Morphology, chemical composition, and structure}

\begin{figure*}
    \centering
    \includegraphics[width=\linewidth]{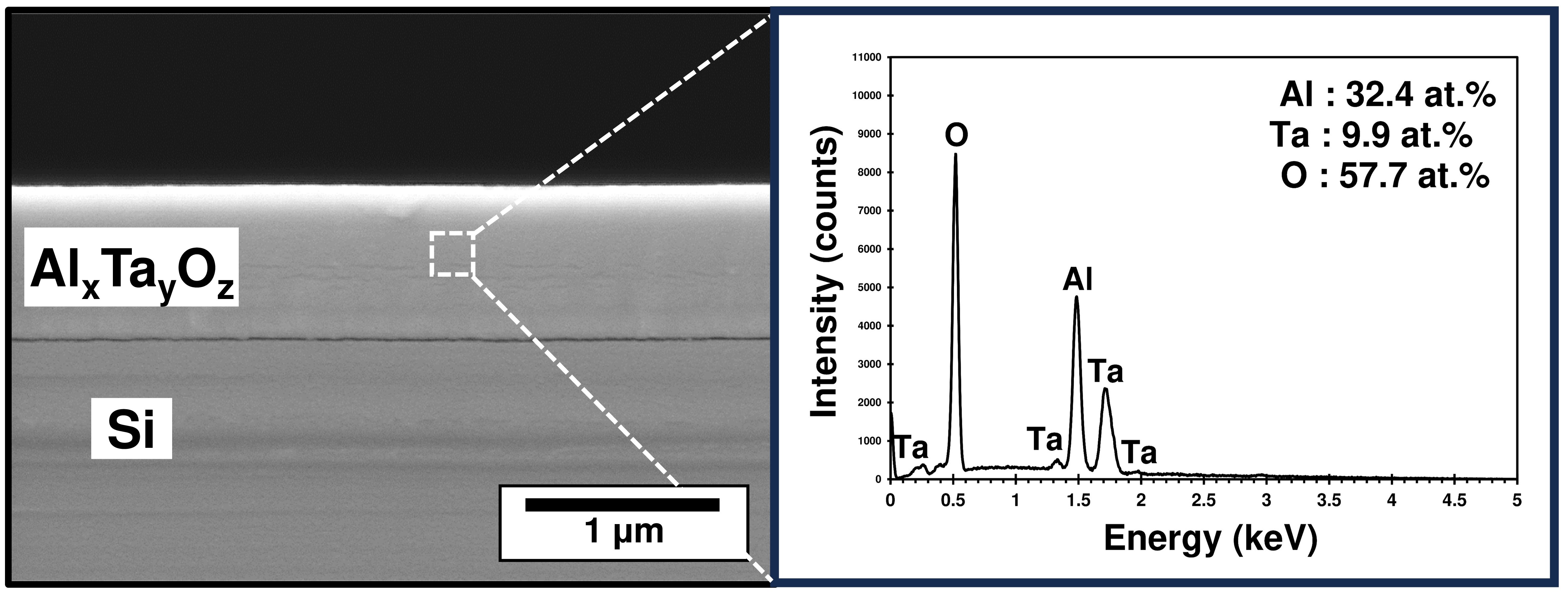}
    \caption{\label{fig:SEM}Typical cross-section SEM image and EDX spectrum of \AlTaO{} thin films deposited on a silicon substrate by pulsed-DC reactive magnetron sputtering at an oxygen flow of \qty{5}{\sccm}. The EDX spectrum corresponds to the indicated region.}
\end{figure*}

The cross-section SEM image of Figure~\ref{fig:SEM} shows the typical morphology of the thin films deposited on a silicon substrate by pulsed direct current reactive magnetron sputtering. They are dense and uniform with no specific growth direction observable at this scale. The EDX measurements indicate that the films contain aluminum, tantalum, and oxygen in different concentrations as a function of the oxygen flow used during the deposition (Table~\ref{tab:compositions}). The oxygen concentration in the thin film increases with the oxygen flow introduced in the deposition chamber. The tantalum concentration is barely impacted by the oxygen flow whereas the aluminum concentration decreases when the oxygen flow increases.
For all these experiments, the Al/Ta atomic ratio in the thin film is lower than that in the target. This result is due to the much lower atomic mass of aluminum ($m_\mathrm{Al} = \qty{26.98}{\atomicmassunit}$, where the unified atomic mass unit $\qty{1}{\atomicmassunit} = \qty{1.66e-27}{\kilogram}$) compared to that of tantalum ($m_\mathrm{Ta} = \qty{180.95}{\atomicmassunit}$)~\cite{Prohaska2022}. 
During deposition, tantalum, owing to its high atomic mass, follows a more ballistic (line-of-sight) trajectory from the target to the substrate~\cite{Feng2013}. In contrast, lighter aluminum atoms experience stronger scattering due to collisions with argon, leading to a broader spatial distribution within the chamber.
\change{
Consequently, the Al/Ta atomic ratio in the resulting films is systematically lower than in the target. The oxygen-flow dependent modulation could be explained on the basis of different deposition rates seen in Table~\ref{tab:compositions}. Target poisoning increases with increasing oxygen flow and thus the flux of fast sputtered neutrals decreases. This can reduce gas rarefaction (local depletion of Ar/O$_2$ near the target), which increases the local gas density and collision probability during transport. That preferentially penalizes Al as the lighter element and could lead to the observed further reduction in the relative Al/Ta content at higher O$_2$ flows.
}

\begin{table}
    \centering
     \caption{\label{tab:compositions}Chemical compositions and deposition rates of \AlTaO{} thin films deposited by pulsed direct current reactive magnetron sputtering at different oxygen flows ($\phi_{\mathrm{O_2}}$).}
    \begin{tabular}{cccccc}
    \hline
    $\phi_{\mathrm{O_2}}$ & Al      & Ta      & O       & Al / Ta & dep. rate\\
    (sccm)              & (at.\%) & (at.\%) & (at.\%) & & (\qty{}{\nm\per\minute})\\
    \hline\hline
    5& 32.4& 9.9& 57.7&3.3 & 55\\
    6& 29.3& 9.3& 61.4&3.1 & 54\\
    10& 25.4& 10.0& 64.6&2.5 & 38\\
    15& 19.9& 11.1& 69.0&1.8 & 18\\
    20& 19.3& 11.1& 69.6&1.7 & 15\\
    \hline
    \end{tabular}
\end{table}

\begin{figure}
    \centering
    \includegraphics[width=0.5\linewidth]{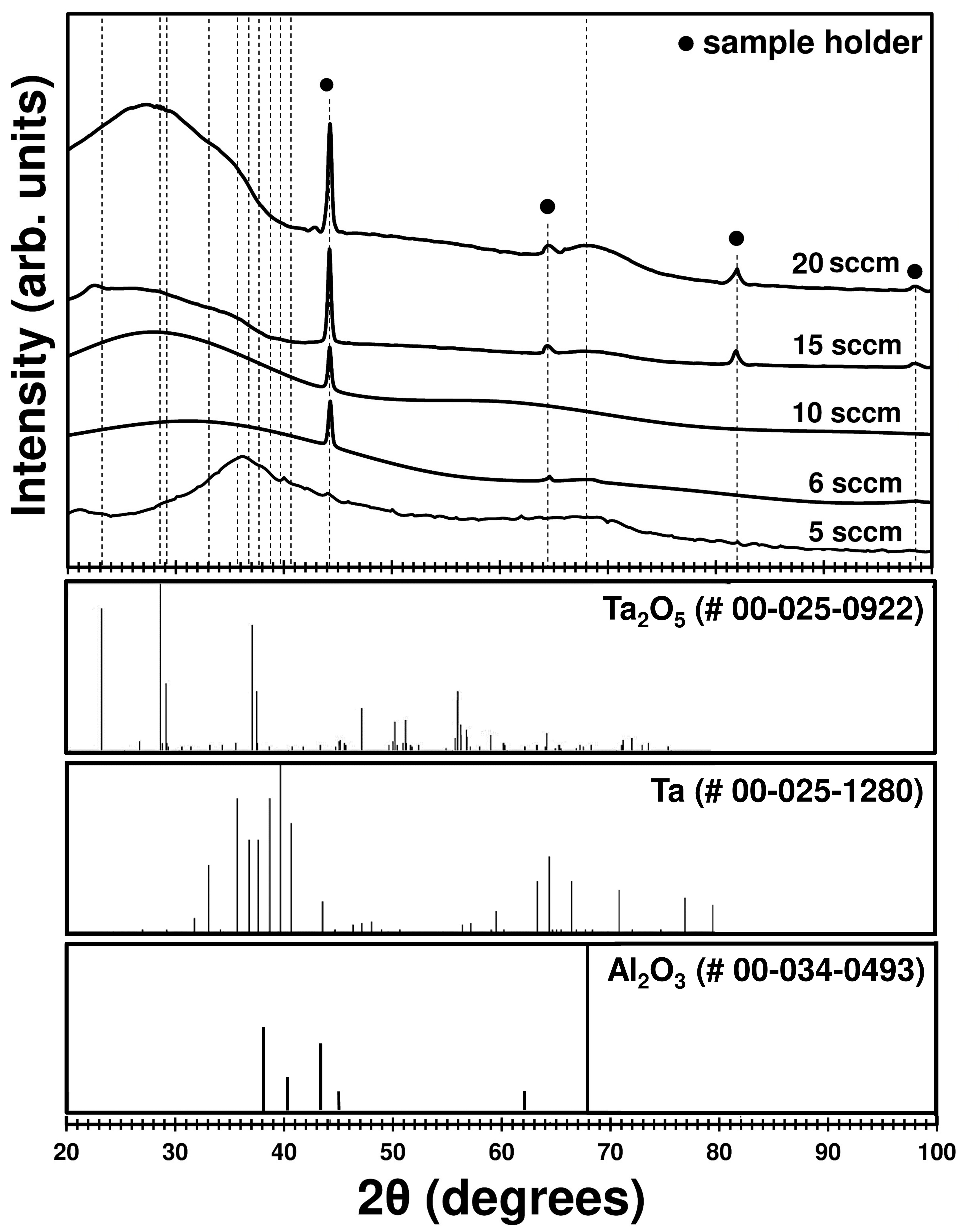}
    \caption{\label{fig:XRD}XRD patterns of \AlTaO{} thin films deposited by pulsed-DC reactive magnetron sputtering at different oxygen flows. Black dots indicate reflections from the sample holder; reference patterns are shown below.}
\end{figure}

The XRD patterns of the \AlTaO{} thin films are plotted in Figure~\ref{fig:XRD}. They show only very broad features at diffraction angles ranging from \qty{20}{\degree} to \qty{40}{\degree}. They indicate the very low degree of crystallization of the synthesized oxide thin films due to the low temperature used during the magnetron sputtering deposition~\cite{Frunz2014}. However, broad maxima are observable, providing information on the nature of the films as a function of the oxygen flow used for the deposition, i.e., as a function of the oxygen content in the oxide thin film. At low oxygen content, the main signal is in the region of diffraction angles ranging from \qty{35}{\degree} to \qty{40}{\degree}, which is the region of the seven main diffraction peaks of tantalum (PDF card \# 00-025-1280). This could be an indication of increased probability of metal–metal bonding motifs in the thin film because of the too-low oxygen flow used during the process~\cite{Panjan2020}. On the other hand, at high oxygen contents, the broad signal becomes more intense between \qty{25}{\degree} and \qty{30}{\degree} where the two main diffraction peaks of \TaO{} are positioned (PDF card \# 00-025-0922). This signal is barely observable by XRD, indicating that the \AlTaO{} thin films are mainly amorphous.

\begin{figure}
    \centering
    \includegraphics[width=0.5\linewidth]{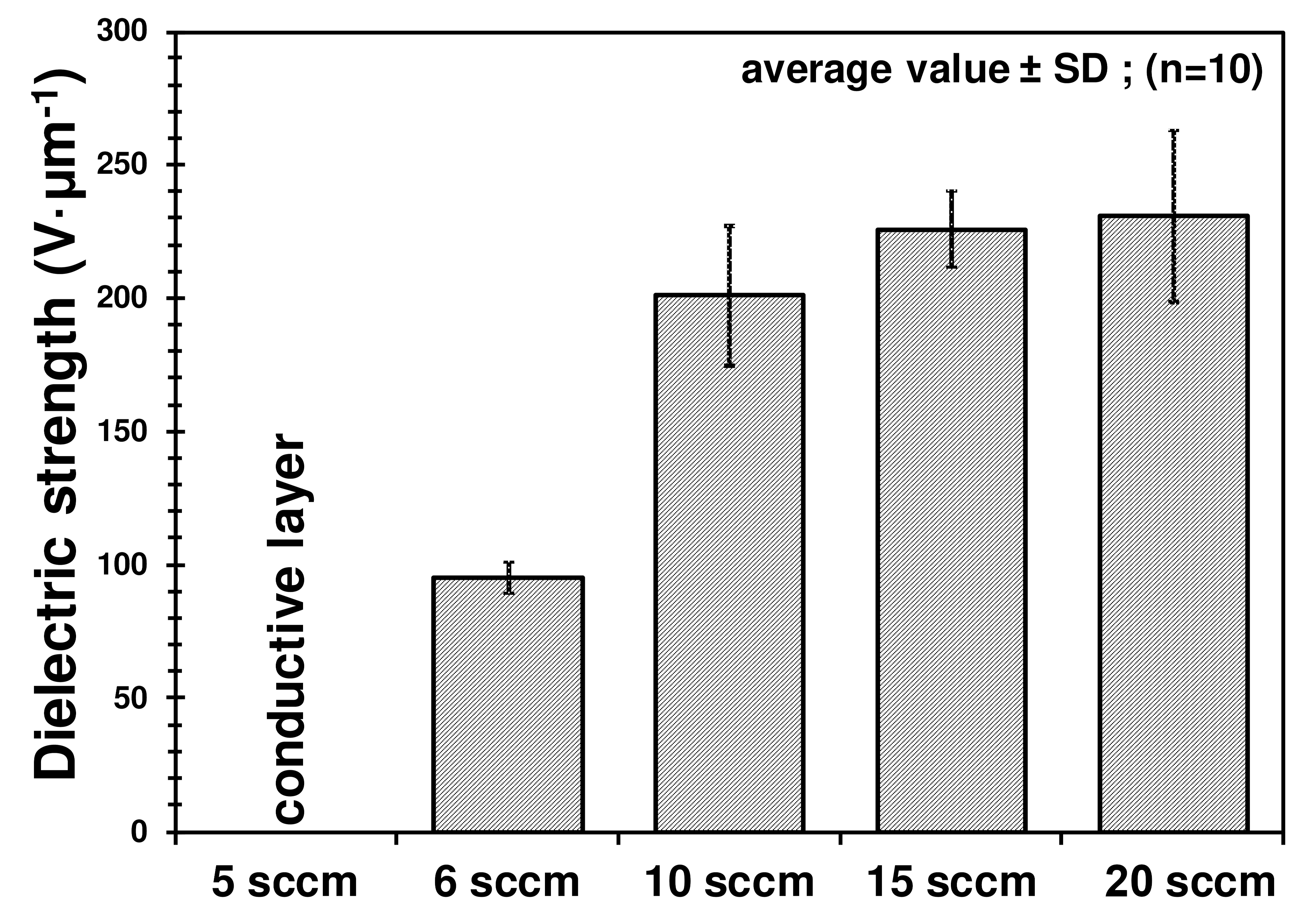}
    \caption{\label{fig:diel-strength}Dielectric strengths of \AlTaO{} thin films deposited by pulsed-DC reactive magnetron sputtering at different oxygen flow.}
\end{figure}

The dielectric strengths of all the \AlTaO{} thin films are reported in Figure~\ref{fig:diel-strength}.
\change{
Under our measurement protocol, the dielectric strength could not be determined for the film deposited at an oxygen flow of \qty{5}{\sccm} because current flowed immediately upon voltage application. This observation is consistent with the XRD patterns in Figure~\ref{fig:XRD}, which show features in the region of Ta reflections at low oxygen content.}
The presence of metallic tantalum clusters was reported to promote the propagation of electrical breakdown through the thin film although it is mainly made of an amorphous oxide~\cite{Budenstein1980, Wu2019, McPherson2002}. Dielectric strengths are measurable at higher oxygen content, and the value increases with the concentration of oxygen in the thin film. The dielectric strength of the thin film deposited at low oxygen flow (6\,sccm) is \qty{96}{\volt\per\micro\metre}, and this value increases to \qty{201}{\volt\per\micro\metre}, \qty{226}{\volt\per\micro\metre}, \qty{231}{\volt\per\micro\metre} for the three thin films deposited at 10\,sccm, 15\,sccm, and 20\,sccm, respectively.

\subsection{XPS}

\begin{figure}[htb!]
    \centering
    \includegraphics{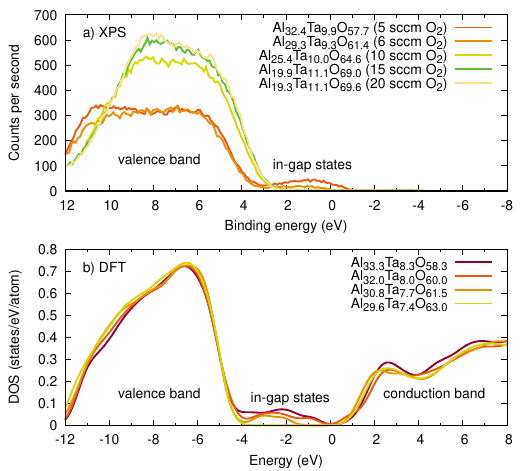}
    \caption{\label{fig:DOSXPS} a) XPS valence band spectra. b) Density of states calculated for the amorphous \AlTaO{} structures with variable O content. Gaussian broadening of \qty{0.3}{\eV} was applied. DOS plots were aligned by the valence band edge for visualization purposes.}
\end{figure}

\change{
XPS core-level and valence-band spectra were measured for all films. The complete core-level data (O 1s, Ta 4f, Al 2p and C\,1s) are provided in the Supporting Information (Fig. SI1); they show no additional features beyond those previously reported for oxygen-deficient \AlTaO{}  films, where a metallic Ta component in Ta 4f appears at low oxygen flows and correlates with degraded dielectric performance and, in some cases, increased conductivity ~\cite{Drevet2024,Drevet2025}. Consistent with these reports, a metallic Ta contribution is present in \Ttwo{} and \Tthree{}, while the O\,1s and Al\,2p spectra remain largely unchanged across samples. Here, we therefore focus on the valence band (Fig.~\ref{fig:DOSXPS}a) to identify additional spectral signatures associated with the degraded dielectric response of oxygen-deficient films.
}

The most notable feature in the valence band XPS spectra is the emergence of in-gap states extending above the valence-band edge for the two samples with the lowest O content, i.e., \Ttwo{} and \Tthree{}.
The states are very broad and stretch up to $\sim\!\qty{4}{\eV}$ and $\sim\!\qty{3}{\eV}$ above the valence band edge for samples \Ttwo{} and \Tthree{} respectively. A maximum in the defect state intensity is observed for \Tthree{} at $\sim\!\qty{2.5}{\eV}$ from the valence band edge. However, no pronounced Fermi edge is observed in either case. In the spectrum of \Ttwo{}, nonzero intensity is observed above the nominal \qty{0}{\eV} level in our plot, indicating uncertainty in the binding-energy referencing based on adventitious carbon. Previous work has shown that adventitious-carbon referencing may not reliably align spectra to the Fermi level and can instead align closer to the vacuum level~\cite{Greczynski2020a,Greczynski2020b}. However, even the sample \Ttwo{}, despite not showing any measurable dielectric strength during the dielectric testing, was not conductive enough to prevent charging during XPS measurements, and no better charge referencing option was available. We furthermore reiterate that the energy alignment was done mostly for the visualization purposes and that no later conclusions are based on the exact energy positions. 
The valence-band shape and intensity vary between samples, which we attribute to overall intensity fluctuations (e.g., differences in adventitious carbon content) and to composition changes, specifically the Ta/Al ratio. 
The origin and properties of the defect states in the band gap are examined in more detail using the ab initio results in Section~\ref{sec:ab}.

\subsection{Optical properties}

\begin{figure}[hbt!]
\centering
\includegraphics[width=0.6\textwidth]{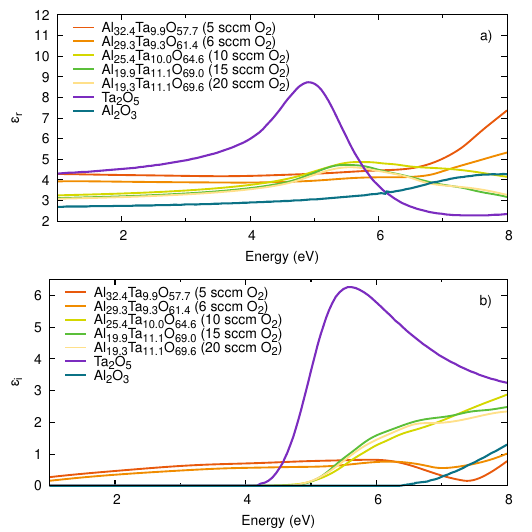}
\caption{\label{fig:opticsvis} Real a) and imaginary b) parts of the complex dielectric function in the NIR-VUV spectral region with reference spectra of amorphous \TaO{} and \AlO{}.}
\end{figure}

The measured complex dielectric functions for all coatings in the NIR-VUV spectral region are shown in Figure~\ref{fig:opticsvis} with spectra of amorphous \TaO{}\cite{Franta2025-Ta2O5} and amorphous \AlO{}~\cite{Franta2015-Al2O3} added for comparison. The absorption in this region is caused by electronic transitions between the occupied valence and empty conduction bands.
\change{
A notable difference in the visible region between samples deposited at different oxygen flows is the presence of absorption at energies below $\sim\!\qty{4.5}{\eV}$. Specifically, the coatings deposited at 5 and 6\,sccm O$_2$ exhibit significant absorption across the visible range, whereas the other films show an absorption onset --- determined from the $E_\mathrm{g}$ parameter of the dispersion model --- in the \qtyrange{4.6}{4.8}{\eV} range. The absorption in the visible range can therefore be attributed to transitions involving low-lying states in the band gap observed in the XPS valence-band measurements. Consistent with XPS, only the two samples with the lowest oxygen content, which exhibit states within the band gap, also show visible-range absorption.}
The absorption in the visible range also leads to a significantly higher real part of the dielectric function in the visible range and an anomalous dispersion that is only observed for the samples deposited at 5 and 6 sccm O$_2$ flow rate.
For three samples with high O content, the optical band gap, as determined from the dispersion model parameter $E_\mathrm{g}$ value, lies between the values reported for amorphous \TaO{} (\qty{4.1}{\eV}~\cite{Franta2025-Ta2O5}, \qty{4.2}{\eV}~\cite{Shvets2008}, \qtyrange{4.2}{4.3}{\eV}~\cite{Sertel2019}) and amorphous \AlO{} (\qty{6.44}{\eV}~\cite{Franta2015-Al2O3}); however, it is closer to the amorphous tantala value, even though the tantalum content is only $\sim$\qty{20}{\percent} of metal atoms.
This is expected, because the lowest-lying conduction-band states are predominantly Ta-derived.
Similar behavior was for example shown for amorphous \ce{Ti_{x}Si_{1-x}O_2} mixed oxide, where the band gap deviates from the band gap of \ce{TiO2} by less than \qty{0.2}{\eV} in most parts of the compositional range (i.e., for $x \ge 0.33$ concentrations)~\cite{Ondraka2017}.

\begin{figure}[hbt!]
\centering
\includegraphics[width=0.6\textwidth]{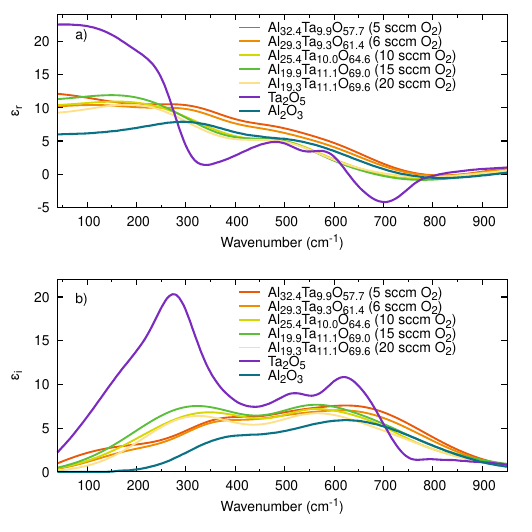}
\caption{\label{fig:opticsir} Real a) and imaginary b) parts of the complex dielectric function in the IR spectral region.}
\end{figure}

The complex dielectric functions of all coatings in the infrared spectral region are shown in Figure~\ref{fig:opticsir} with spectra of \TaO{}\cite{Franta2025-Ta2O5} and \AlO{}~\cite{Franta2015-Al2O3} added again for comparison.
Absorption in this region originates from lattice vibrations.
The spectra of all samples exhibit very broad features with
only minor differences.
Based on comparison with the \TaO{} and \AlO{} spectra, we can loosely attribute the broad absorption peak around \qty{300}{\per\cm} to vibrations of \ce{Ta-O} groups, absorption above $\sim\!\qty{700}{\per\cm}$ to \ce{Al-O} groups, and the absorption in between to mixed \ce{Ta-O} and \ce{Al-O} modes.
The changes between samples mostly pertain to an overall shift of the spectral weight to higher wavenumbers and to much less pronounced \TaO{}-like features for the two films with low O content. This can be explained simply by the changes in the Al/Ta ratio at different O flows, specifically by the lower Ta content in the two films with low O content. 

We note that additional vibrational frequencies might be expected from the vibrations of the metal atoms in O-deficient environment in the O-deficient samples. Such vibrations would be theoretically expected at lower wavenumbers, since the metals have heavier mass as compared to oxygen and the metallic bonds are expected to be weaker, however, no such features are visible in the spectra. It is possible such modes are not visible due to the small number of metal–metal bonds and large overall broadening of the IR spectra. Additionally, both fcc Al and bcc Ta have no IR active phonons (even disregarding the fact that they would be screened by the free electrons), therefore, even with the symmetry breakage in the amorphous structure, the IR optical activity of the metallic clusters might be low in the first place.

\change{
Importantly, no free-electron absorption was detected in the far-IR region, even for the sample with the lowest oxygen content that appeared conductive in the dielectric-strength measurements. This is consistent with the valence-band XPS, where no pronounced Fermi edge was observed. One possible explanation is that the film remains overall insulating but its breakdown strength is below the reliable operating range of our setup, which is mainly intended for high-breakdown materials and may be less accurate at low applied voltages. Another possible explanation is that conductive regions exist but occupy a low volume fraction, so they do not produce a clear Drude-like response in the volume-sensitive optical measurements.}

While it is not possible to directly estimate the static dielectric function $\varepsilon_\mathrm{r,0}$ from the optical measurements, the dielectric function at the lowest measured wavenumber (\qty{40}{\per\cm}) $\varepsilon_\mathrm{r,lf}$ can be taken as its indicator and an approximation (albeit possibly underestimated) of it. All the here measured coatings have the low frequency dielectric function around 10 and no trends with respect to the oxygen content can be observed. While the two films with lowest O content have higher electronic contribution to the $\varepsilon_\mathrm{r,lf}$ due to the absorption in the visible range, their spectral weight in the IR region is shifted to higher energies and \TaO{}-like feature around \qty{300}{\per\cm} is much less pronounced, resulting in comparable $\varepsilon_\mathrm{r,lf}$ for all coatings.
Therefore, we estimate that the static dielectric constant is comparable for all of the films.

\subsection{{\it Ab initio} calculations}
\label{sec:ab}

\begin{figure}
    \centering
    \includegraphics[width=16cm]{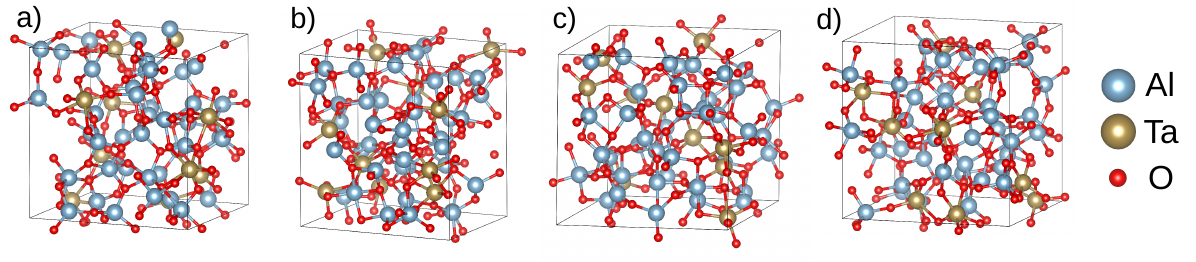}
    \caption{\label{fig:struct} Amorphous structures of a) \DFTone{} b) \DFTtwo{} c) \DFTthree{} d) \DFTfour{} with highlighted metal--O bonds. Visualized with VESTA~\cite{momma_vesta_2011}.}
\end{figure}

\begin{figure}
    \centering
    \includegraphics[width=10cm]{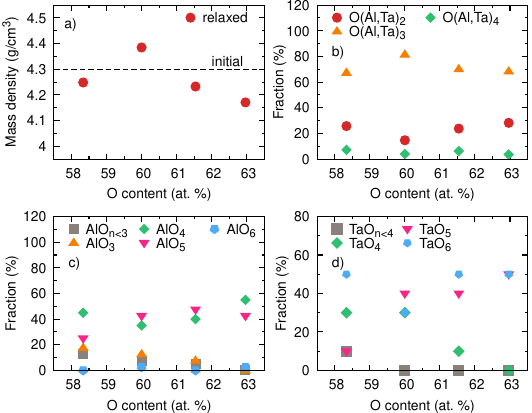}
    \caption{\label{fig:coord} a) Mass density for the DFT models. Distribution of local coordination states for b) O, c) Al, d) Ta atoms.}
\end{figure}

The amorphous \AlTaO{} structures prepared using the melt \& quench simulated annealing procedure are visualized in Figure~\ref{fig:struct}.

Mass density for the most common \AlO{} crystalline phases spans the \qtyrange{3.6}{4.0}{\gram\per\cm\cubed} range~\cite{horton_accelerated_2025} and \qtyrange{2.66}{3.4}{\gram\per\cm\cubed} for the amorphous phase~\cite{Shi2019}.
\TaO{} mass density for the ambient pressure crystalline phases is reported in the range \qtyrange{8.18}{8.41}{\gram\per\cm\cubed}~\cite{askeljung_effect_2003} and in the \qtyrange{7.0}{8.2}{\gram\per\cm\cubed} range for the amorphous phase~\cite{Capilla2011}. Using the reported density ranges for both \AlO{} and \TaO{} and making a simple interpolation with Vegard's law, the estimated mass density range for the calculated amorphous compositions with the 0.2 Ta/Ta+Al ratio would be \qtyrange{3.5}{4.4}{\gram\per\cm\cubed}. This is indeed in good agreement with the mass densities of the final relaxed simulated cells shown in Figure~\ref{fig:coord}a) which spans the range \qtyrange{4.16}{4.39}{\gram\per\cm\cubed}. The simulated cells are in the higher part of the density range, which is expected, as in most cases the low values of reported experimental mass density of the amorphous phase correspond to porosity, which is not present in the simulated structures due to the melt \& quench scheme specifics. There is no clear trend of the simulated mass density with respect to the O content and in fact it appears mostly constant with some random fluctuations due to the random nature of the melt \& quench scheme.

The predicted distributions of the local coordination numbers for the amorphous structures are shown in Figure~\ref{fig:coord} b), c) and d) for O, Al, and Ta atoms respectively. The Al--O, Ta--O, and O--Al/Ta coordination numbers were calculated based on cutoff distances corresponding to the first minima in the partial pair distribution functions. The used cutoffs were \qty{2.5}{\angstrom} and \qty{2.3}{\angstrom} for \ce{Ta-O} and \ce{Al-O}, respectively. We note that for this analysis the metal-metal coordination in the O-deficient structures was not considered. This is due to the difficulty of determining a metal-metal bond based on the distance only, for example, there are cases where \ce{Al-Al} at distance of $\sim\!\qty{2.6}{\angstrom}$ form a metallic bond (as will be shown using more sophisticated analysis later), however there are other Al atoms with the same distance which are in fact not forming direct bond but are connected by a bridging oxygen instead. Therefore, in the following analysis only metal--O coordination numbers are discussed, while the metal-metal interactions will be discussed later.

Crystalline \AlO{} forms a large variety of different structures~\cite{Levin1998}, with the most stable being corundum-structured $\alpha$-alumina. Most of the polymorphs are composed of face, edge or corner sharing AlO$_6$ octahedra (distorted to various extends) or, in some cases, a mix of AlO$_6$ octahedrons and AlO$_4$ tetrahedrons. In amorphous \AlO{}, the local structure is dominated by 4- and 5-coordinated Al atoms, with a minor fraction of 6-coordinated ones. For example, vapor-deposited thin films show $\sim$\qty{55}{\percent} AlO$_4$, $\sim$\qty{42}{\percent} AlO$_5$, and only $\sim$\qty{3}{\percent} AlO$_6$ by $^{27}$Al NMR~\cite{lee_structure_2009}. In this regard, the stoichiometric \AlTaO{} model with 63\, at.\% of O is consistent with these Al coordination statistics. However, at lower O contents we see an increase of Al atoms with a smaller number of O neighbors, particularly AlO$_3$, AlO$_2$, AlO$_1$, and even Al atoms with a metal first coordination shell only. While the some AlO$_3$ building blocks are still part of the fully metal-O bridging networks, others, as well as all Al atoms with one or two O neighbors are forming metal–metal bonds to compensate.

\TaO{} does not form crystalline structures as readily, so the experimentally-reported structures consist of $\alpha$ and $\beta$-\TaO{} and the Z high pressure phase. Those structures are composed either of TaO$_6$ octahedra or TaO$_7$ pentagonal bipyramids with 3 or 4-coordinated O atoms~\cite{horton_accelerated_2025}. Recent X-ray pair distribution and PDF studies find that Ta coordination in a-\TaO{} is reduced to about \numrange{4}{5}, versus \numrange{6}{7} in crystalline phases~\cite{bassiri_order_2015, martinelli_deep_2021} and report 2 or 3-coordinated O atoms. Therefore Ta atomic environment in the modeled \AlTaO{} cells at the simulated densities with significant amount of Ta atoms with 6 oxygen neighbors and similarly the non-negligible amount of 4-coordinated O falls toward the higher-coordination end of reported ranges. However, it must be noted that the Ta atom statistics have large uncertainty due to the small amount of Ta atoms in the simulated structures.

\change{
Overall, the coordination-number distributions are consistent with reported local motifs in amorphous \AlO{} and \TaO{}, providing a basic structural consistency check for the melt–quench models prior to the electronic-structure analysis below.}

Calculated density of states with the PBE0 hybrid functional for different O contents at the fixed Ta/(Al + Ta) ratio of 0.2 is presented in Figure~\ref{fig:DOSXPS}~b). The overall shape and position of both the valence and conduction band are mostly constant with changes in the O content. However, a pronounced change is the appearance of in-gap states near the valence-band edge for structures with O content lower than what would correspond to the (\AlO{})$_{0.8}$(\TaO{})$_{0.2}$ mixture, i.e., less than $\sim\!\qty{63.0}{\atpercent}$.
As the O content decreases, the states appear deeper within the band gap and the difference between the highest occupied and lowest unoccupied state is \qtylist{3.3;1.5;1.7}{\eV} for models with \qtylist{61.5;60.0;58.3}{\atpercent}~O, respectively. This is in stark contrast to the value of \qty{5.7}{\eV} for the stoichiometric \DFTfour, although it must be noted that the HOMO--LUMO gap is not a good approximation of the band gap in amorphous materials in general due to localized tail states near the band edges and finite-size statistics in small supercells~\cite{Ondraka2017, Dicks2017}.
Additionally, due to the small cell sizes, there are just a few states in the band gap and, therefore, the obtained HOMO--LUMO gap values are to a large extent influenced by the random nature of the amorphous cells.
The calculated HOMO--LUMO gap values both for the substoichiometric and stoichiometric compositions are slightly larger than the experimental optical band gap values discussed previously. This is not unexpected, as the optical absorption onset is shifted to smaller energies due to excitonic effects. Specifically, the binding energy of the lowest exciton was found to be \qty{0.3}{\eV} in amorphous tantala~\cite{Lee2014} and \qty{0.4}{\eV} in $\alpha$-\AlO{}~\cite{Marinopoulos2011}.


\begin{figure}
    \centering
    \includegraphics[width=0.65\linewidth]{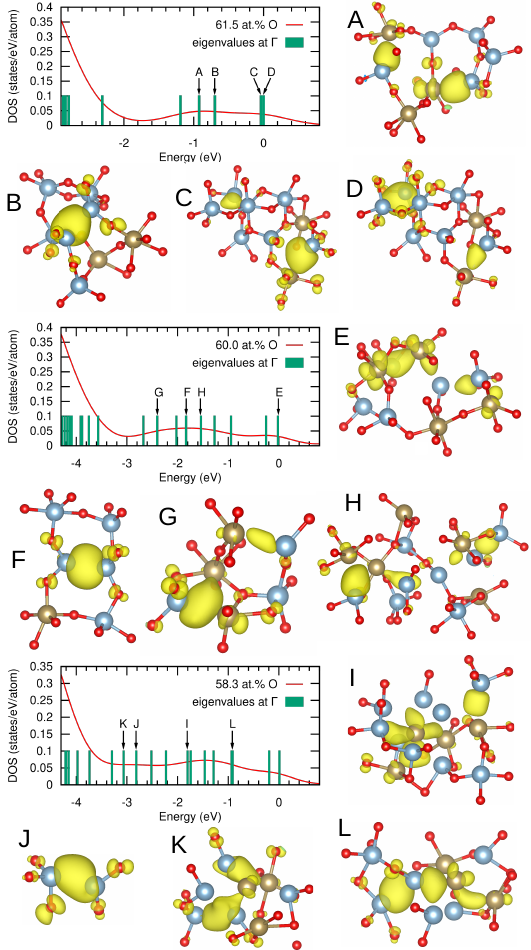}
    \caption{\label{fig:states-details} Details of the density of the defect states within the band gap for the three sub-stoichiometric models. Gaussian broadening of \qty{0.3}{\eV} was applied. Zero energy is at the position of the highest occupied state. Bars denote the Kohn-Sham eigenvalues at the $\Gamma$ point. Charge density distribution corresponding to selected states are also plotted. The yellow-colored isosurfaces are drawn at \qty{0.025}{\elementarycharge\per\angstrom\cubed}. }
\end{figure}

The distribution of the states within the band gap for the three substoichiometric compositions is shown in Figure~\ref{fig:states-details}. The eigenvalues shown correspond to the $\Gamma$ point. For the structures with \qtylist{61.5;60.0;58.3}{\atpercent}~O (missing 5, 10 and 15 O atoms with respect to the perfect mixture of stoichiometric \TaO{} and \AlO{}) this corresponds to 5, 9 and 13 states in the band gap.
Therefore, the number of defect states appearing is consistent with the expectation from the simple bonding model, i.e., each missing oxygen introduces two electrons; in our non-spin-polarized models this corresponds approximately to one additional occupied metal-derived state in the gap, consistent with the observed state counts.
For the case of \qty{61.5}{\atpercent} of O, the defects are located within the \qtyrange{1.6}{1.8}{\eV} range from the edge of the valence band (although the edge is also not that well defined). The state at \qty{-2.3}{\eV} is not a metal--metal-bonding state and is just strained valence O state, i.e., a localized valence-band tail state. As the amount of oxygen reduces, the distribution of the defect states broadens and shifts slightly deeper into band gap. As the amount of oxygen is further reduced, the state distribution appears to be more homogeneously spaced, with the highest in-gap states extending up to $\sim\!\qty{3.6}{\eV}$ above the valence band edge. The DFT results are therefore qualitatively consistent with the XPS valence-band measurements, reproducing the emergence of in-gap states for the oxygen-deficient compositions.

Charge density iso-surfaces corresponding to the specific Kohn-Sham states lying within the band gap are shown in the Figure~\ref{fig:states-details} as well. In all cases, the states lying in the band gap are located between neighboring metal atoms; thus, they effectively contribute to metal--metal bonding. They can be either localized on a single pair or triplet, like in Figure~\ref{fig:states-details}~B, F or J, however, that seems to be relatively rare. The defect states are most often delocalized on two or more metal pairs/triplets like in Figure~\ref{fig:states-details}~C or D. In fact with the decreasing O content the metal states become even more delocalized, effectively being located on a chain of atoms connected by metal bonds or a metal atoms cluster, e.g.,  Figure~\ref{fig:states-details}~E, G, H, I, K, or L. If we define number of metal atoms where the state is localized as number of metal atoms in the vicinity of a significant charge density of the specific state (closer than the covalent bond radius and density larger than \qty{0.025}{\elementarycharge\per\angstrom\cubed} which is the density value used for the isosurfaces plotting), then the average number of metal atoms a defect state is localized onto is \qtylist{4.0;5.4;6.3} for structures with \qtylist{61.5;60.0;58.3}{\atpercent}~O, respectively.
Therefore, we propose the creation of metallic chains/clusters and the presence of increasingly delocalized metal-derived in-gap states close to the top of the valence band as a likely origin of the earlier dielectric failure at lower oxygen contents

Regarding the specific atoms associated with the defect states, where ''associated'' refers to atoms in the vicinity of pronounced defect-state charge-density localization, the Ta to Ta+Al ratio is \qtylist{0.27;0.44;0.23} for the structures with \qtylist{61.5;60.0;58.3}{\atpercent}~O, respectively.
Considering that the overall Ta to Ta+Al ratio is \qty{0.20}, the Ta atoms appear over-represented in these cells when it comes to probability that a specific metal atom will be associated with the defect-state localization. This would suggest that Ta atoms are more likely to contribute to the defect states and thus the defect states (and the dielectric breakdown in general) may be associated with Ta-rich metallic motifs. This is consistent with the XPS observation that no metallic component could be observed in the Al\,2p XPS core-level spectra, as opposed to the Ta\,4f where the Ta metallic bonds are clearly visible. However, given the random nature of the amorphous structures, small atomic sizes of our models (only 40 Al and 10 Ta atoms) and also limited cooling speeds, we believe larger simulation models are needed to definitely confirm this.

\section{Conclusions}
Amorphous oxide thin films of \AlTaO{} were synthesized by pulsed direct current reactive magnetron sputtering. They are dense and uniform with various concentrations in aluminum, tantalum, and oxygen as a function of the oxygen flow inside the chamber during the deposition. The dielectric strength of the \AlTaO{} thin films is impacted by their chemical composition. In the here studied range the highest dielectric strength was observed for the sample composition \Tfive{} deposited at \qty{20}{\sccm} O$_2$ flow with a value of \qty{231}{\volt\per\micro\metre}.
\change{
The reduced dielectric performance of the low-oxygen films was linked to defect states near the top of the valence band extending into the band gap, as directly observed by valence-band XPS. In the dielectric-strength test, \Ttwo{} exhibited immediate current flow (no measurable breakdown field), whereas \Tthree{} showed a substantially reduced but measurable dielectric strength.
However, neither a Drude-like free-carrier response in the far-IR nor a pronounced Fermi edge in valence-band XPS is observed, suggesting that \Ttwo{} is not a uniformly metallic conductor.}
{\it Ab initio} calculations revealed that those states are metal states---localized on metal atoms---and we also show that they are present for all models where the O content is below the O content corresponding to mixture of \AlO{} and \TaO{}, in good agreement with the experimental results. The defect states and their properties were further confirmed with optical measurements, where they cause pronounced absorption over in the NIR-VIS range.

\section*{Acknowledgment}
This work was supported by the Ministry of Education, Youth and Sports of the Czech Republic (MEYS CR) through the projects LM2023039 and e-INFRA CZ (ID:90254). This research was supported by the TREND project FW06010462 from the Technology Agency of the Czech Republic. CzechNanoLab project LM2023051 funded by MEYS CR is gratefully acknowledged for the financial support of the XPS and optical measurements at CEITEC Nano Research Infrastructure.

\section*{Data availability}
\change{
All ab initio calculation input and output files, as well as raw XPS data and the corresponding processed datasets in NeXus and VAMAS formats, are available in the NOMAD Archive under the CC BY 4.0 license~\cite{NOMADdata, NOMADdataXPS, Scheidgen2023}. All other experimental data supporting this study are available from the corresponding author upon reasonable request.}

\section*{CRediT authorship contribution statement}
Pavel Ondračka: Conceptualization, Formal analysis, Investigation, Methodology, Writing - original draft, review \& editing, Visualization.
Richard Drevet: Conceptualization, Formal analysis, Investigation, Methodology, Writing - original draft, review \& editing, Visualization.
Daniel Franta: Investigation, Formal analysis, Methodology, Writing - original draft, review \& editing, Visualization.
Jan Dvořák: Investigation, Formal analysis, Writing - original draft, review \& editing.
Ivan Ohlídal: Supervision, Writing - review \& editing.
Petr Vašina: Supervision, Writing - review \& editing, Funding acquisition.

\bibliographystyle{elsarticle-num} 
\bibliography{bib-db}

\begin{thebibliography}{10}
\expandafter\ifx\csname url\endcsname\relax
  \def\url#1{\texttt{#1}}\fi
\expandafter\ifx\csname urlprefix\endcsname\relax\def\urlprefix{URL }\fi
\expandafter\ifx\csname href\endcsname\relax
  \def\href#1#2{#2} \def\path#1{#1}\fi

\bibitem{Arunadevi2022}
N.~Arunadevi, S.~J. Kirubavathy, Chapter 2 - metal oxides: Advanced inorganic
  materials, in: C.~Verma, J.~Aslam, C.~M. Hussain (Eds.), Inorganic
  Anticorrosive Materials, Elsevier, 2022, pp. 21--54.
\newblock \href
  {https://doi.org/https://doi.org/10.1016/B978-0-323-90410-0.00002-7}
  {\path{doi:https://doi.org/10.1016/B978-0-323-90410-0.00002-7}}.

\bibitem{Chavali2019}
M.~S. Chavali, M.~P. Nikolova, Metal oxide nanoparticles and their applications
  in nanotechnology, SN Applied Sciences 1~(6) (2019).
\newblock \href {https://doi.org/10.1007/s42452-019-0592-3}
  {\path{doi:10.1007/s42452-019-0592-3}}.

\bibitem{Yang2019}
L.~Yang, X.~Kong, F.~Li, H.~Hao, Z.~Cheng, H.~Liu, J.-F. Li, S.~Zhang,
  Perovskite lead-free dielectrics for energy storage applications, Progress in
  Materials Science 102 (2019) 72–108.
\newblock \href {https://doi.org/10.1016/j.pmatsci.2018.12.005}
  {\path{doi:10.1016/j.pmatsci.2018.12.005}}.

\bibitem{Palumbo2019}
F.~Palumbo, C.~Wen, S.~Lombardo, S.~Pazos, F.~Aguirre, M.~Eizenberg, F.~Hui,
  M.~Lanza, A review on dielectric breakdown in thin dielectrics: Silicon
  dioxide, high‐k, and layered dielectrics, Advanced Functional Materials
  30~(18) (2019).
\newblock \href {https://doi.org/10.1002/adfm.201900657}
  {\path{doi:10.1002/adfm.201900657}}.

\bibitem{AnjuBalaraman2022}
A.~Anju~Balaraman, S.~Dutta, Inorganic dielectric materials for energy storage
  applications: a review, Journal of Physics D: Applied Physics 55~(18) (2022)
  183002.
\newblock \href {https://doi.org/10.1088/1361-6463/ac46ed}
  {\path{doi:10.1088/1361-6463/ac46ed}}.

\bibitem{Devaray2022}
P.~Devaray, S.~F. W.~M. Hatta, Y.~H. Wong, An overview of conventional and new
  advancements in high kappa thin film deposition techniques in metal oxide
  semiconductor devices, Journal of Materials Science: Materials in Electronics
  33~(10) (2022) 7313–7348.
\newblock \href {https://doi.org/10.1007/s10854-022-07975-7}
  {\path{doi:10.1007/s10854-022-07975-7}}.

\bibitem{Robertson2015}
J.~Robertson, R.~M. Wallace, {High-K materials and metal gates for CMOS
  applications}, Materials Science and Engineering: R: Reports 88 (2015)
  1–41.
\newblock \href {https://doi.org/10.1016/j.mser.2014.11.001}
  {\path{doi:10.1016/j.mser.2014.11.001}}.

\bibitem{Robertson2004}
J.~Robertson, High dielectric constant oxides, The European Physical Journal
  Applied Physics 28~(3) (2004) 265–291.
\newblock \href {https://doi.org/10.1051/epjap:2004206}
  {\path{doi:10.1051/epjap:2004206}}.

\bibitem{McPherson2003}
J.~McPherson, J.~Kim, A.~Shanware, H.~Mogul, J.~Rodriguez, Trends in the
  ultimate breakdown strength of high dielectric-constant materials, IEEE
  Transactions on Electron Devices 50~(8) (2003) 1771–1778.
\newblock \href {https://doi.org/10.1109/ted.2003.815141}
  {\path{doi:10.1109/ted.2003.815141}}.

\bibitem{Verwey1990}
J.~F. Verwey, E.~A. Amerasekera, J.~Bisschop, {The physics of SiO$_2$ layers},
  Reports on Progress in Physics 53~(10) (1990) 1297–1331.
\newblock \href {https://doi.org/10.1088/0034-4885/53/10/002}
  {\path{doi:10.1088/0034-4885/53/10/002}}.

\bibitem{Pawowski1988}
L.~Pawłowski, The relationship between structure and dielectric properties in
  plasma-sprayed alumina coatings, Surface and Coatings Technology 35~(3–4)
  (1988) 285–298.
\newblock \href {https://doi.org/10.1016/0257-8972(88)90042-4}
  {\path{doi:10.1016/0257-8972(88)90042-4}}.

\bibitem{Musil2010}
J.~Musil, J.~Blažek, P.~Zeman, S.~Prokšová, M.~Šašek, R.~Čerstvý,
  {Thermal stability of alumina thin films containing $\gamma$-Al$_2$O$_3$
  phase prepared by reactive magnetron sputtering}, Applied Surface Science
  257~(3) (2010) 1058–1062.
\newblock \href {https://doi.org/10.1016/j.apsusc.2010.07.107}
  {\path{doi:10.1016/j.apsusc.2010.07.107}}.

\bibitem{Eklund2009}
P.~Eklund, M.~Sridharan, G.~Singh, J.~Bøttiger, Thermal stability and phase
  transformations of $\gamma$‐/amorphous‐{Al$_2$O$_3$} thin films, Plasma
  Processes and Polymers 6~(S1) (2009).
\newblock \href {https://doi.org/10.1002/ppap.200932301}
  {\path{doi:10.1002/ppap.200932301}}.

\bibitem{Drevet2023}
R.~Drevet, P.~Souček, P.~Mareš, M.~Dubau, Z.~Czigány, K.~Balázsi,
  P.~Vašina, Multilayer thin films of aluminum oxide and tantalum oxide
  deposited by pulsed direct current magnetron sputtering for dielectric
  applications, Vacuum 210 (2023) 111870.
\newblock \href {https://doi.org/10.1016/j.vacuum.2023.111870}
  {\path{doi:10.1016/j.vacuum.2023.111870}}.

\bibitem{Chaneliere1998}
C.~Chaneliere, J.~Autran, R.~Devine, B.~Balland, Tantalum pentoxide
  ({Ta$_2$O$_5$}) thin films for advanced dielectric applications, Materials
  Science and Engineering: R: Reports 22~(6) (1998) 269–322.
\newblock \href {https://doi.org/10.1016/s0927-796x(97)00023-5}
  {\path{doi:10.1016/s0927-796x(97)00023-5}}.

\bibitem{AlmeidaAlves2018}
C.~Almeida~Alves, C.~Mansilla, L.~Pereira, F.~Paumier, T.~Girardeau,
  S.~Carvalho, {Influence of magnetron sputtering conditions on the chemical
  bonding, structural, morphological and optical behavior of Ta$_{1-x}$O$_x$
  coatings}, Surface and Coatings Technology 334 (2018) 105–115.
\newblock \href {https://doi.org/10.1016/j.surfcoat.2017.11.001}
  {\path{doi:10.1016/j.surfcoat.2017.11.001}}.

\bibitem{Ezhilvalavan1999}
S.~Ezhilvalavan, T.~Y. Tseng, {Preparation and properties of tantalum pentoxide
  (Ta$_2$O$_5$) thin films for ultra large scale integrated circuits (ULSIs)
  application – A review}, Journal of Materials Science: Materials in
  Electronics 10~(1) (1999) 9–31.
\newblock \href {https://doi.org/10.1023/a:1008970922635}
  {\path{doi:10.1023/a:1008970922635}}.

\bibitem{Chandra2008}
S.~J. Chandra, S.~Uthanna, G.~M. Rao, {Effect of substrate temperature on the
  structural, optical and electrical properties of DC magnetron sputtered
  tantalum oxide films}, Applied Surface Science 254~(7) (2008) 1953–1960.
\newblock \href {https://doi.org/10.1016/j.apsusc.2007.08.005}
  {\path{doi:10.1016/j.apsusc.2007.08.005}}.

\bibitem{Dong2023}
D.~Dong, S.~S. Dhanabalan, P.~F.~M. Elango, M.~Yang, S.~Walia, S.~Sriram,
  M.~Bhaskaran, Emerging applications of metal-oxide thin films for flexible
  and stretchable electronic devices, Applied Physics Reviews 10~(3) (2023).
\newblock \href {https://doi.org/10.1063/5.0151297}
  {\path{doi:10.1063/5.0151297}}.

\bibitem{Sivaperuman2024}
K.~Sivaperuman, A.~Thomas, R.~Thangavel, L.~Thirumalaisamy, S.~Palanivel,
  S.~Pitchaimuthu, N.~Ahsan, Y.~Okada, Binary and ternary metal oxide
  semiconductor thin films for effective gas sensing applications: A
  comprehensive review and future prospects, Progress in Materials Science 142
  (2024) 101222.
\newblock \href {https://doi.org/10.1016/j.pmatsci.2023.101222}
  {\path{doi:10.1016/j.pmatsci.2023.101222}}.

\bibitem{Drevet2024}
R.~Drevet, P.~Souček, P.~Mareš, P.~Ondračka, M.~Dubau, T.~Kolonits,
  Z.~Czigány, K.~Balázsi, P.~Vašina, Aluminum tantalum oxide thin films
  deposited at low temperature by pulsed direct current reactive magnetron
  sputtering for dielectric applications, Vacuum 221 (2024) 112881.
\newblock \href {https://doi.org/10.1016/j.vacuum.2023.112881}
  {\path{doi:10.1016/j.vacuum.2023.112881}}.

\bibitem{Chen2015}
D.~Chen, Q.~Wang, R.~Wang, G.~Shen, Ternary oxide nanostructured materials for
  supercapacitors: a review, Journal of Materials Chemistry A 3~(19) (2015)
  10158–10173.
\newblock \href {https://doi.org/10.1039/c4ta06923d}
  {\path{doi:10.1039/c4ta06923d}}.

\bibitem{Qu2020}
J.~Qu, D.~Zagaceta, W.~Zhang, Q.~Zhu, High dielectric ternary oxides from
  crystal structure prediction and high-throughput screening, Scientific Data
  7~(1) (2020).
\newblock \href {https://doi.org/10.1038/s41597-020-0418-6}
  {\path{doi:10.1038/s41597-020-0418-6}}.

\bibitem{Gupta2017}
C.~Gupta, S.~H. Chan, A.~Agarwal, N.~Hatui, S.~Keller, U.~K. Mishra, First
  demonstration of {AlSiO} as gate dielectric in {GaN} {FETs}; applied to a
  high performance {OG-FET}, IEEE Electron Device Letters 38~(11) (2017)
  1575–1578.
\newblock \href {https://doi.org/10.1109/led.2017.2756926}
  {\path{doi:10.1109/led.2017.2756926}}.

\bibitem{Rebouta2015}
L.~Rebouta, A.~Sousa, M.~Andritschky, F.~Cerqueira, C.~Tavares, P.~Santilli,
  K.~Pischow, Solar selective absorbing coatings based on
  {AlSiN/AlSiON/AlSiO$_y$} layers, Applied Surface Science 356 (2015)
  203–212.
\newblock \href {https://doi.org/10.1016/j.apsusc.2015.07.193}
  {\path{doi:10.1016/j.apsusc.2015.07.193}}.

\bibitem{ALRjoub2020}
A.~AL-Rjoub, L.~Rebouta, N.~Cunha, F.~Fernandes, N.~Barradas, E.~Alves,
  {W/AlSiTiN$_x$/SiAlTiO$_y$N$_x$/SiAlO$_x$} multilayered solar thermal
  selective absorber coating, Solar Energy 207 (2020) 192–198.
\newblock \href {https://doi.org/10.1016/j.solener.2020.06.094}
  {\path{doi:10.1016/j.solener.2020.06.094}}.

\bibitem{Xiao2018}
S.~Xiao, M.~Yao, W.~Gao, Z.~Su, X.~Yao, Significantly enhanced dielectric
  constant and breakdown strength in crystalline@amorphous core-shell
  structured {SrTiO$_3$} nanocomposite thick films, Journal of Alloys and
  Compounds 762 (2018) 370–377.
\newblock \href {https://doi.org/10.1016/j.jallcom.2018.05.221}
  {\path{doi:10.1016/j.jallcom.2018.05.221}}.

\bibitem{Franta2025-Ta2O5}
D.~Franta, J.~Vohánka, J.~Dvořák, P.~Franta, I.~Ohlídal, P.~Klapetek,
  J.~Březina, D.~Škoda, Wide spectral range optical characterization of
  tantalum pentoxide ({Ta$_2$O$_5$}) films by the universal dispersion model,
  Optical Materials Express 15~(4) (2025) 903.
\newblock \href {https://doi.org/10.1364/ome.550708}
  {\path{doi:10.1364/ome.550708}}.

\bibitem{Franta2018}
D.~Franta, J.~Vohánka, M.~Čermák, Universal Dispersion Model for
  Characterization of Thin Films Over Wide Spectral Range, Springer
  International Publishing, 2018, p. 31–82.
\newblock \href {https://doi.org/10.1007/978-3-319-75325-6\_3}
  {\path{doi:10.1007/978-3-319-75325-6\_3}}.

\bibitem{Baer2020}
D.~R. Baer, K.~Artyushkova, H.~Cohen, C.~D. Easton, M.~Engelhard, T.~R.
  Gengenbach, G.~Greczynski, P.~Mack, D.~J. Morgan, A.~Roberts, {XPS} guide:
  Charge neutralization and binding energy referencing for insulating samples,
  Journal of Vacuum Science \& Technology A 38~(3) (2020).
\newblock \href {https://doi.org/10.1116/6.0000057}
  {\path{doi:10.1116/6.0000057}}.

\bibitem{Greczynski2020a}
G.~Greczynski, L.~Hultman, X-ray photoelectron spectroscopy: Towards reliable
  binding energy referencing, Progress in Materials Science 107 (2020) 100591.
\newblock \href {https://doi.org/10.1016/j.pmatsci.2019.100591}
  {\path{doi:10.1016/j.pmatsci.2019.100591}}.

\bibitem{Greczynski2020b}
G.~Greczynski, L.~Hultman, Compromising science by ignorant instrument
  calibration—need to revisit half a century of published {XPS} data,
  Angewandte Chemie International Edition 59~(13) (2020) 5002–5006.
\newblock \href {https://doi.org/10.1002/anie.201916000}
  {\path{doi:10.1002/anie.201916000}}.

\bibitem{Hohenberg1964}
P.~Hohenberg, W.~Kohn, Inhomogeneous electron gas, Physical Review 136~(3B)
  (1964) B864--B871.
\newblock \href {https://doi.org/10.1103/PhysRev.136.B864}
  {\path{doi:10.1103/PhysRev.136.B864}}.

\bibitem{Kohn1965}
W.~Kohn, L.~J. Sham, Self-consistent equations including exchange and
  correlation effects, Physical Review 140~(4A) (1965) A1133--A1138.
\newblock \href {https://doi.org/10.1103/PhysRev.140.A1133}
  {\path{doi:10.1103/PhysRev.140.A1133}}.

\bibitem{Nose1984}
S.~Nos{\'{e}}, {A unified formulation of the constant temperature molecular
  dynamics methods}, The Journal of Chemical Physics 81~(1) (1984) 511--519.
\newblock \href {https://doi.org/10.1063/1.447334}
  {\path{doi:10.1063/1.447334}}.

\bibitem{Perdew1996}
J.~P. Perdew, K.~Burke, M.~Ernzerhof, Generalized gradient approximation made
  simple, Physical Review Letters 77~(18) (1996) 3865--3868.
\newblock \href {https://doi.org/10.1103/PhysRevLett.77.3865}
  {\path{doi:10.1103/PhysRevLett.77.3865}}.

\bibitem{Adamo1999}
C.~Adamo, V.~Barone, {Toward reliable density functional methods without
  adjustable parameters: The PBE0 model}, The Journal of Chemical Physics
  110~(13) (1999) 6158--6170.
\newblock \href {https://doi.org/10.1063/1.478522}
  {\path{doi:10.1063/1.478522}}.

\bibitem{NOMADdata}
P.~Ondračka, {NOMAD dataset: Effect of oxygen content on optical, structural,
  and dielectric properties of {Al$_x$Ta$_y$O$_z$} thin films - {\it ab initio}
  data} (2026).
\newblock \href {https://doi.org/10.17172/NOMAD/2026.02.05-1}
  {\path{doi:10.17172/NOMAD/2026.02.05-1}}.

\bibitem{Scheidgen2023}
M.~Scheidgen, L.~Himanen, A.~N. Ladines, D.~Sikter, M.~Nakhaee, A.~Fekete,
  T.~Chang, A.~Golparvar, J.~A. Márquez, S.~Brockhauser, S.~Br\"{u}ckner,
  L.~M. Ghiringhelli, F.~Dietrich, D.~Lehmberg, T.~Denell, A.~Albino,
  H.~N\"{a}sstr\"{o}m, S.~Shabih, F.~Dobener, M.~K\"{u}hbach, R.~Mozumder,
  J.~F. Rudzinski, N.~Daelman, J.~M. Pizarro, M.~Kuban, C.~Salazar,
  P.~Ondračka, H.-J. Bungartz, C.~Draxl, {NOMAD: A distributed web-based
  platform for managing materials science research data}, Journal of Open
  Source Software 8~(90) (2023) 5388.
\newblock \href {https://doi.org/10.21105/joss.05388}
  {\path{doi:10.21105/joss.05388}}.

\bibitem{Prohaska2022}
T.~Prohaska, J.~Irrgeher, J.~Benefield, J.~K. B\"{o}hlke, L.~A. Chesson, T.~B.
  Coplen, T.~Ding, P.~J.~H. Dunn, M.~Gr\"{o}ning, N.~E. Holden, H.~A.~J.
  Meijer, H.~Moossen, A.~Possolo, Y.~Takahashi, J.~Vogl, T.~Walczyk, J.~Wang,
  M.~E. Wieser, S.~Yoneda, X.-K. Zhu, J.~Meija, Standard atomic weights of the
  elements 2021 ({IUPAC} technical report), Pure and Applied Chemistry 94~(5)
  (2022) 573–600.
\newblock \href {https://doi.org/10.1515/pac-2019-0603}
  {\path{doi:10.1515/pac-2019-0603}}.

\bibitem{Feng2013}
X.~Feng, G.~Tang, M.~Sun, X.~Ma, L.~Wang, K.~Yukimura, Structure and properties
  of multi-targets magnetron sputtered {ZrNbTaTiW} multi-elements alloy thin
  films, Surface and Coatings Technology 228 (2013) S424–S427.
\newblock \href {https://doi.org/10.1016/j.surfcoat.2012.05.038}
  {\path{doi:10.1016/j.surfcoat.2012.05.038}}.

\bibitem{Frunz2014}
R.~C. Frunză, B.~Kmet, M.~Jankovec, M.~Topič, B.~Malič, {Ta$_2$O$_5$}-based
  high-{K} dielectric thin films from solution processed at low temperatures,
  Materials Research Bulletin 50 (2014) 323–328.
\newblock \href {https://doi.org/10.1016/j.materresbull.2013.11.025}
  {\path{doi:10.1016/j.materresbull.2013.11.025}}.

\bibitem{Panjan2020}
P.~Panjan, A.~Drnovšek, P.~Gselman, M.~Čekada, M.~Panjan, Review of growth
  defects in thin films prepared by {PVD} techniques, Coatings 10~(5) (2020)
  447.
\newblock \href {https://doi.org/10.3390/coatings10050447}
  {\path{doi:10.3390/coatings10050447}}.

\bibitem{Budenstein1980}
P.~Budenstein, On the mechanism of dielectric breakdown of solids, IEEE
  Transactions on Electrical Insulation EI-15~(3) (1980) 225–240.

\bibitem{Wu2019}
E.~Y. Wu, Facts and myths of dielectric breakdown processes—part {I}:
  Statistics, experimental, and physical acceleration models, IEEE Transactions
  on Electron Devices 66~(11) (2019) 4523–4534.
\newblock \href {https://doi.org/10.1109/ted.2019.2933612}
  {\path{doi:10.1109/ted.2019.2933612}}.

\bibitem{McPherson2002}
J.~McPherson, J.~Kim, A.~Shanware, H.~Mogul, J.~Rodriguez, Proposed universal
  relationship between dielectric breakdown and dielectric constant, in:
  Digest. International Electron Devices Meeting,, IEDM-02, IEEE, 2002, pp.
  633--636.
\newblock \href {https://doi.org/10.1109/iedm.2002.1175919}
  {\path{doi:10.1109/iedm.2002.1175919}}.

\bibitem{Drevet2025}
R.~Drevet, P.~Souček, P.~Mareš, P.~Ondračka, M.~Fekete, M.~Dubau,
  P.~Vašina, Influence of oxygen flow on the structure, chemical composition,
  and dielectric strength of al$_x$ta$_y$o$_z$ thin films deposited by
  pulsed-dc reactive magnetron sputtering, Surface and Coatings Technology 498
  (2025) 131865.
\newblock \href {https://doi.org/10.1016/j.surfcoat.2025.131865}
  {\path{doi:10.1016/j.surfcoat.2025.131865}}.

\bibitem{Franta2015-Al2O3}
D.~Franta, D.~Nečas, I.~Ohlídal, A.~Giglia, Dispersion model for optical thin
  films applicable in wide spectral range, in: A.~Duparré, R.~Geyl (Eds.),
  Optical Systems Design 2015: Optical Fabrication, Testing, and Metrology V,
  Vol. 9628, SPIE, 2015, p. 96281U.
\newblock \href {https://doi.org/10.1117/12.2190104}
  {\path{doi:10.1117/12.2190104}}.

\bibitem{Shvets2008}
V.~Shvets, V.~Aliev, D.~Gritsenko, S.~Shaimeev, E.~Fedosenko, S.~Rykhlitski,
  V.~Atuchin, V.~Gritsenko, V.~Tapilin, H.~Wong, Electronic structure and
  charge transport properties of amorphous {Ta$_2$O$_5$} films, Journal of
  Non-Crystalline Solids 354~(26) (2008) 3025–3033.
\newblock \href {https://doi.org/10.1016/j.jnoncrysol.2007.12.013}
  {\path{doi:10.1016/j.jnoncrysol.2007.12.013}}.

\bibitem{Sertel2019}
T.~Sertel, N.~A. Sonmez, S.~S. Cetin, S.~Ozcelik, Influences of annealing
  temperature on anti-reflective performance of amorphous {Ta$_2$O$_5$} thin
  films, Ceramics International 45~(1) (2019) 11–18.
\newblock \href {https://doi.org/10.1016/j.ceramint.2018.09.237}
  {\path{doi:10.1016/j.ceramint.2018.09.237}}.

\bibitem{Ondraka2017}
P.~Ondračka, D.~Holec, D.~Nečas, E.~Kedroňová, S.~Elisabeth, A.~Goullet,
  L.~Zajíčková, Optical properties of {Ti$x$Si$_{1-x}$O$_2$} solid
  solutions, Physical Review B 95~(19) (2017).
\newblock \href {https://doi.org/10.1103/physrevb.95.195163}
  {\path{doi:10.1103/physrevb.95.195163}}.

\bibitem{momma_vesta_2011}
K.~Momma, F.~Izumi, {VESTA} 3 for three-dimensional visualization of crystal,
  volumetric and morphology data, Journal of Applied Crystallography 44~(6)
  (2011) 1272--1276.
\newblock \href {https://doi.org/10.1107/S0021889811038970}
  {\path{doi:10.1107/S0021889811038970}}.

\bibitem{horton_accelerated_2025}
M.~K. Horton, P.~Huck, R.~X. Yang, J.~M. Munro, S.~Dwaraknath, A.~M. Ganose,
  R.~S. Kingsbury, M.~Wen, J.~X. Shen, T.~S. Mathis, A.~D. Kaplan, K.~Berket,
  J.~Riebesell, J.~George, A.~S. Rosen, E.~W.~C. Spotte-Smith, M.~J. McDermott,
  O.~A. Cohen, A.~Dunn, M.~C. Kuner, G.-M. Rignanese, G.~Petretto,
  D.~Waroquiers, S.~M. Griffin, J.~B. Neaton, D.~C. Chrzan, M.~Asta,
  G.~Hautier, S.~Cholia, G.~Ceder, S.~P. Ong, A.~Jain, K.~A. Persson,
  Accelerated data-driven materials science with the {Materials} {Project},
  Nature Materials (2025).
\newblock \href {https://doi.org/10.1038/s41563-025-02272-0}
  {\path{doi:10.1038/s41563-025-02272-0}}.

\bibitem{Shi2019}
C.~Shi, O.~L.~G. Alderman, D.~Berman, J.~Du, J.~Neuefeind, A.~Tamalonis,
  J.~K.~R. Weber, J.~You, C.~J. Benmore, The structure of amorphous and deeply
  supercooled liquid alumina, Frontiers in Materials 6 (2019).
\newblock \href {https://doi.org/10.3389/fmats.2019.00038}
  {\path{doi:10.3389/fmats.2019.00038}}.

\bibitem{askeljung_effect_2003}
C.~Askeljung, B.-O. Marinder, M.~Sundberg, Effect of heat treatment on the
  structure of {L}-{Ta$_2$O$_5$}:, Journal of Solid State Chemistry 176~(1)
  (2003) 250--258.
\newblock \href {https://doi.org/10.1016/j.jssc.2003.07.003}
  {\path{doi:10.1016/j.jssc.2003.07.003}}.

\bibitem{Capilla2011}
J.~Capilla, J.~Olivares, M.~Clement, J.~Sangrador, E.~Iborra, A.~Devos,
  Characterization of amorphous tantalum oxide for insulating acoustic mirrors,
  in: 2011 Joint Conference of the IEEE International Frequency Control and the
  European Frequency and Time Forum (FCS) Proceedings, IEEE, 2011.
\newblock \href {https://doi.org/10.1109/fcs.2011.5977833}
  {\path{doi:10.1109/fcs.2011.5977833}}.

\bibitem{Levin1998}
I.~Levin, D.~Brandon, Metastable alumina polymorphs: Crystal structures and
  transition sequences, Journal of the American Ceramic Society 81~(8) (1998)
  1995–2012.
\newblock \href {https://doi.org/10.1111/j.1151-2916.1998.tb02581.x}
  {\path{doi:10.1111/j.1151-2916.1998.tb02581.x}}.

\bibitem{lee_structure_2009}
S.~K. Lee, S.~B. Lee, S.~Y. Park, Y.~S. Yi, C.~W. Ahn, Structure of {Amorphous}
  {Aluminum} {Oxide}, Physical Review Letters 103~(9) (2009) 095501.
\newblock \href {https://doi.org/10.1103/PhysRevLett.103.095501}
  {\path{doi:10.1103/PhysRevLett.103.095501}}.

\bibitem{bassiri_order_2015}
R.~Bassiri, F.~Liou, M.~R. Abernathy, A.~C. Lin, N.~Kim, A.~Mehta, B.~Shyam,
  R.~L. Byer, E.~K. Gustafson, M.~Hart, I.~MacLaren, I.~W. Martin, R.~K. Route,
  S.~Rowan, J.~F. Stebbins, M.~M. Fejer, Order within disorder: {The} atomic
  structure of ion-beam sputtered amorphous tantala (a-{Ta$_2$O$_5$}), APL
  Materials 3~(3) (2015) 036103.
\newblock \href {https://doi.org/10.1063/1.4913586}
  {\path{doi:10.1063/1.4913586}}.

\bibitem{martinelli_deep_2021}
A.~Martinelli, M.~Giovannini, M.~Neri, G.~Gemme, Deep insights into the local
  structure of amorphous {Ta$_2$O$_5$} thin films by {X}-ray pair distribution
  function analysis, Physical Review Materials 5~(11) (2021) 115603--115603,
  publisher: American Physical Society.
\newblock \href {https://doi.org/10.1103/PhysRevMaterials.5.115603}
  {\path{doi:10.1103/PhysRevMaterials.5.115603}}.

\bibitem{Dicks2017}
O.~A. Dicks, A.~L. Shluger, Theoretical modeling of charge trapping in
  crystalline and amorphous {Al$_2$O$_3$}, Journal of Physics: Condensed Matter
  29~(31) (2017) 314005.
\newblock \href {https://doi.org/10.1088/1361-648x/aa7767}
  {\path{doi:10.1088/1361-648x/aa7767}}.

\bibitem{Lee2014}
J.~Lee, W.~Lu, E.~Kioupakis, Electronic properties of tantalum pentoxide
  polymorphs from first-principles calculations, Applied Physics Letters
  105~(20) (2014).
\newblock \href {https://doi.org/10.1063/1.4901939}
  {\path{doi:10.1063/1.4901939}}.

\bibitem{Marinopoulos2011}
A.~G. Marinopoulos, M.~Gr\"{u}ning, Local-field and excitonic effects in the
  optical response of $\alpha$-alumina, Physical Review B 83~(19) (2011).
\newblock \href {https://doi.org/10.1103/physrevb.83.195129}
  {\path{doi:10.1103/physrevb.83.195129}}.

\bibitem{NOMADdataXPS}
P.~Ondračka, {NOMAD} dataset: Effect of oxygen content on optical, structural,
  and dielectric properties of {Al$_x$Ta$_y$O$_z$} thin films - {XPS} data
  (2026).
\newblock \href {https://doi.org/10.17172/NOMAD/2026.02.09-1}
  {\path{doi:10.17172/NOMAD/2026.02.09-1}}.

\end{thebibliography}


\begin{thebibliography}{10}
\expandafter\ifx\csname url\endcsname\relax
  \def\url#1{\texttt{#1}}\fi
\expandafter\ifx\csname urlprefix\endcsname\relax\def\urlprefix{URL }\fi

\bibitem{Ozaki2003}
T.~Ozaki, Variationally optimized atomic orbitals for large-scale electronic
  structures, Physical Review B 67~(15).

\bibitem{Ozaki2004}
T.~Ozaki, H.~Kino, Numerical atomic basis orbitals from h to kr, Physical
  Review B 69~(19).

\bibitem{Ozaki2005}
T.~Ozaki, H.~Kino, Efficient projector expansion for theab initiolcao method,
  Physical Review B 72~(4).

\bibitem{Giannozzi2009}
P.~Giannozzi, S.~Baroni, N.~Bonini, M.~Calandra, R.~Car, C.~Cavazzoni,
  D.~Ceresoli, G.~L. Chiarotti, M.~Cococcioni, I.~Dabo, A.~Dal~Corso,
  S.~de~Gironcoli, S.~Fabris, G.~Fratesi, R.~Gebauer, U.~Gerstmann,
  C.~Gougoussis, A.~Kokalj, M.~Lazzeri, L.~Martin-Samos, N.~Marzari, F.~Mauri,
  R.~Mazzarello, S.~Paolini, A.~Pasquarello, L.~Paulatto, C.~Sbraccia,
  S.~Scandolo, G.~Sclauzero, A.~P. Seitsonen, A.~Smogunov, P.~Umari, R.~M.
  Wentzcovitch, {QUANTUM ESPRESSO: a modular and open-source software project
  for quantum simulations of materials}, Journal of Physics: Condensed Matter
  21~(39) (2009) 395502.

\bibitem{Giannozzi2017}
P.~Giannozzi, O.~Andreussi, T.~Brumme, O.~Bunau, M.~Buongiorno~Nardelli,
  M.~Calandra, R.~Car, C.~Cavazzoni, D.~Ceresoli, M.~Cococcioni, N.~Colonna,
  I.~Carnimeo, A.~Dal~Corso, S.~de~Gironcoli, P.~Delugas, R.~A. DiStasio,
  A.~Ferretti, A.~Floris, G.~Fratesi, G.~Fugallo, R.~Gebauer, U.~Gerstmann,
  F.~Giustino, T.~Gorni, J.~Jia, M.~Kawamura, H.-Y. Ko, A.~Kokalj,
  E.~K\"{u}\c{c}\"{u}kbenli, M.~Lazzeri, M.~Marsili, N.~Marzari, F.~Mauri,
  N.~L. Nguyen, H.-V. Nguyen, A.~Otero-de-la Roza, L.~Paulatto, S.~Poncé,
  D.~Rocca, R.~Sabatini, B.~Santra, M.~Schlipf, A.~P. Seitsonen, A.~Smogunov,
  I.~Timrov, T.~Thonhauser, P.~Umari, N.~Vast, X.~Wu, S.~Baroni, {Advanced
  capabilities for materials modelling with Quantum ESPRESSO}, Journal of
  Physics: Condensed Matter 29~(46) (2017) 465901.

\bibitem{Giannozzi2020}
P.~Giannozzi, O.~Baseggio, P.~Bonfà, D.~Brunato, R.~Car, I.~Carnimeo,
  C.~Cavazzoni, S.~de~Gironcoli, P.~Delugas, F.~Ferrari~Ruffino, A.~Ferretti,
  N.~Marzari, I.~Timrov, A.~Urru, S.~Baroni, {Quantum ESPRESSO toward the
  exascale}, The Journal of Chemical Physics 152~(15).

\bibitem{Hamann2013}
D.~R. Hamann, Optimized norm-conserving vanderbilt pseudopotentials, Physical
  Review B 88~(8).

\bibitem{Scheidgen2023}
M.~Scheidgen, L.~Himanen, A.~N. Ladines, D.~Sikter, M.~Nakhaee, A.~Fekete,
  T.~Chang, A.~Golparvar, J.~A. Márquez, S.~Brockhauser, S.~Br\"{u}ckner,
  L.~M. Ghiringhelli, F.~Dietrich, D.~Lehmberg, T.~Denell, A.~Albino,
  H.~N\"{a}sstr\"{o}m, S.~Shabih, F.~Dobener, M.~K\"{u}hbach, R.~Mozumder,
  J.~F. Rudzinski, N.~Daelman, J.~M. Pizarro, M.~Kuban, C.~Salazar,
  P.~Ondračka, H.-J. Bungartz, C.~Draxl, {NOMAD: A distributed web-based
  platform for managing materials science research data}, Journal of Open
  Source Software 8~(90) (2023) 5388.

\bibitem{NOMADdata}
P.~Ondračka, {NOMAD dataset: Effect of oxygen content on optical, structural,
  and dielectric properties of {Al$_x$Ta$_y$O$_z$} thin films - {\it ab initio}
  data} (2026).

\bibitem{newAD}
D.~Franta, D.~Nečas, J.~Vohánka, {Software for optical characterization
  newAD2}, \url{newad.physics.muni.cz}.

\bibitem{Franta25b}
D.~Franta, J.~Vohánka, J.~Dvořák, P.~Franta, I.~Ohlídal, P.~Klapetek,
  J.~Březina, D.~Škoda, {Wide spectral range optical characterization of
  tantalum pentoxide (Ta\textsubscript{2}O\textsubscript{5}) films by the
  universal dispersion model}, {Opt. Mater. Express} 15 (2025) 903--919.

\bibitem{Franta24a}
D.~Franta, B.~Hroncová, J.~Dvořák, J.~Vohánka, P.~Franta, I.~Ohlídal,
  J.~Březina, D.~Škoda, {Wide spectral range optical characterization of
  niobium pentoxide (Nb\textsubscript{2}O\textsubscript{5}) films by universal
  dispersion model}, {Opt. Mater.} 157 (2024) 116133.

\bibitem{Franta17a}
D.~Franta, M.~Čermák, J.~Vohánka, I.~Ohlídal, {Dispersion models describing
  interband electronic transitions combining Tauc's law and Lorentz model},
  {Thin Solid Films} 631 (2017) 12--22.

\bibitem{Ferlauto02a}
A.~S. Ferlauto, G.~M. Ferreira, J.~M. Pearce, C.~R. Wronski, R.~W. Collins,
  X.~M. Deng, G.~Ganguly, {Analytical model for the optical functions of
  amorphous semiconductors from the near-infrared to ultraviolet: Applications
  in thin film photovoltaics}, {J. Appl. Phys.} 92 (2002) 2424--2436.

\bibitem{Franta18a}
D.~Franta, J.~Vohánka, M.~Čermák, {Universal Dispersion Model for
  Characterization of Thin Films Over Wide Spectral Range}, in: O.~Stenzel,
  M.~Ohlídal (Eds.), {Optical Characterization of Thin Solid Films}, Vol.~64
  of Springer Series in Surface Sciences, Springer, Cham, 2018, pp. 31--82.

\bibitem{Rice51a}
S.~O. Rice, {Reflection of Electromagnetic Waves from Slightly Rough Surfaces},
  Commun. Pure Appl. Math. 4 (1951) 351--378.

\bibitem{Franta05c}
D.~Franta, I.~Ohlídal, {Comparison of Effective Medium Approximation and
  {Rayleigh--Rice} Theory Concerning Ellipsometric Characterization of Rough
  Surfaces}, {Opt. Commun.} 248 (2005) 459--467.

\bibitem{Franta08e}
D.~Franta, I.~Ohlídal, D.~Nečas, {Influence of cross-correlation effects on
  the optical quantities of rough films}, {Opt. Express} 16 (2008) 7789--7803.

\end{thebibliography}

\end{document}


\begin{center} \Large \bf
Effect of oxygen content on optical, structural, and dielectric properties of \AlTaO{} thin films -- supplementary information \\ \bigskip
\small \rm
Pavel Ondračka$^{a}$, Richard Drevet$^{b}$, Daniel Franta$^{a,c}$, Jan Dvořák$^{a}$, Ivan Ohlídal$^{a}$, Petr Vašina$^{a}$\\ \medskip \it
$^{a}$ Department of Plasma Physics and Technology, Faculty of Science, Masaryk University, Kotlářská 2, 61137 Brno, Czechia \\
$^{b}$ Institut de Thermique, Mécanique et Matériaux (ITheMM), EA 7548, Université de Reims Champagne-Ardenne (URCA), Bât.6, Moulin de la Housse, BP 1039, 51687 Reims Cedex 2, France \\
$^{c}$ CEITEC -- Central European Institute of Technology, Brno University of Technology, Purkyňova 123, 612 00 Brno, Czechia
\end{center}

\vspace{5cm}

\tableofcontents

\clearpage

\section{DFT computational details}

The molecular dynamics simulations and subsequent crude relaxations were performed with the numerical atomic basis set pseudo-potential DFT OpenMX code~\cite{Ozaki2003, Ozaki2004, Ozaki2005}. The used numerical atomic bases were \verb|Al7.0-s2p2d1|, \verb|Ta7.0-s3p2d2f1| and \verb|O6.0-s2p2d1| for Al, Ta and O atoms respectively. Energy cutoff of 250\,Ry (400\,Ry for the relaxation), Fermi-Dirac smearing of 300\,K, $k$-point grid of 3$\times$3$\times$3 were used. The timestep for the MD simulations was 2\,fs.
Fine relaxation (final residual forces less than 2.5\,meV/\r{A} and residual stress less than 5\,MPa) and hybrid calculations utilized a more precise plane-wave pseudo-potential DFT package Quantum ESPRESSO~\cite{Giannozzi2009, Giannozzi2017, Giannozzi2020} with norm-conserving pseudopotentials~\cite{Hamann2013}. 4$\times$4$\times$4 $k$-point grid was employed for the fine relaxation, while 3$\times$3$\times$3 $k$-point (and a single $\Gamma$ $q$-point) grid was employed for the subsequent more computationally demanding hybrid calculations with Gaussian smearing of 0.1\,eV. Kinetic energy cutoff was set to 80\,Ry for wavefunctions, 160\,Ry for exact exchange operator in the hybrid calculations and 320\,Ry for charge density and potential.
All DFT calculations with further details are available under Creative Commons license in the NOMAD Archive~\cite{Scheidgen2023, NOMADdata}. 

\section{XPS}

\begin{figure}[h!]
    \centering
    \includegraphics[width=12cm]{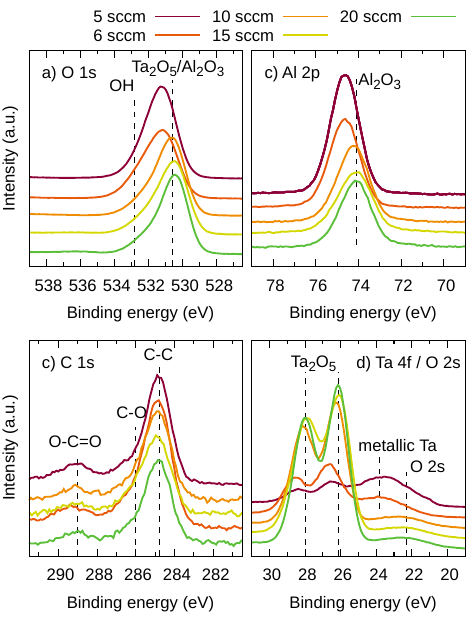}
    \caption{\label{fig:SIspectra} XPS region spectra of a) O\,1s, b) Al\,2p, c) C\,1s, and d) Ta\,2p regions.}
\end{figure}

\section{Optical characterization}

\subsection{Heterogeneous data processing}

\begin{table}[ht!]
\caption{List of experimental data sets used for optical characterization.} \label{tab.Experimental}
\centering
\centering\begin{tblr}{clllll}
\hline
& instrument & quantity & mode & range (resolution), AOI & figs.\\
\hline
1 & VUVAS 1000 & $R$ & front &  115--320\,(1)\,nm, 10$^\circ$ & \ref{fig.R-VUVV-T32}, \ref{fig.R-VUVV-T33}, \ref{fig.R-VUVV-T34}, \ref{fig.R-VUVV-T35}, \ref{fig.R-VUVV-T36} \\
2 & Lambda 1050+ & $R$ & front &  190--860\,(1)\,$\rm nm$, 6$^\circ$ & \ref{fig.R-VUVV-T32}, \ref{fig.R-VUVV-T33}, \ref{fig.R-VUVV-T34}, \ref{fig.R-VUVV-T35}, \ref{fig.R-VUVV-T36} \\
3 & RC2 & $\Ell$ & front &  193--1690\,(2.25, 3.5)\,nm, 45--80$^\circ$ & \ref{fig.E-UVNf-T32}, \ref{fig.E-UVNf-T33}, \ref{fig.E-UVNf-T34}, \ref{fig.E-UVNf-T35}, \ref{fig.E-UVNf-T36} \\
4 & RC2 & $\Ell$ & back &  193--1690\,(2.25, 3.5)\,nm, 45--80$^\circ$ & \ref{fig.E-UVNb-T32}, \ref{fig.E-UVNb-T33}, \ref{fig.E-UVNb-T34}, \ref{fig.E-UVNb-T35}, \ref{fig.E-UVNb-T36} \\
5 & Lambda 1050+ & $T$ &  &  860--2500\,(5)\,$\rm nm$, 0$^\circ$ & \ref{fig.TR-NIR-T32}, \ref{fig.TR-NIR-T33}, \ref{fig.TR-NIR-T34}, \ref{fig.TR-NIR-T35}, \ref{fig.TR-NIR-T36} \\
6 & Lambda 1050+ & $R^{\rm f}$ & front &  860--2500\,(5)\,$\rm nm$, 6$^\circ$ & \ref{fig.TR-NIR-T32}, \ref{fig.TR-NIR-T33}, \ref{fig.TR-NIR-T34}, \ref{fig.TR-NIR-T35}, \ref{fig.TR-NIR-T36} \\
7 & Lambda 1050+ & $R^{\rm b}$ & back &  860--2500\,(5)\,$\rm nm$, 6$^\circ$ & \ref{fig.TR-NIR-T32}, \ref{fig.TR-NIR-T33}, \ref{fig.TR-NIR-T34}, \ref{fig.TR-NIR-T35}, \ref{fig.TR-NIR-T36} \\
8 & Lambda 1050+ & $R^{\rm f}/R^{\rm b}$ & relative &  860--2500\,(5)\,$\rm nm$, 6$^\circ$ & \ref{fig.TR-NIR-T32}, \ref{fig.TR-NIR-T33}, \ref{fig.TR-NIR-T34}, \ref{fig.TR-NIR-T35}, \ref{fig.TR-NIR-T36} \\
9 & Vertex 80v & $T$ &  & 370--7500\,(8)\,$\rm cm^{-1}$, 0$^\circ$ & \ref{fig.TR-MIR-T32}, \ref{fig.TR-MIR-T33}, \ref{fig.TR-MIR-T34}, \ref{fig.TR-MIR-T35}, \ref{fig.TR-MIR-T36} \\
10 & Vertex 80v & $R^{\rm f}$ & front & 370--7500\,(8)\,$\rm cm^{-1}$, 10$^\circ$ & \ref{fig.TR-MIR-T32}, \ref{fig.TR-MIR-T33}, \ref{fig.TR-MIR-T34}, \ref{fig.TR-MIR-T35}, \ref{fig.TR-MIR-T36} \\
11 & Vertex 80v & $R^{\rm b}$ & back & 370--7500\,(8)\,$\rm cm^{-1}$, 10$^\circ$ & \ref{fig.TR-MIR-T32}, \ref{fig.TR-MIR-T33}, \ref{fig.TR-MIR-T34}, \ref{fig.TR-MIR-T35}, \ref{fig.TR-MIR-T36} \\
12 & Vertex 80v & $R^{\rm f}/R^{\rm b}$ & relative & 370--7500\,(8)\,$\rm cm^{-1}$, 10$^\circ$ & \ref{fig.TR-MIR-T32}, \ref{fig.TR-MIR-T33}, \ref{fig.TR-MIR-T34}, \ref{fig.TR-MIR-T35}, \ref{fig.TR-MIR-T36} \\
13 & Vertex 80v & $T$ &  & 70--680\,(8)\,$\rm cm^{-1}$, 0$^\circ$ & \ref{fig.TR-FIR-T32}, \ref{fig.TR-FIR-T33}, \ref{fig.TR-FIR-T34}, \ref{fig.TR-FIR-T35}, \ref{fig.TR-FIR-T36} \\
14 & Vertex 80v & $R$ & front & 70--680\,(8)\,$\rm cm^{-1}$, 10$^\circ$ & \ref{fig.TR-FIR-T32}, \ref{fig.TR-FIR-T33}, \ref{fig.TR-FIR-T34}, \ref{fig.TR-FIR-T35}, \ref{fig.TR-FIR-T36} \\
15 & Vertex 80v & $R$ & back & 70--680\,(8)\,$\rm cm^{-1}$, 10$^\circ$ & \ref{fig.TR-FIR-T32}, \ref{fig.TR-FIR-T33}, \ref{fig.TR-FIR-T34}, \ref{fig.TR-FIR-T35}, \ref{fig.TR-FIR-T36} \\
16 & Vertex 80v & $R^{\rm f}/R^{\rm b}$ & relative & 70--680\,(8)\,$\rm cm^{-1}$, 10$^\circ$ & \ref{fig.TR-FIR-T32}, \ref{fig.TR-FIR-T33}, \ref{fig.TR-FIR-T34}, \ref{fig.TR-FIR-T35}, \ref{fig.TR-FIR-T36} \\
17 & Vertex 70v & $T$ &  & 25--680\,(1)\,$\rm cm^{-1}$, 0$^\circ$ & \ref{fig.T-FIRHR-T32}, \ref{fig.T-FIRHR-T33}, \ref{fig.T-FIRHR-T34}, \ref{fig.T-FIRHR-T35}, \ref{fig.T-FIRHR-T36} \\
\hline
\end{tblr}
\end{table}

Optical characterization was based on a method of processing a set of heterogeneous data from several optical instruments in a wide spectral range (see Table~\ref{tab.Experimental}).
This unique method is implemented in the newAD2 program we are developing~\cite{newAD}.
The algorithm we developed minimizes the product of the sums of squares $S_k(\vec p)$ of the deviations between the model and the experimental data of individual measurements:
\begin{equation}
 \chi(\vec p) = \left( \prod_{k=1}^N \chi_k(\vec p) \right)^{1/N}, \qquad \chi_k(\vec p) = \sqrt{\frac{S_k(\vec p)}{N_k}},
\end{equation}
where $\vec p$ is vector of model parameters, $N_k$ is number of experimental points corresponding to $k$-th set of measurement and $N$ is number of experimental data sets (in our case $N=17$).
The individual $\chi_k$ values are introduced in the Figures.
The algorithm for fitting experimental data is described in detail in our previous works~\cite{Franta25b}.

\clearpage

\subsection{Dispersion models}

All materials forming the samples are modeled using the Universal Dispersion Model (UDM). The parameters of the crystalline float zone of silicon and the native layer were fixed during the fitting at the values determined earlier based on our database. Here, the UDM replaces the tabulated values of the optical constants.

The optical constants of the studied layers are also calculated using the UDM. In this case, however, the dispersion parameters of the model are sought within the framework of the optical characterization together with the structural parameters of the layers.

The dispersion models of the studied layers used different numbers of individual contributions describing individual elementary excitations depending on the oxygen flow rate during deposition.
Interband electron excitations describing electron transitions from extended occupied states in the valence band and extended unoccupied states in the conduction band were modeled by
one broad band~\cite{Franta24a} and two to six absorption peaks described by the ASF model~\cite{Franta17a} (this model is based on the Ferlauto \etal model~\cite{Ferlauto02a}).
Electron excitations involving localized states were modeled by an Urbach tail and three to five Gaussian-broadened discrete transitions representing absorption at localized states~\cite{Franta18a}.
Phonon excitations were modeled by five to eight discrete transitions also with Gaussian broadening.

\clearpage

\subsection{Structural model}

\begin{figure}[h!]
\includegraphics[width=\textwidth]{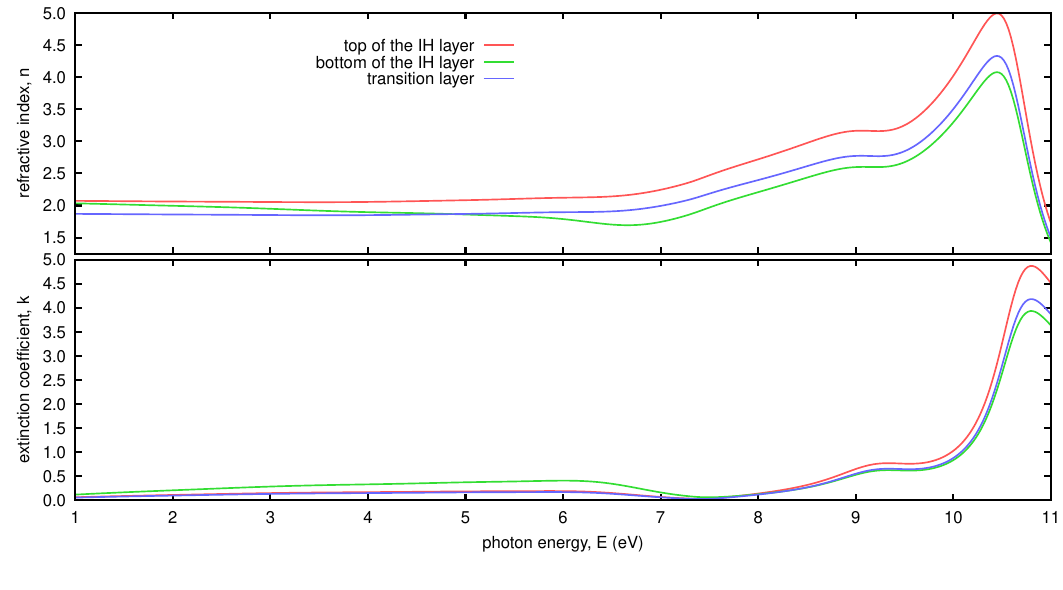}
\caption{Spectral dependencies of optical constants of \SampleTThirtyTwo.
} \label{fig.nk-T32}
\end{figure}

\begin{figure}[h!]
\includegraphics[width=\textwidth]{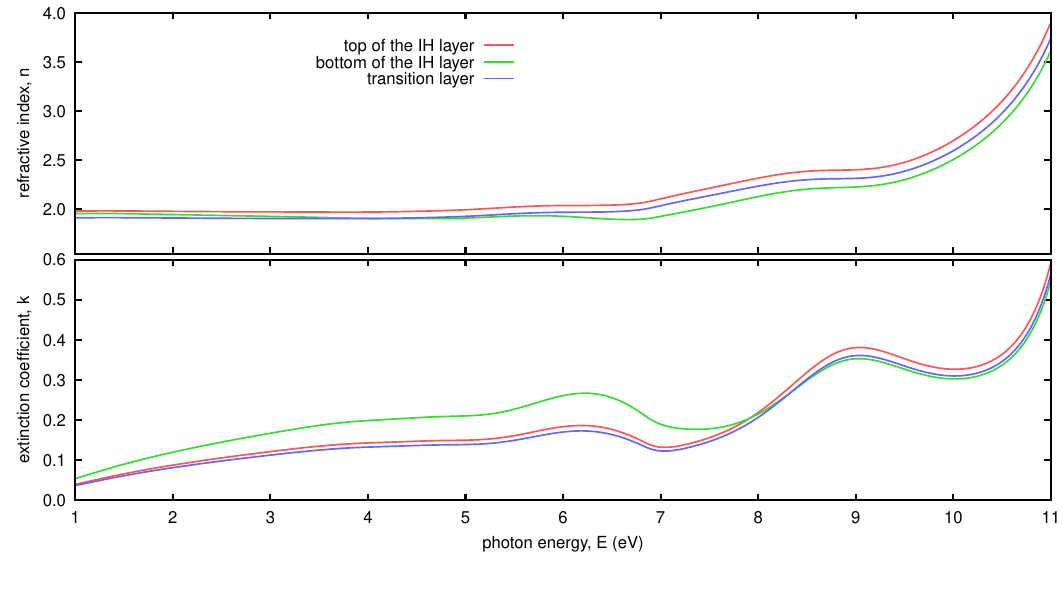}
\caption{Spectral dependencies of optical constants of \SampleTThirtyThree.
} \label{fig.nk-T33}
\end{figure}

\begin{figure}[h!]
\includegraphics[width=\textwidth]{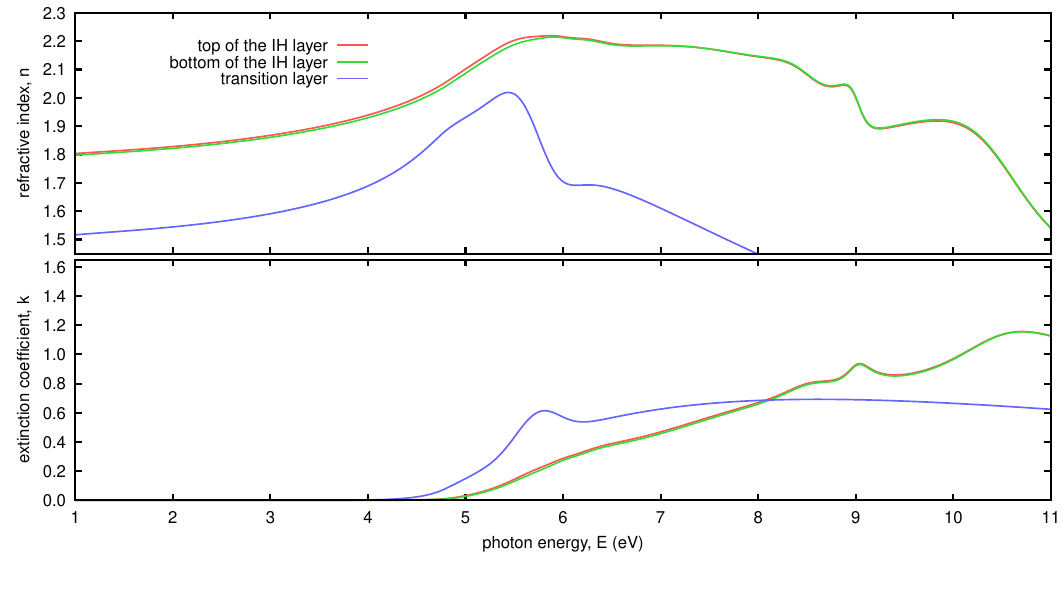}
\caption{Spectral dependencies of optical constants of \SampleTThirtyFour.
} \label{fig.nk-T34}
\end{figure}

\begin{figure}[h!]
\includegraphics[width=\textwidth]{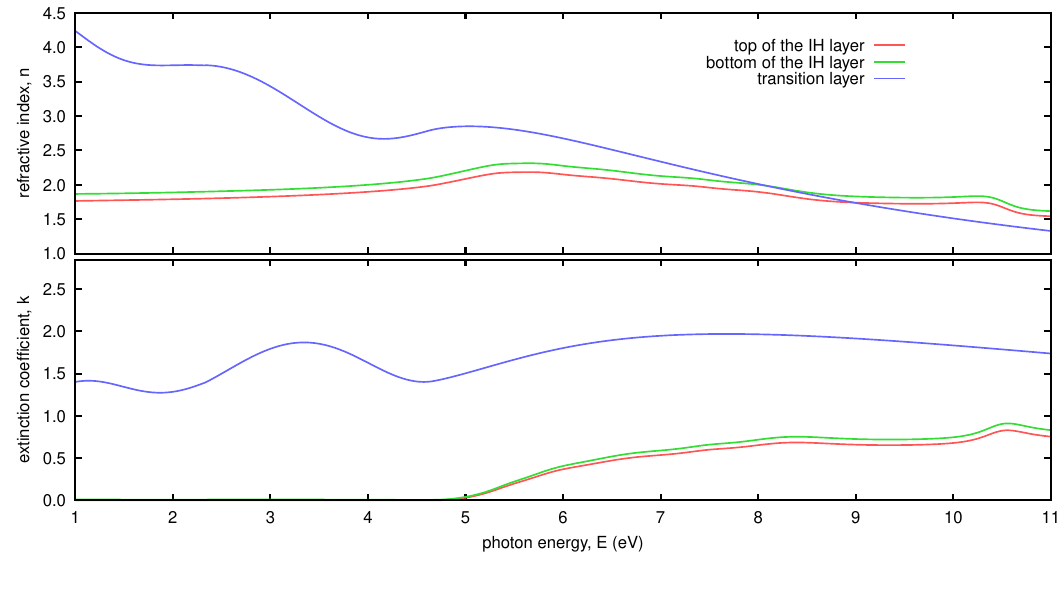}
\caption{Spectral dependencies of optical constants of \SampleTThirtyFive.
} \label{fig.nk-T35}
\end{figure}

\begin{figure}[h!]
\includegraphics[width=\textwidth]{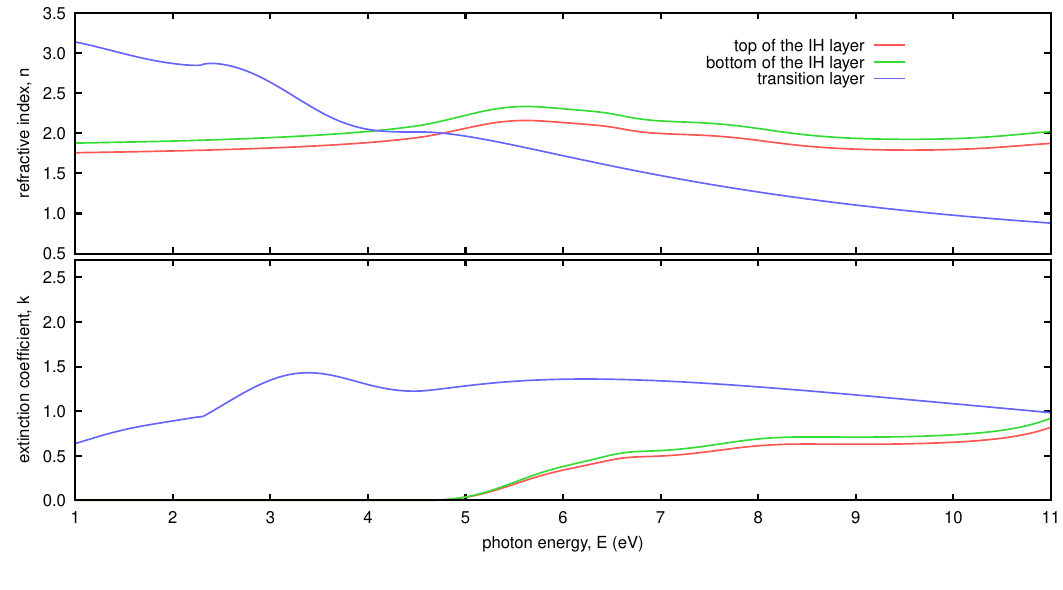}
\caption{Spectral dependencies of optical constants of \SampleTThirtySix.
} \label{fig.nk-T36}
\end{figure}

The \AlTaO layer is modeled by two sublayers.
The lower sublayer is homogeneous and the upper sublayer is assumed to have a linear response function profile.
Thus, the \AlTaO layer can be characterized by three response functions, which are shown in Figures~\ref{fig.nk-T32}, \ref{fig.nk-T33}, \ref{fig.nk-T34}, \ref{fig.nk-T35}, \ref{fig.nk-T36} and two thicknesses given in Tables~\ref{tab.par-T32}, \ref{tab.par-T33}, \ref{tab.par-T34}, \ref{tab.par-T35}, \ref{tab.par-T36}.

The three response functions describing the inhomogeneous \AlTaO layer are modeled using identical parameters except for the transition strength parameters describing the electronic excitations of extended states and the transition strength parameters describing the electronic excitations involving localized states.
The response function profile expresses the inhomogeneity of the density of localized states along the layer deposition.

Moreover, the upper boundary of the layer is assumed to be rough. The roughness is included using Rayleigh--Rice theory~\cite{Rice51a,Franta05c} in a simplified matrix formalism~\cite{Franta08e},
here we assume that the layer thickness is much larger than the roughness autocorrelation length. The roughness parameters are also given in the tables.

\clearpage

\subsection{Results of optical characterization of \SampleTThirtyTwo}

\begin{table}[h!]
\caption{Basic parameter characterizing \SampleTThirtyTwo.}\label{tab.par-T32}
\centering\begin{tblr}{lc}
 \hline
 parameter & value \\
 \hline
 total thickness of the \SampleTThirtyTwo & $771.7$\,nm \\
 $d_{\rm f}$ thickness of the upper inhomogeneous layer & $684.7\pm0.5$\,nm \\
 $d_{\rm b}$ thickness of the bottom homogeneous layer & $87.0\pm0.4$\,nm \\
 $E_{\rm g}$ bandgap of the \AlTaO & $7.32\pm0.09$\,nm \\
 $\sigma$ rms of the heights of roughness & $10.78\pm0.05$\,nm \\
 $\tau$ autocorrelation length of roughness & $39.7\pm0.2$\,nm \\
 \hline
\end{tblr}
\end{table}

\begin{figure}[h!]
\includegraphics[width=\textwidth]{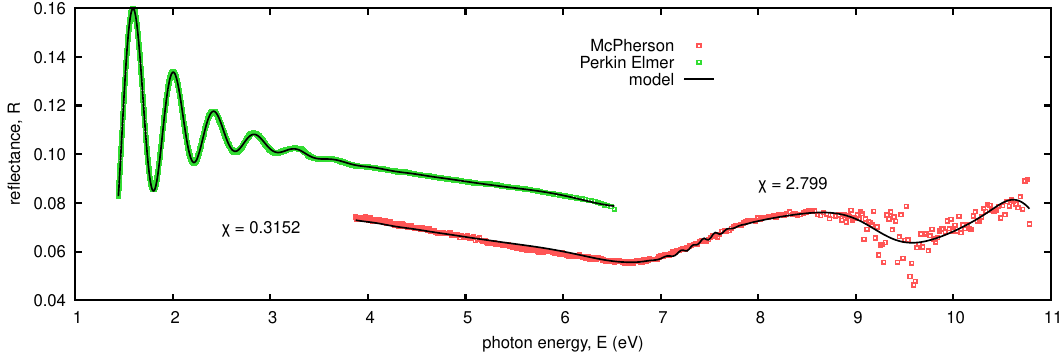}
\caption{Spectral dependencies of near-normal reflectance $R$ of \SampleTThirtyTwo.
} \label{fig.R-VUVV-T32}
\end{figure}

\begin{figure}[h!]
\includegraphics[width=\textwidth]{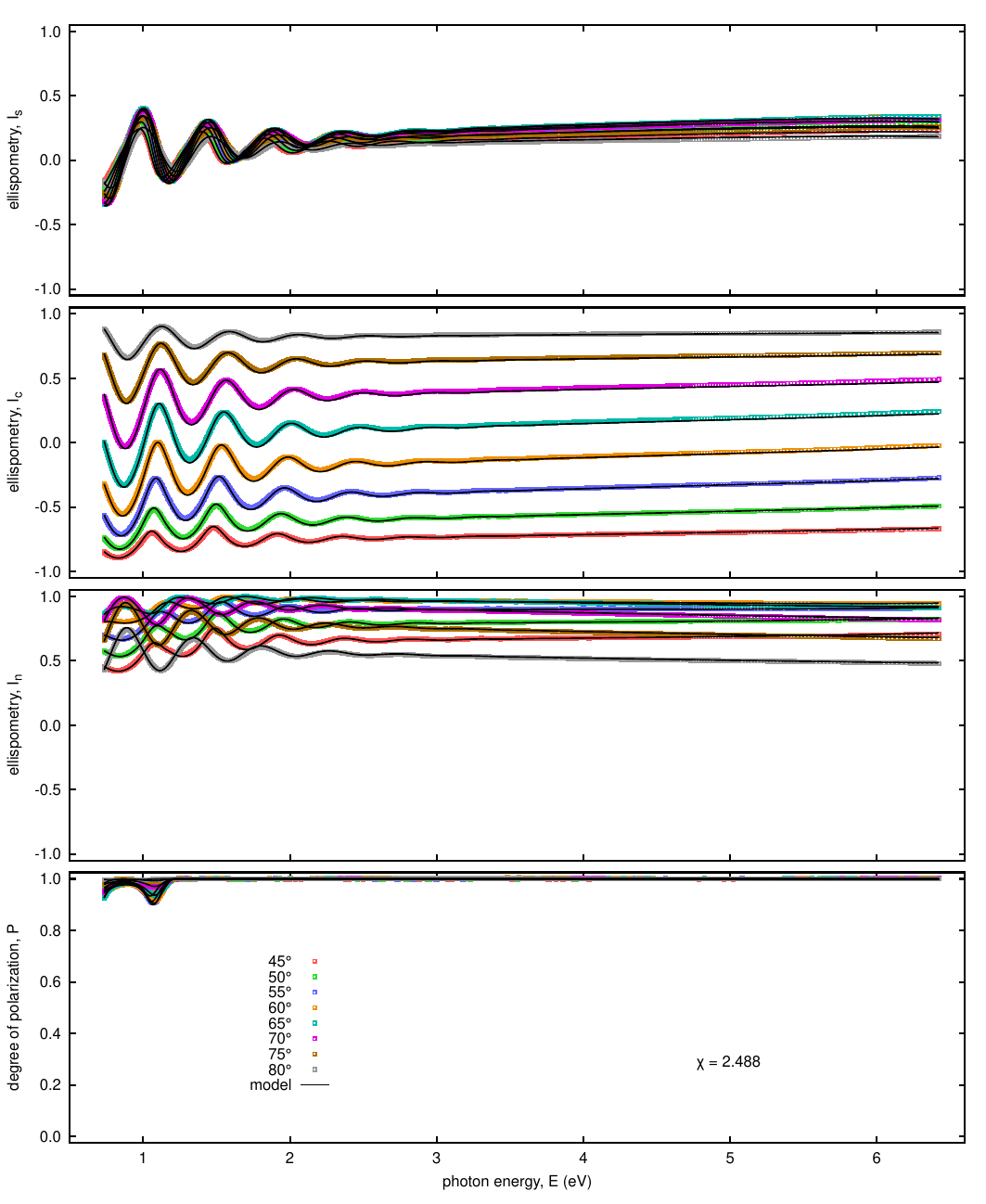}
\caption{Spectral dependencies of independent normalized Mueller matrix elements $\Ell$ in reflected light from front side of \SampleTThirtyTwo.
} \label{fig.E-UVNf-T32}
\end{figure}

\begin{figure}[h!]
\includegraphics[width=\textwidth]{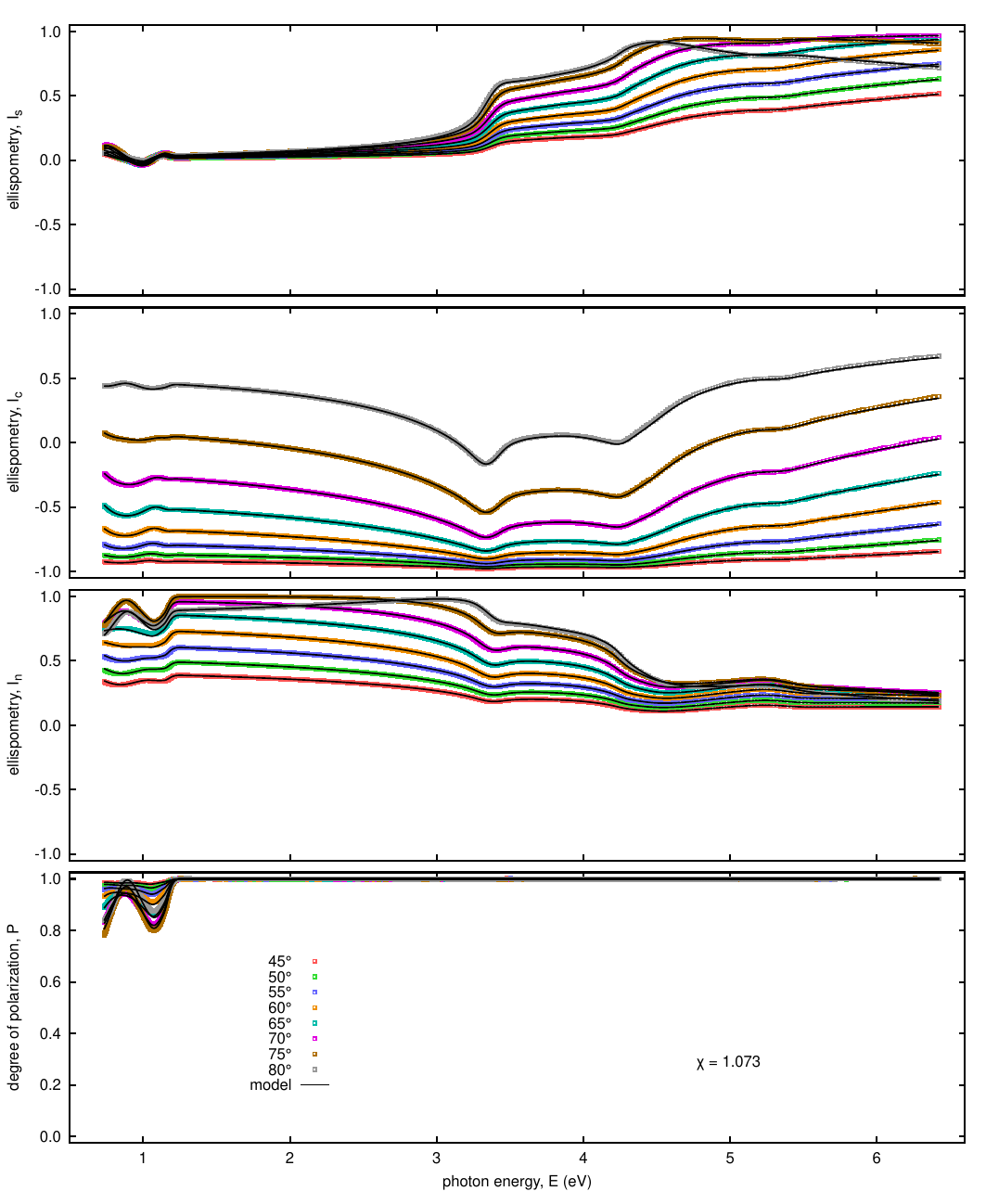}
\caption{Spectral dependencies of independent normalized Mueller matrix elements $\Ell$ in reflected light from back side of \SampleTThirtyTwo.
} \label{fig.E-UVNb-T32}
\end{figure}

\begin{figure}[h!]
\includegraphics[width=\textwidth]{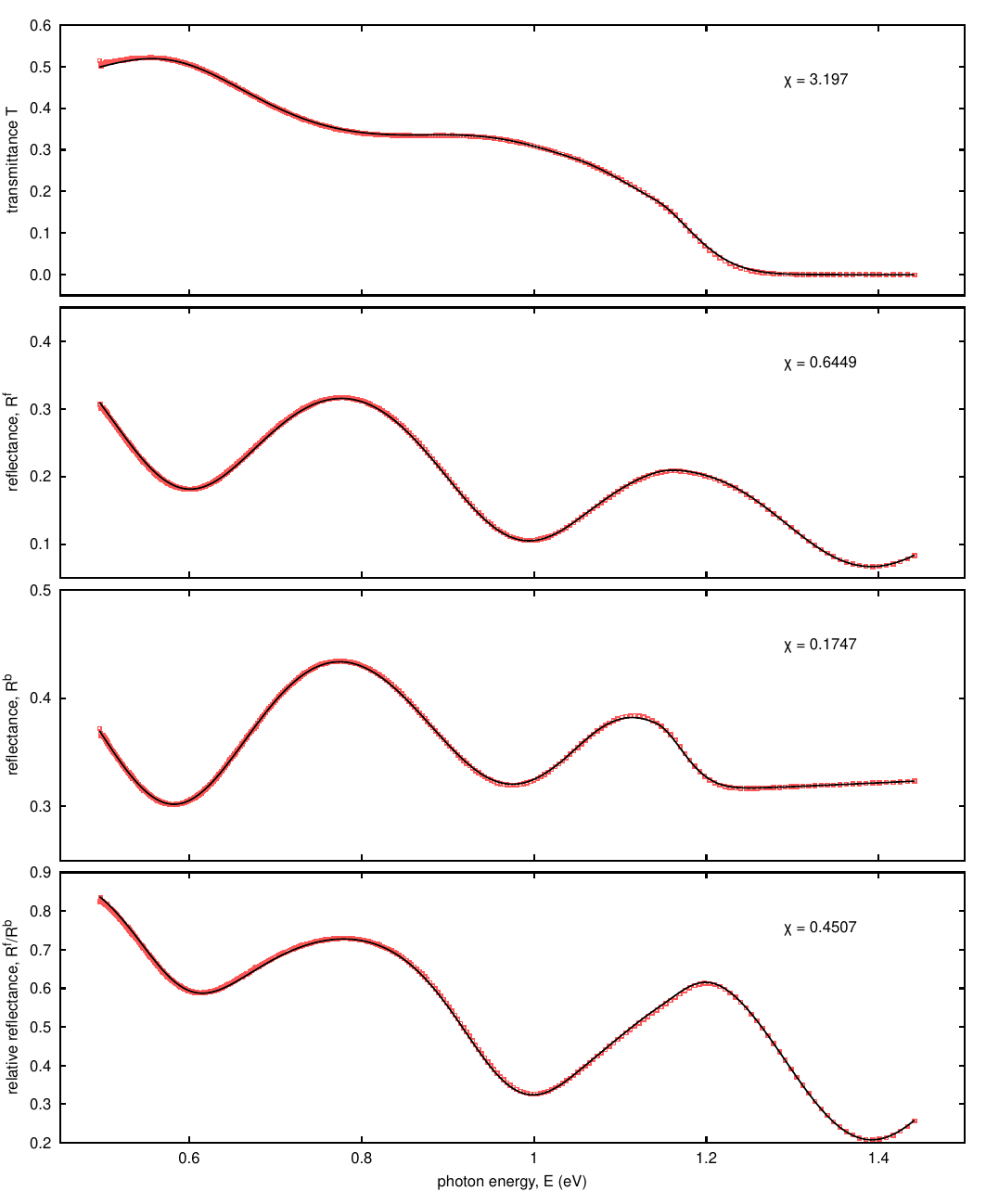}
\caption{Spectral dependencies of transmittance $T$, reflectance from front side $R^{\rm f}$, reflectance from back side $R^{\rm b}$ and relative reflectance $R^{\rm f}/R^{\rm b}$ of \SampleTThirtyTwo.
Measured in near-IR region by Perkin Elmer Lambda 1050+ spectrophotometer.
} \label{fig.TR-NIR-T32}
\end{figure}

\begin{figure}[h!]
\includegraphics[width=\textwidth]{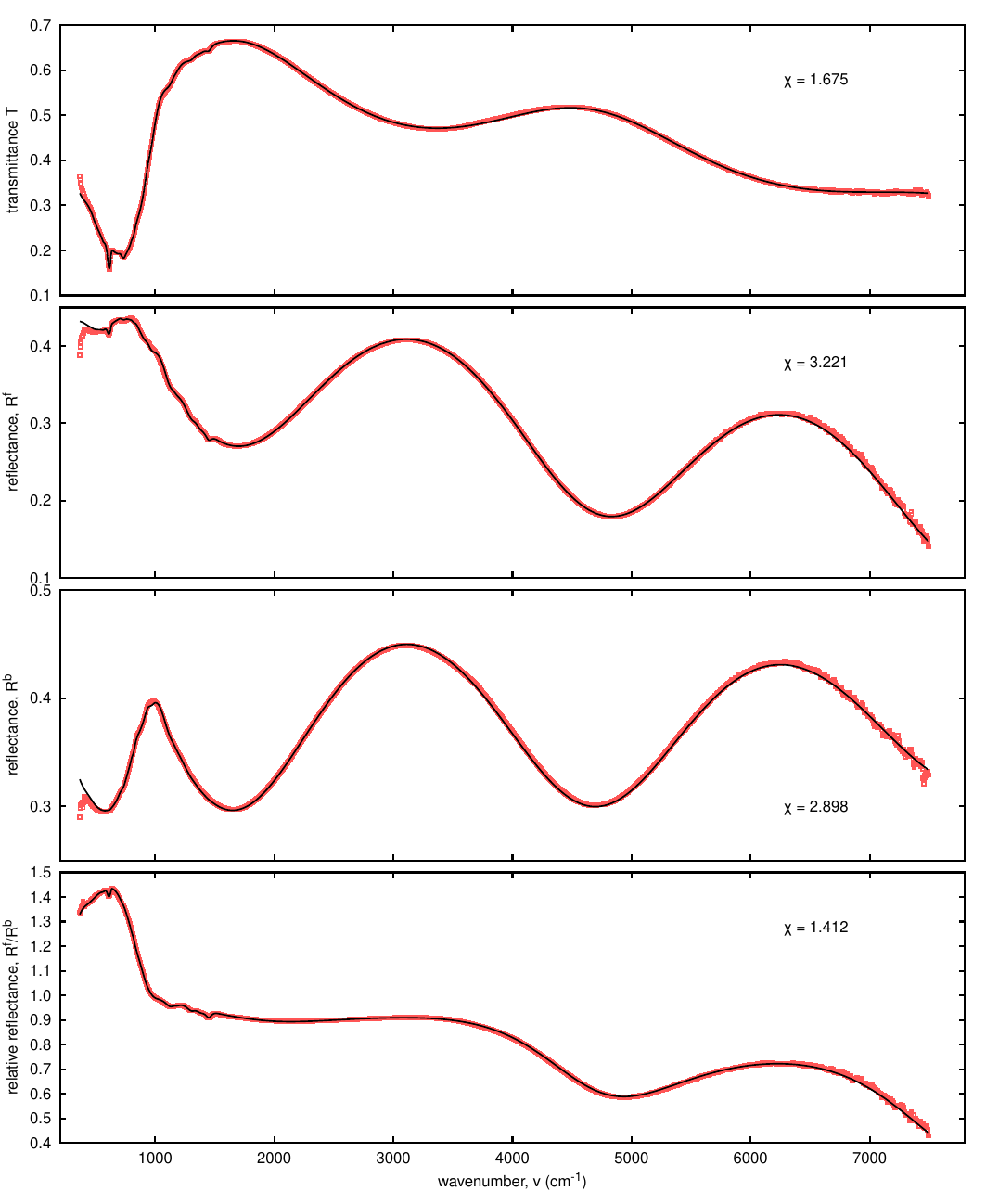}
\caption{Spectral dependencies of transmittance $T$, reflectance from front side $R^{\rm f}$, reflectance from back side $R^{\rm b}$ and relative reflectance $R^{\rm f}/R^{\rm b}$ of \SampleTThirtyTwo.
Measured in mid-IR region by Bruker Vertex 80v spectrophotometer.
} \label{fig.TR-MIR-T32}
\end{figure}

\begin{figure}[h!]
\includegraphics[width=\textwidth]{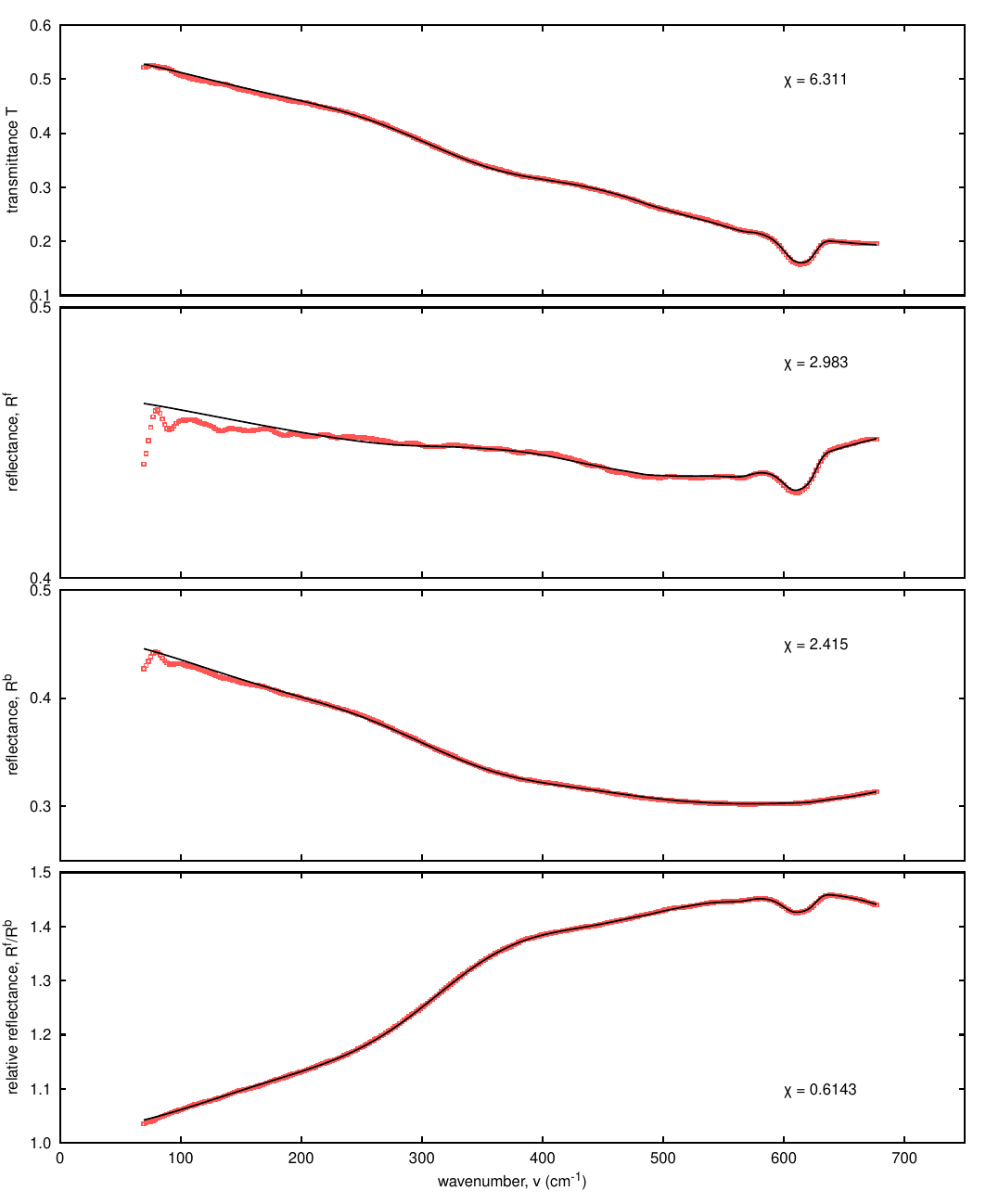}
\caption{Spectral dependencies of transmittance $T$, reflectance from front side $R^{\rm f}$, reflectance from back side $R^{\rm b}$ and relative reflectance $R^{\rm f}/R^{\rm b}$ of \SampleTThirtyTwo.
Measured in far-IR region by Bruker Vertex 80v spectrophotometer.
} \label{fig.TR-FIR-T32}
\end{figure}

\begin{figure}[h!]
\includegraphics[width=\textwidth]{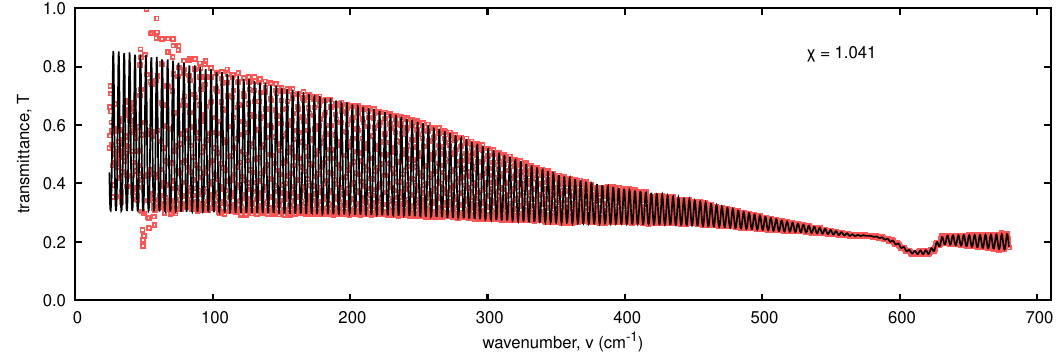}
\caption{Spectral dependencies of high-resolution transmittance $T$ of \SampleTThirtyTwo.
Measured in far-IR region by Bruker Vertex 70v spectrophotometer.
} \label{fig.T-FIRHR-T32}
\end{figure}

\clearpage
\subsection{Results of optical characterization of \SampleTThirtyThree}

\begin{table}[h!]
\caption{Basic parameter characterizing \SampleTThirtyThree.}\label{tab.par-T33}
\centering\begin{tblr}{lc}
 \hline
 parameter & value \\
 \hline
 total thickness of the \SampleTThirtyThree & $724.4$\,nm \\
 $d_{\rm f}$ thickness of the upper inhomogeneous layer & $595.0\pm0.6$\,nm \\
 $d_{\rm b}$ thickness of the bottom homogeneous layer & $129.4\pm0.6$\,nm \\
 $E_{\rm g}$ bandgap of the \AlTaO & $6.8\pm0.2$\,nm \\
 $\sigma$ rms of the heights of roughness & $9.36\pm0.09$\,nm \\
 $\tau$ autocorrelation length of roughness & $34.7\pm0.5$\,nm \\
 \hline
\end{tblr}
\end{table}

\begin{figure}[h!]
\includegraphics[width=\textwidth]{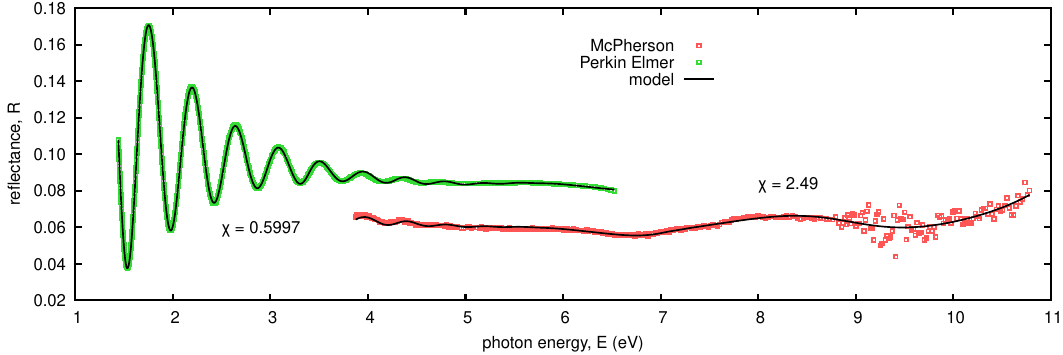}
\caption{Spectral dependencies of near-normal reflectance $R$ of \SampleTThirtyThree.
} \label{fig.R-VUVV-T33}
\end{figure}

\begin{figure}[h!]
\includegraphics[width=\textwidth]{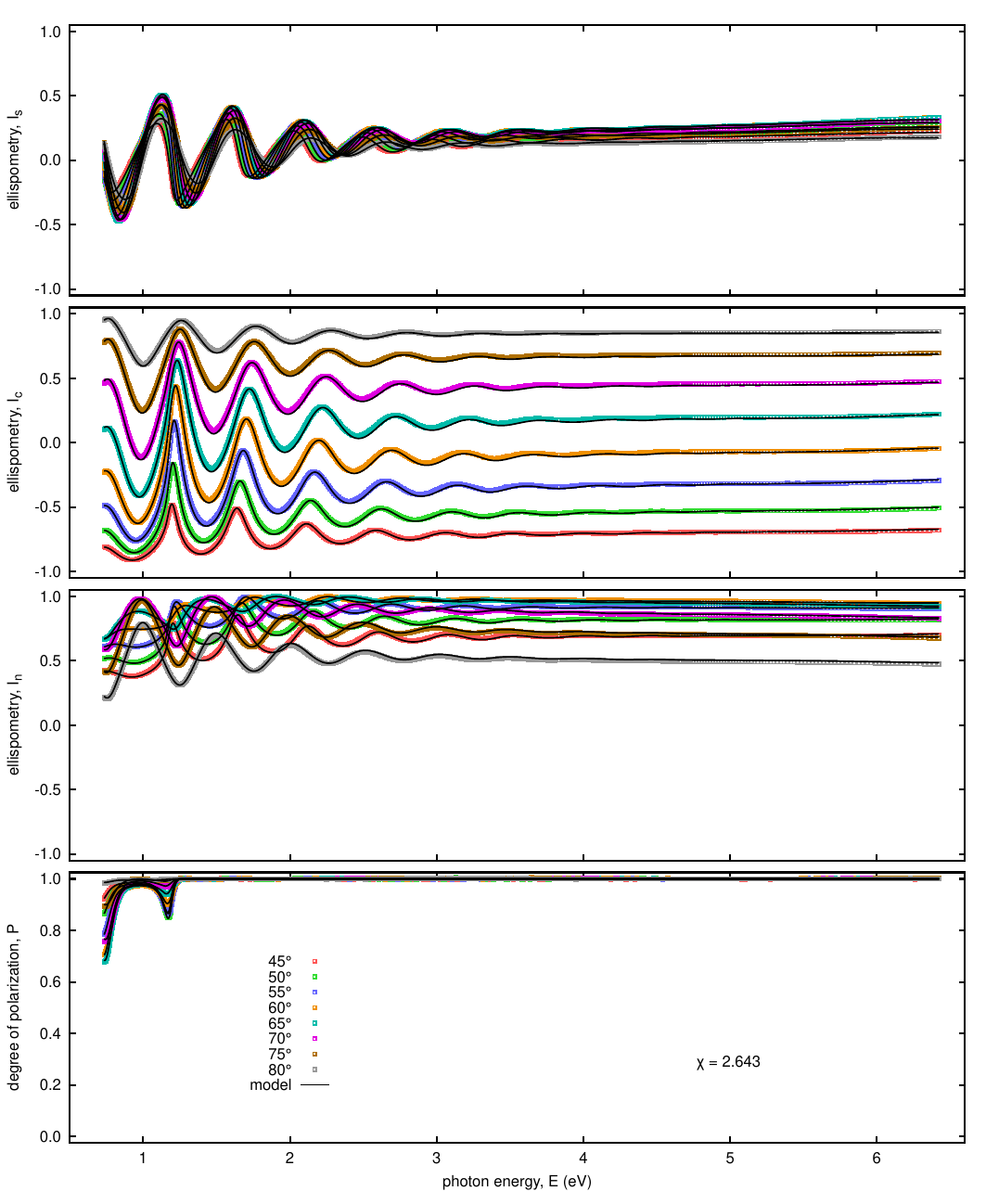}
\caption{Spectral dependencies of independent normalized Mueller matrix elements $\Ell$ in reflected light from front side of \SampleTThirtyThree.
} \label{fig.E-UVNf-T33}
\end{figure}

\begin{figure}[h!]
\includegraphics[width=\textwidth]{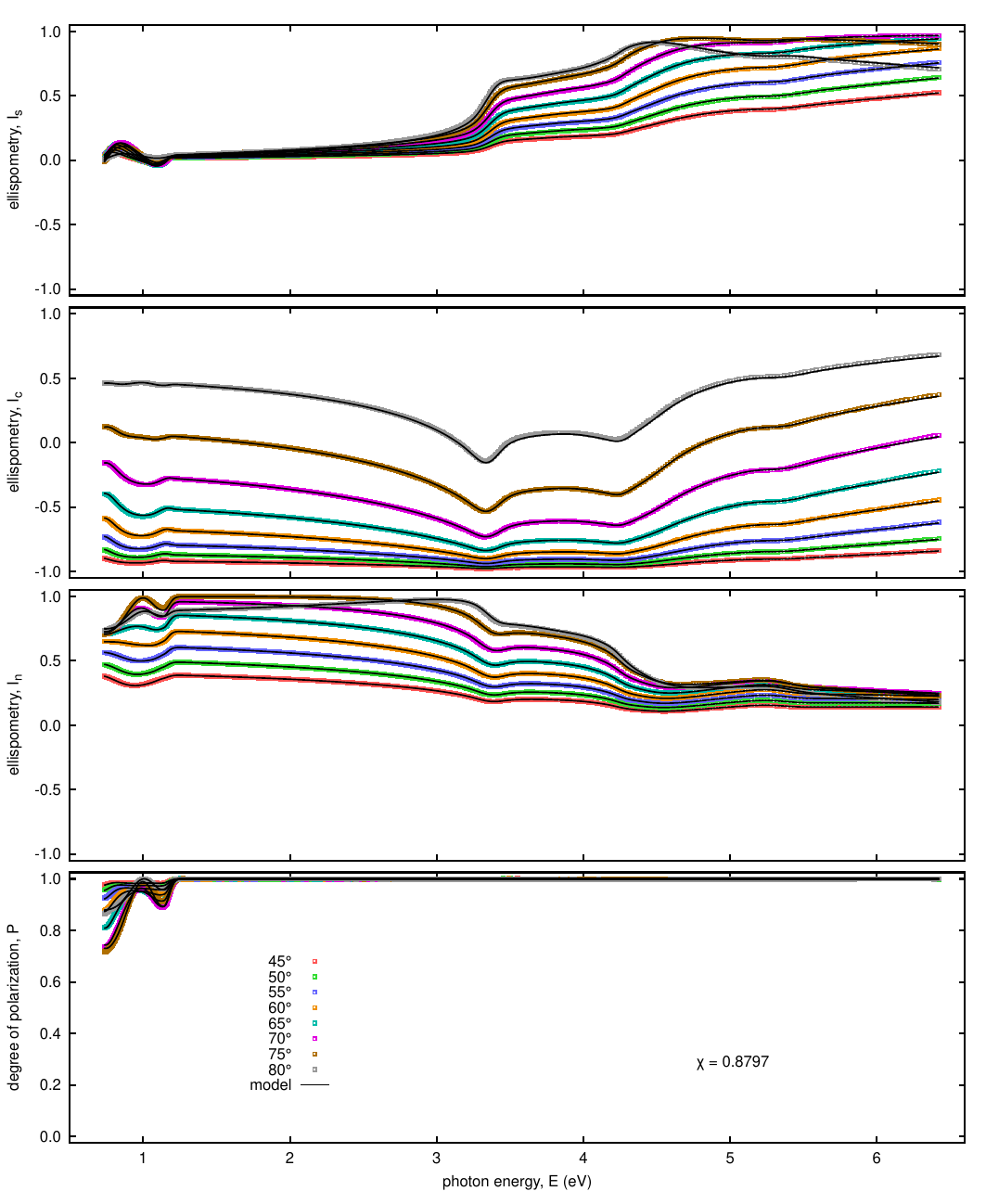}
\caption{Spectral dependencies of independent normalized Mueller matrix elements $\Ell$ in reflected light from back side of \SampleTThirtyThree.
} \label{fig.E-UVNb-T33}
\end{figure}

\begin{figure}[h!]
\includegraphics[width=\textwidth]{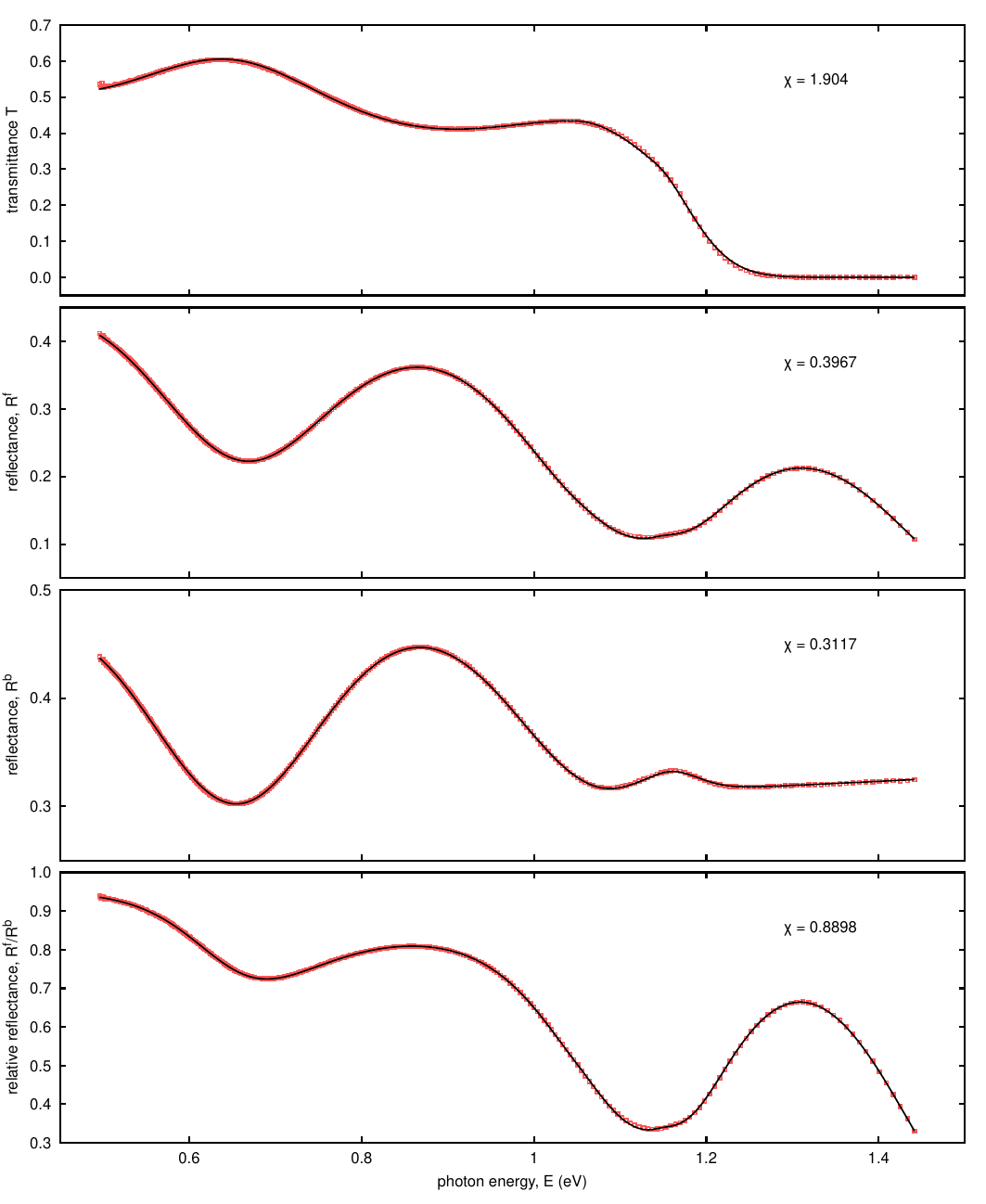}
\caption{Spectral dependencies of transmittance $T$, reflectance from front side $R^{\rm f}$, reflectance from back side $R^{\rm b}$ and relative reflectance $R^{\rm f}/R^{\rm b}$ of \SampleTThirtyThree.
Measured in near-IR region by Perkin Elmer Lambda 1050+ spectrophotometer.
} \label{fig.TR-NIR-T33}
\end{figure}

\begin{figure}[h!]
\includegraphics[width=\textwidth]{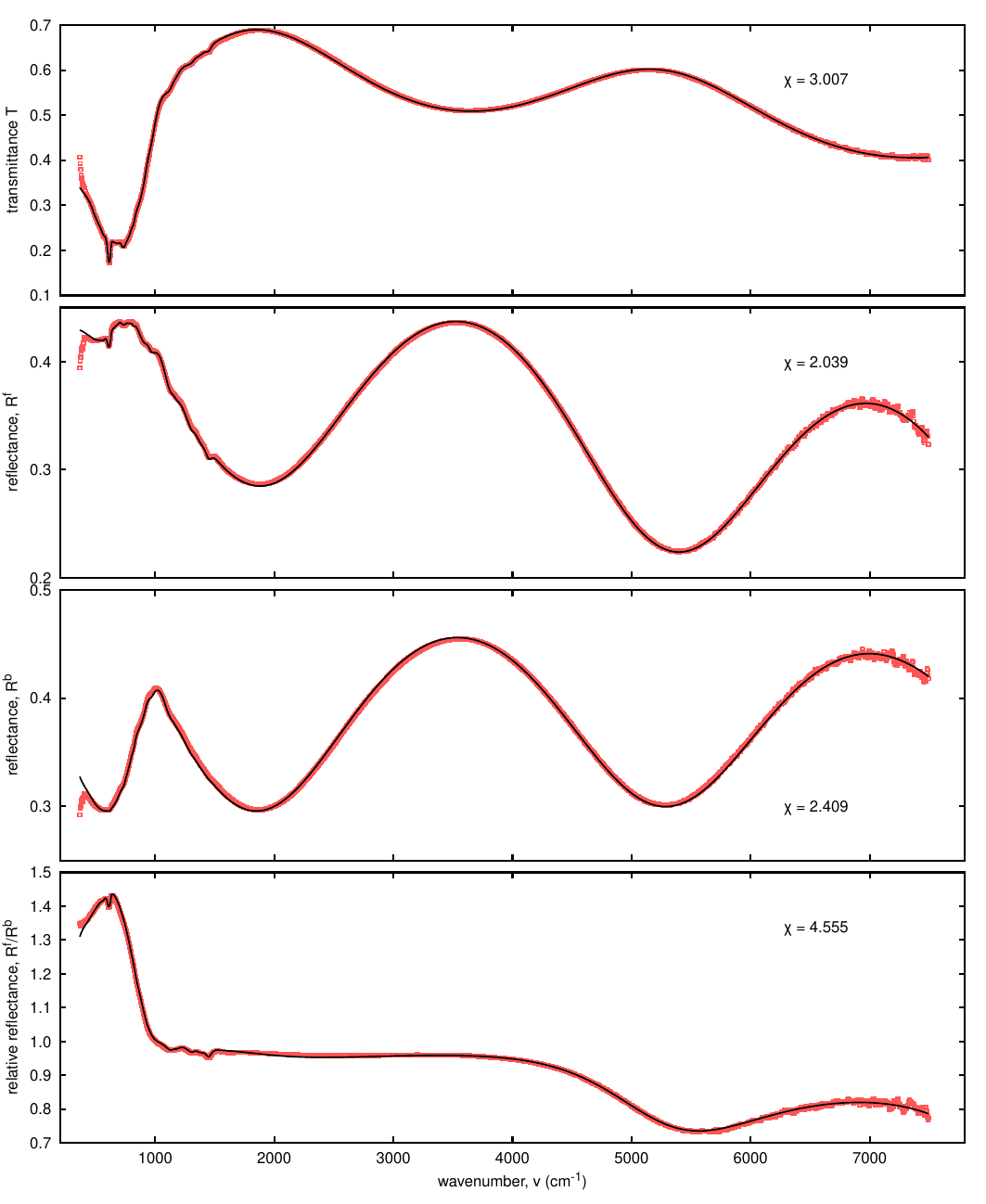}
\caption{Spectral dependencies of transmittance $T$, reflectance from front side $R^{\rm f}$, reflectance from back side $R^{\rm b}$ and relative reflectance $R^{\rm f}/R^{\rm b}$ of \SampleTThirtyThree.
Measured in mid-IR region by Bruker Vertex 80v spectrophotometer.
} \label{fig.TR-MIR-T33}
\end{figure}

\begin{figure}[h!]
\includegraphics[width=\textwidth]{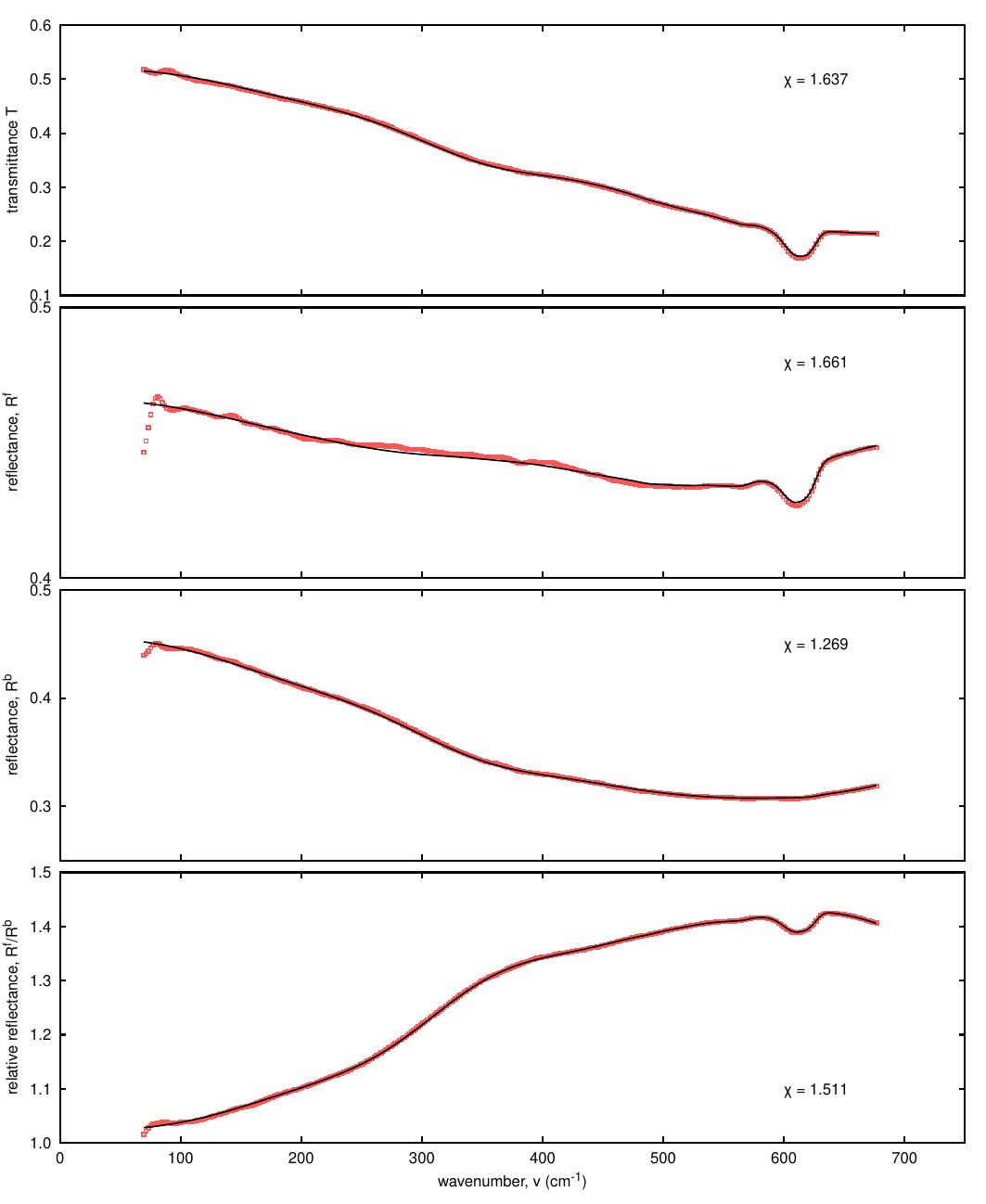}
\caption{Spectral dependencies of transmittance $T$, reflectance from front side $R^{\rm f}$, reflectance from back side $R^{\rm b}$ and relative reflectance $R^{\rm f}/R^{\rm b}$ of \SampleTThirtyThree.
Measured in far-IR region by Bruker Vertex 80v spectrophotometer.
} \label{fig.TR-FIR-T33}
\end{figure}

\begin{figure}[h!]
\includegraphics[width=\textwidth]{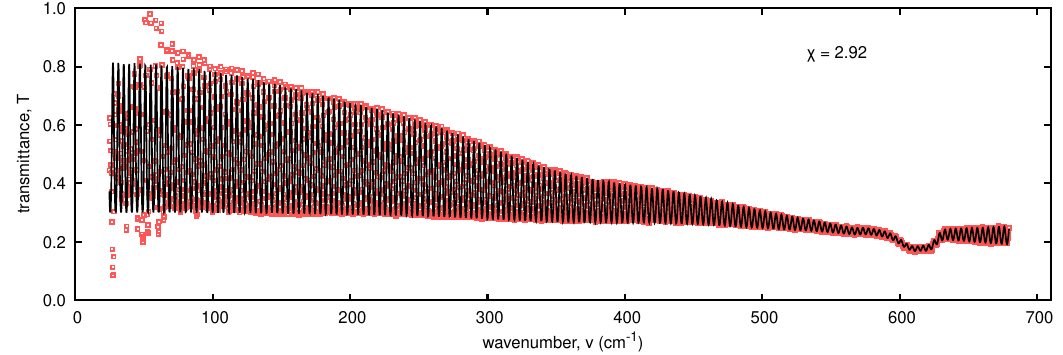}
\caption{Spectral dependencies of high-resolution transmittance $T$ of \SampleTThirtyThree.
Measured in far-IR region by Bruker Vertex 70v spectrophotometer.
} \label{fig.T-FIRHR-T33}
\end{figure}

\clearpage
\subsection{Results of optical characterization of \SampleTThirtyFour}

\begin{table}[h!]
\caption{Basic parameter characterizing \SampleTThirtyFour.}\label{tab.par-T34}
\centering\begin{tblr}{lc}
 \hline
 parameter & value \\
 \hline
 total thickness of the \SampleTThirtyFour & $848.7$\,nm \\
 $d_{\rm f}$ thickness of the upper inhomogeneous layer & $838.5\pm0.5$\,nm \\
 $d_{\rm b}$ thickness of the bottom homogeneous layer & $10.2\pm0.4$\,nm \\
 $E_{\rm g}$ bandgap of the \AlTaO & $4.778\pm0.005$\,nm \\
 $\sigma$ rms of the heights of roughness & $2.39\pm0.16$\,nm \\
 $\tau$ autocorrelation length of roughness & $6.1\pm0.9$\,nm \\
 \hline
\end{tblr}
\end{table}

\begin{figure}[h!]
\includegraphics[width=\textwidth]{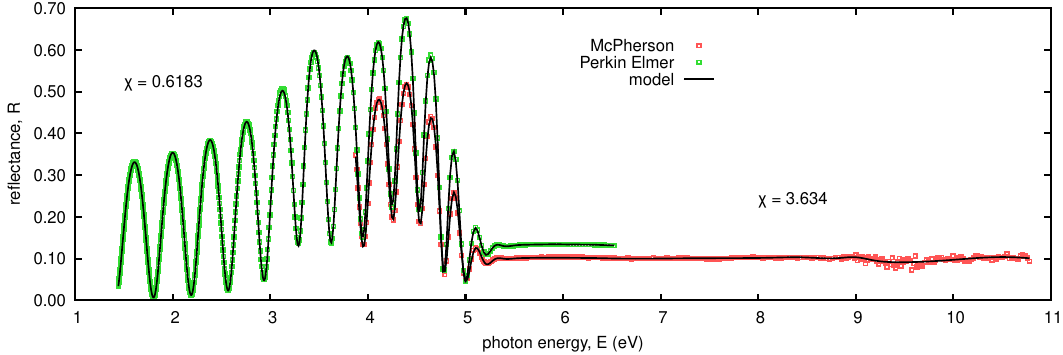}
\caption{Spectral dependencies of near-normal reflectance $R$ of \SampleTThirtyFour.
} \label{fig.R-VUVV-T34}
\end{figure}

\begin{figure}[h!]
\includegraphics[width=\textwidth]{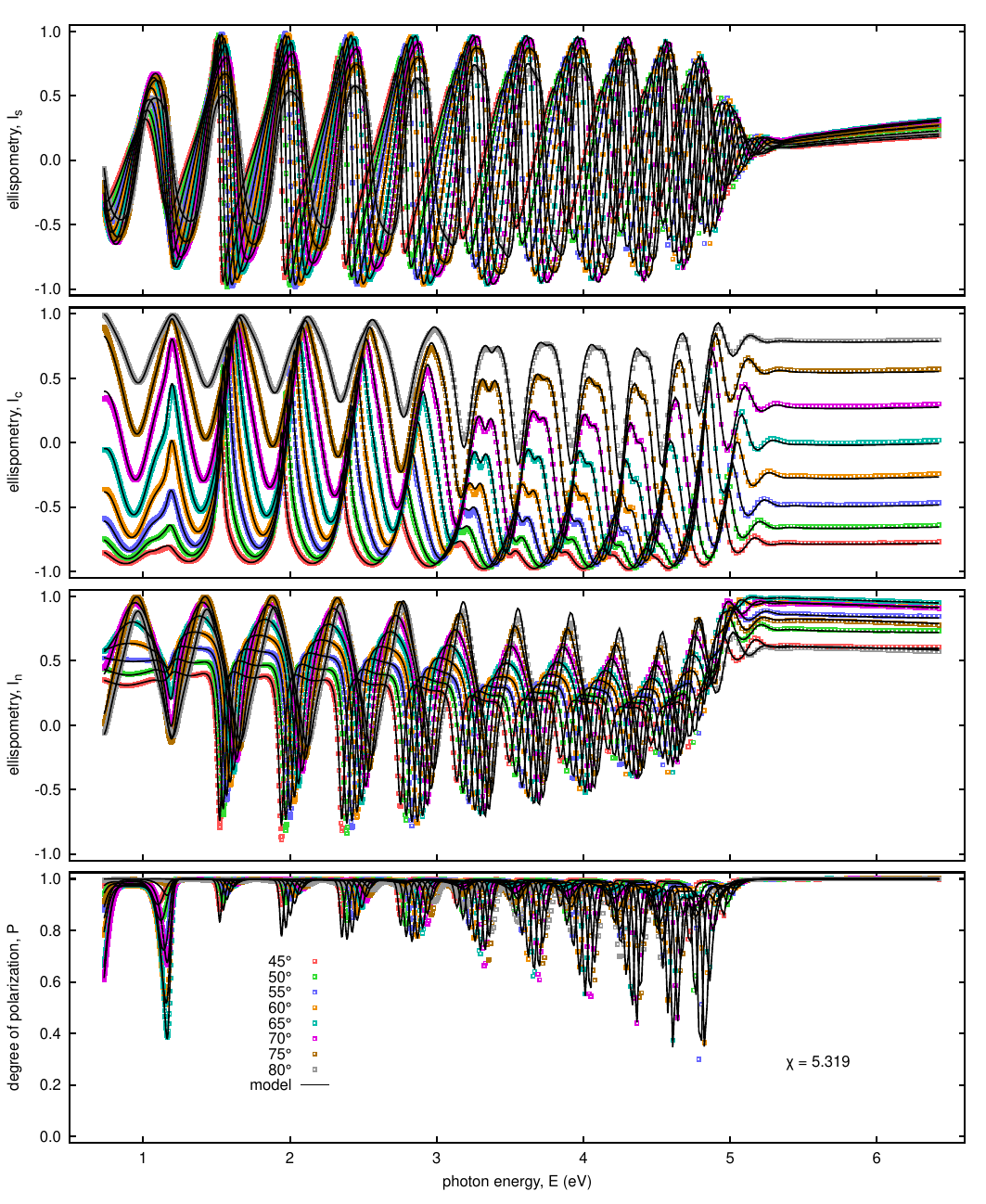}
\caption{Spectral dependencies of independent normalized Mueller matrix elements $\Ell$ in reflected light from front side of \SampleTThirtyFour.
} \label{fig.E-UVNf-T34}
\end{figure}

\begin{figure}[h!]
\includegraphics[width=\textwidth]{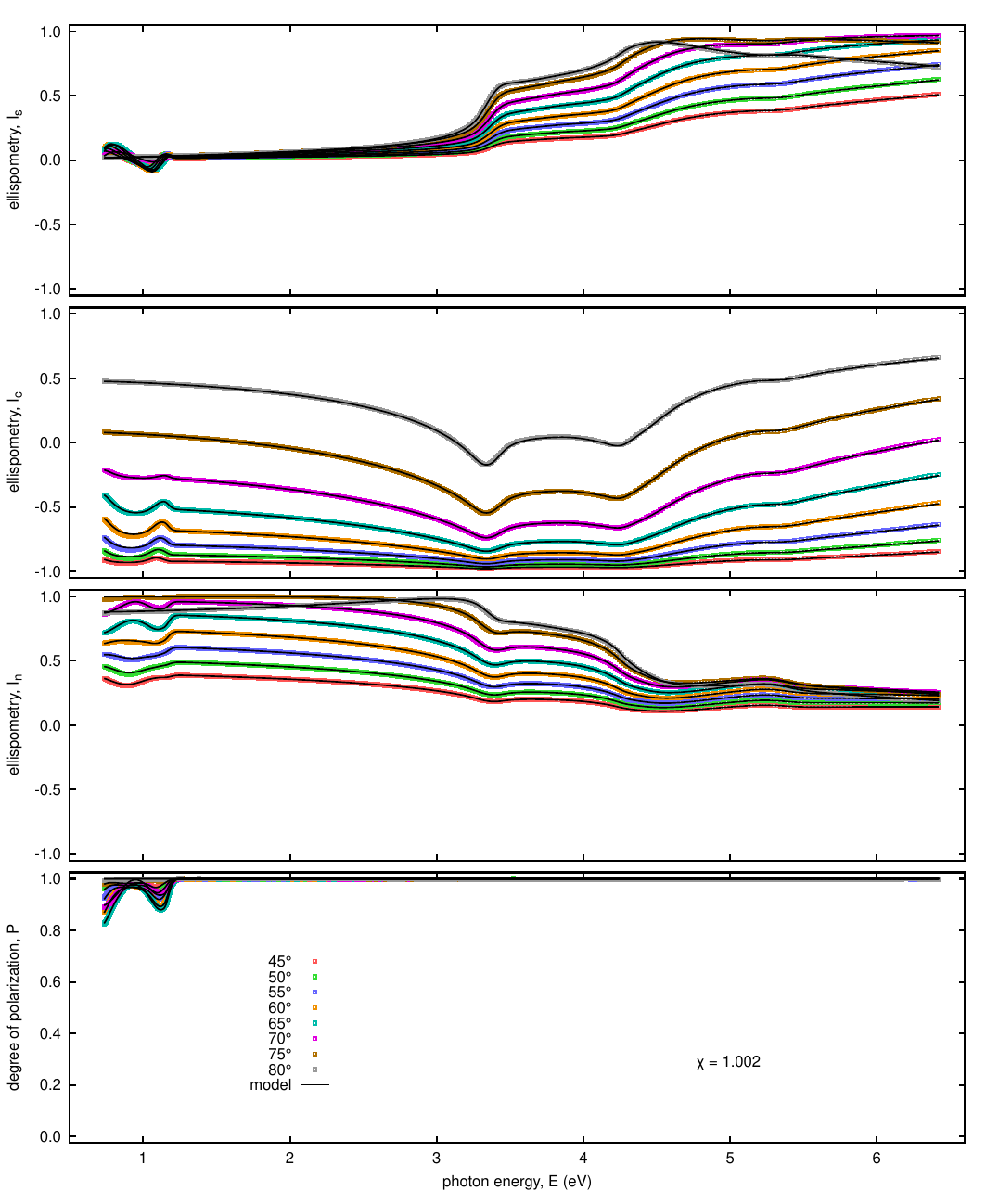}
\caption{Spectral dependencies of independent normalized Mueller matrix elements $\Ell$ in reflected light from back side of \SampleTThirtyFour.
} \label{fig.E-UVNb-T34}
\end{figure}

\begin{figure}[h!]
\includegraphics[width=\textwidth]{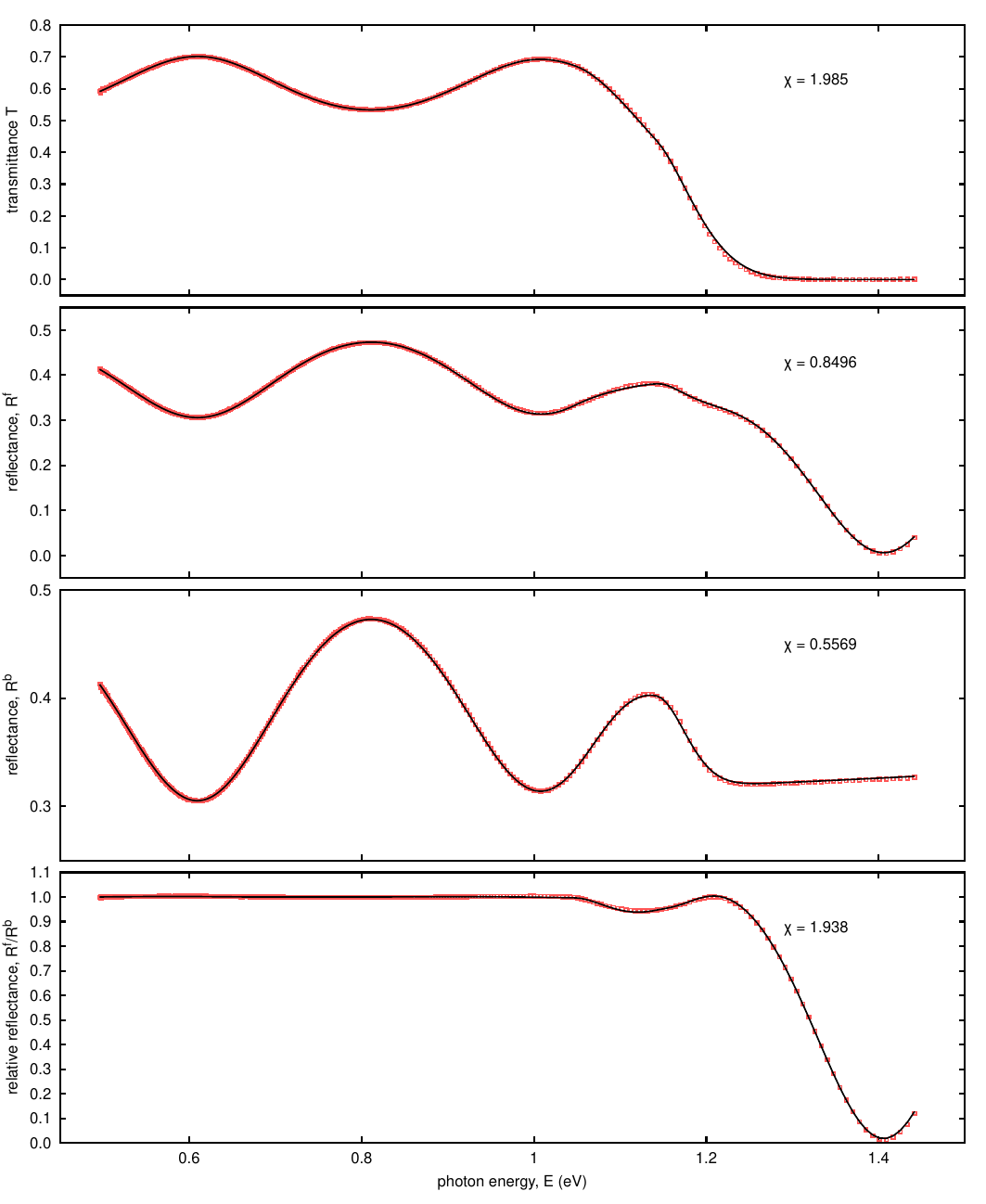}
\caption{Spectral dependencies of transmittance $T$, reflectance from front side $R^{\rm f}$, reflectance from back side $R^{\rm b}$ and relative reflectance $R^{\rm f}/R^{\rm b}$ of \SampleTThirtyFour.
Measured in near-IR region by Perkin Elmer Lambda 1050+ spectrophotometer.
} \label{fig.TR-NIR-T34}
\end{figure}

\begin{figure}[h!]
\includegraphics[width=\textwidth]{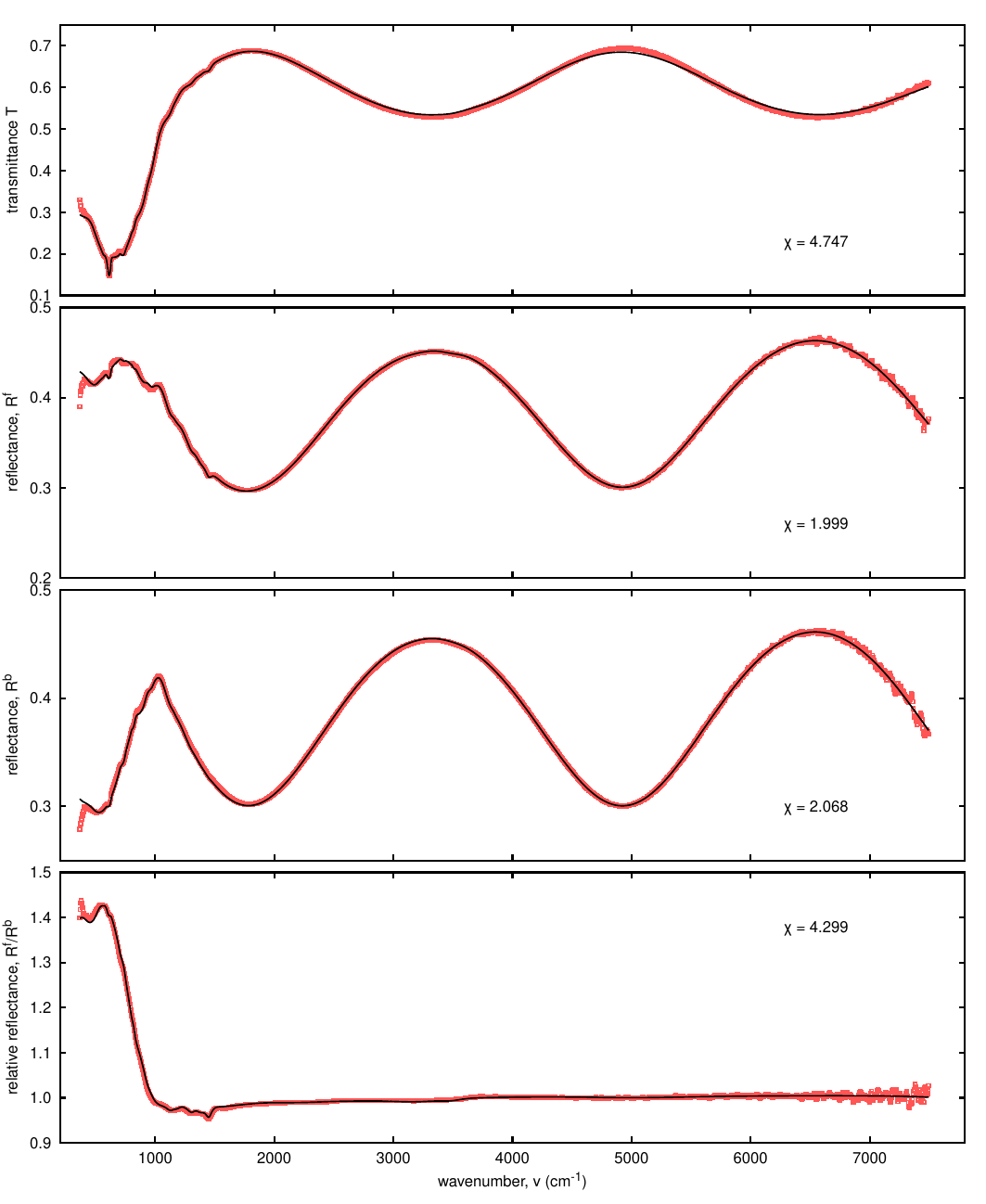}
\caption{Spectral dependencies of transmittance $T$, reflectance from front side $R^{\rm f}$, reflectance from back side $R^{\rm b}$ and relative reflectance $R^{\rm f}/R^{\rm b}$ of \SampleTThirtyFour.
Measured in mid-IR region by Bruker Vertex 80v spectrophotometer.
} \label{fig.TR-MIR-T34}
\end{figure}

\begin{figure}[h!]
\includegraphics[width=\textwidth]{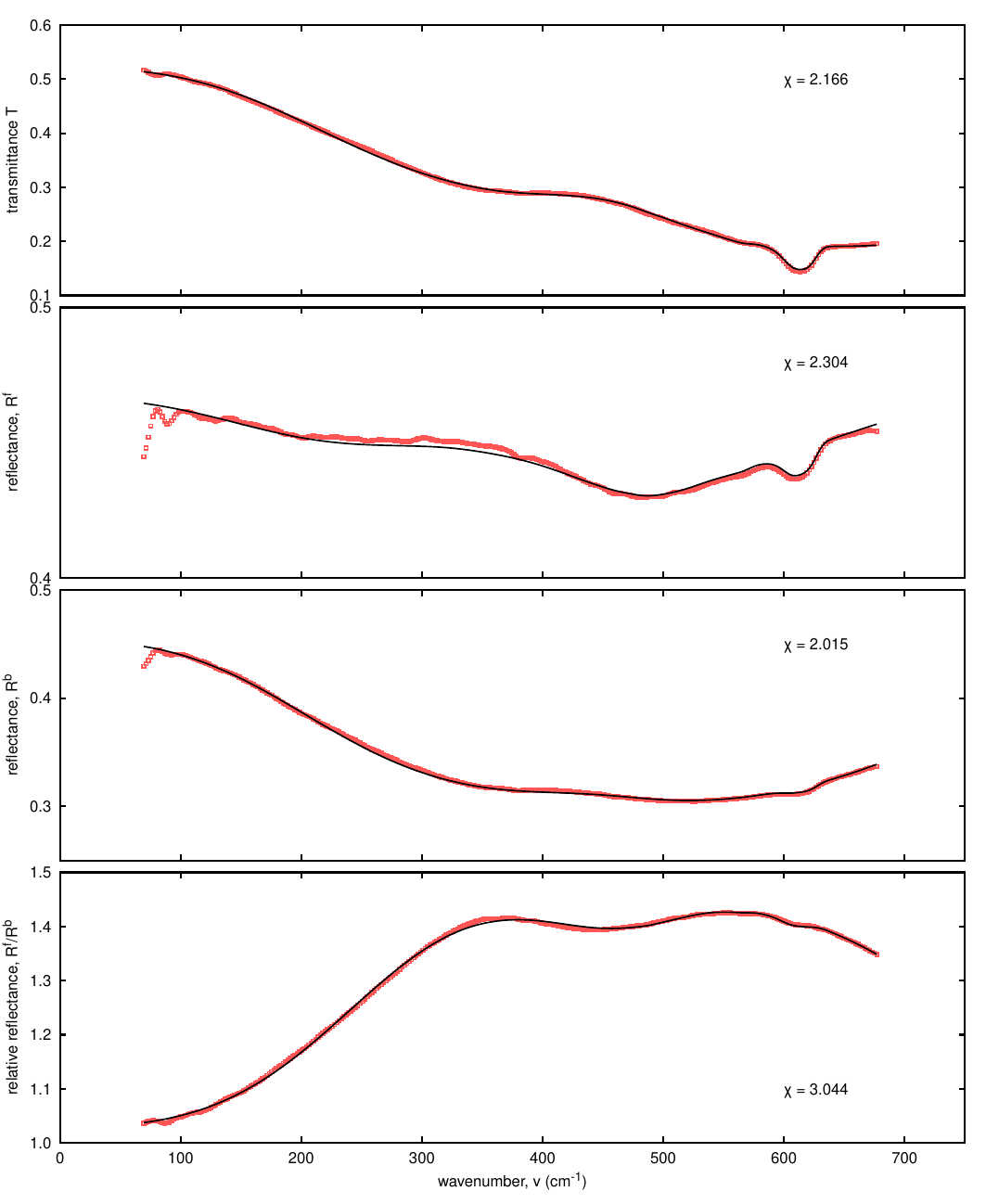}
\caption{Spectral dependencies of transmittance $T$, reflectance from front side $R^{\rm f}$, reflectance from back side $R^{\rm b}$ and relative reflectance $R^{\rm f}/R^{\rm b}$ of \SampleTThirtyFour.
Measured in far-IR region by Bruker Vertex 80v spectrophotometer.
} \label{fig.TR-FIR-T34}
\end{figure}

\begin{figure}[h!]
\includegraphics[width=\textwidth]{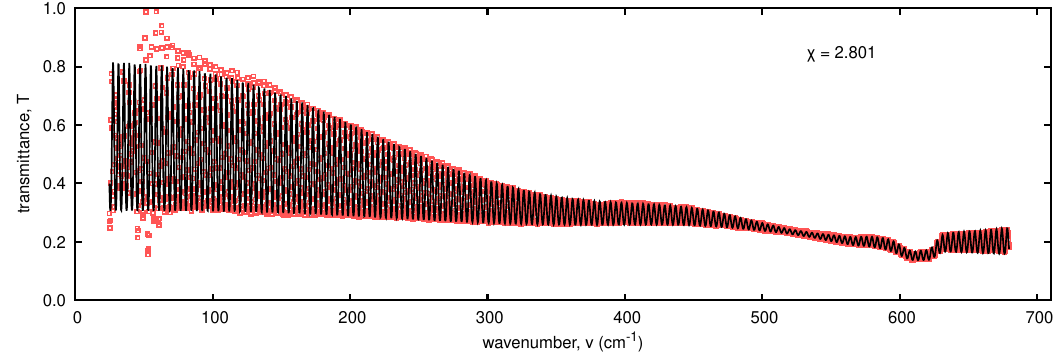}
\caption{Spectral dependencies of high-resolution transmittance $T$ of \SampleTThirtyFour.
Measured in far-IR region by Bruker Vertex 70v spectrophotometer.
} \label{fig.T-FIRHR-T34}
\end{figure}

\clearpage
\subsection{Results of optical characterization of \SampleTThirtyFive}

\begin{table}[h!]
\caption{Basic parameter characterizing \SampleTThirtyFive.}\label{tab.par-T35}
\centering\begin{tblr}{lc}
 \hline
 parameter & value \\
 \hline
 total thickness of the \SampleTThirtyFive & $329.8$\,nm \\
 $d_{\rm f}$ thickness of the upper inhomogeneous layer & $320.9\pm0.3$\,nm \\
 $d_{\rm b}$ thickness of the bottom homogeneous layer & $8.81\pm0.05$\,nm \\
 $E_{\rm g}$ bandgap of the \AlTaO & $4.657\pm0.005$\,nm \\
 $\sigma$ rms of the heights of roughness & $1.1\pm0.3$\,nm \\
 $\tau$ autocorrelation length of roughness & $3.2\pm1.5$\,nm \\
 \hline
\end{tblr}
\end{table}

\begin{figure}[h!]
\includegraphics[width=\textwidth]{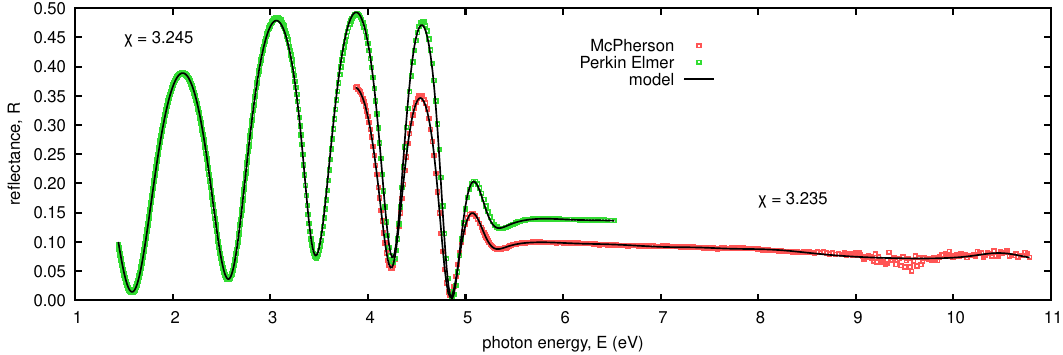}
\caption{Spectral dependencies of near-normal reflectance $R$ of \SampleTThirtyFive.
} \label{fig.R-VUVV-T35}
\end{figure}

\begin{figure}[h!]
\includegraphics[width=\textwidth]{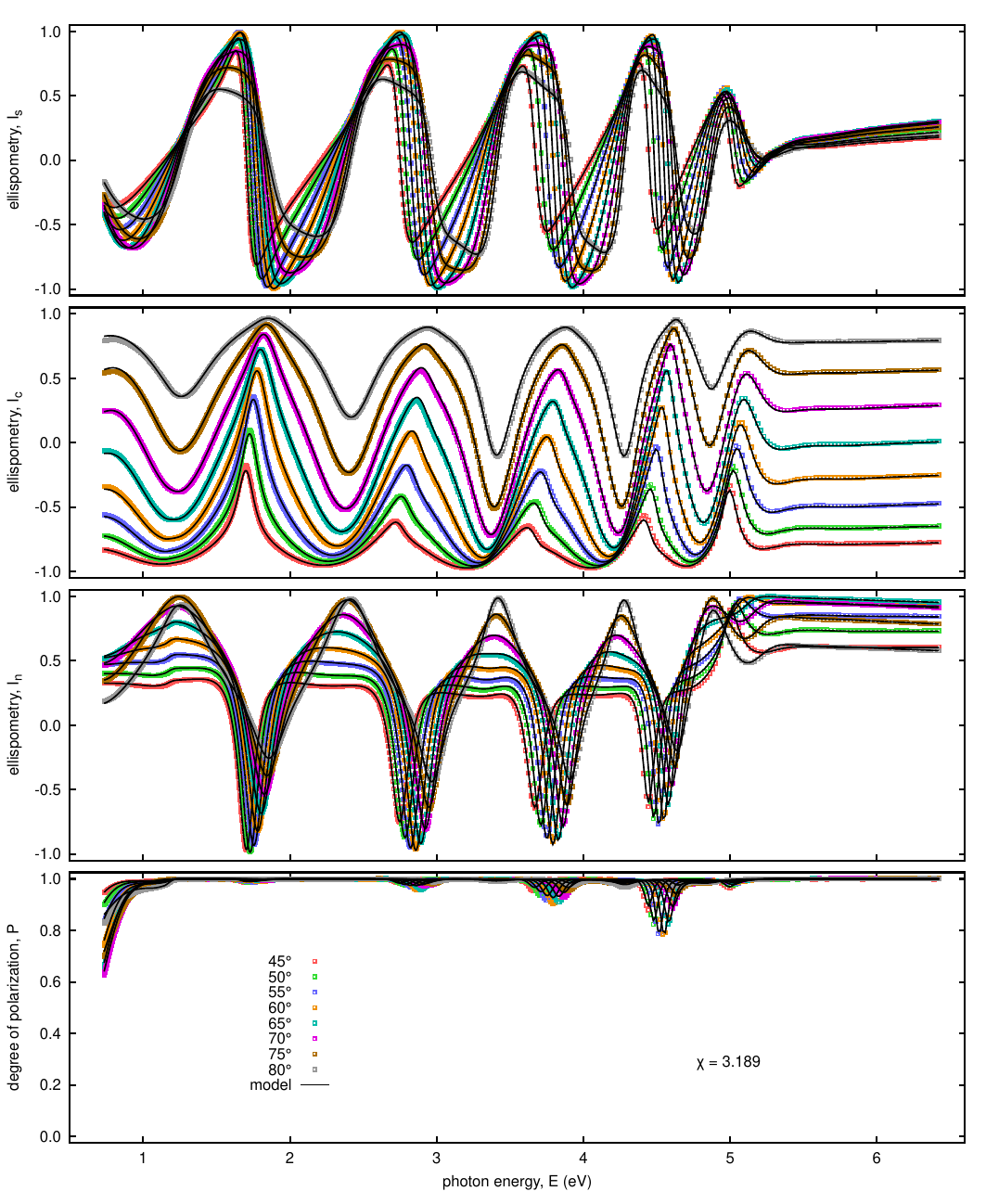}
\caption{Spectral dependencies of independent normalized Mueller matrix elements $\Ell$ in reflected light from front side of \SampleTThirtyFive.
} \label{fig.E-UVNf-T35}
\end{figure}

\begin{figure}[h!]
\includegraphics[width=\textwidth]{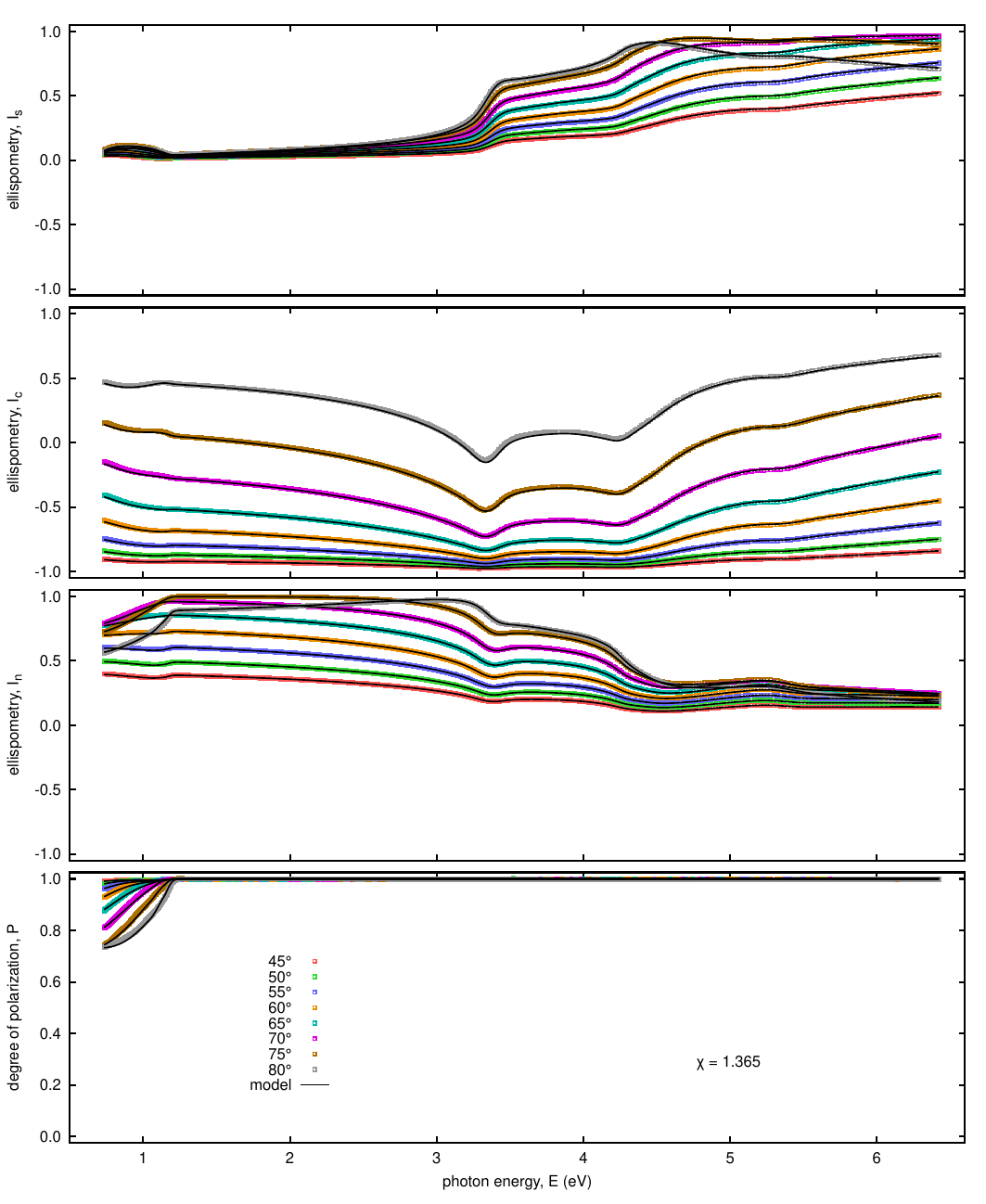}
\caption{Spectral dependencies of independent normalized Mueller matrix elements $\Ell$ in reflected light from back side of \SampleTThirtyFive.
} \label{fig.E-UVNb-T35}
\end{figure}

\begin{figure}[h!]
\includegraphics[width=\textwidth]{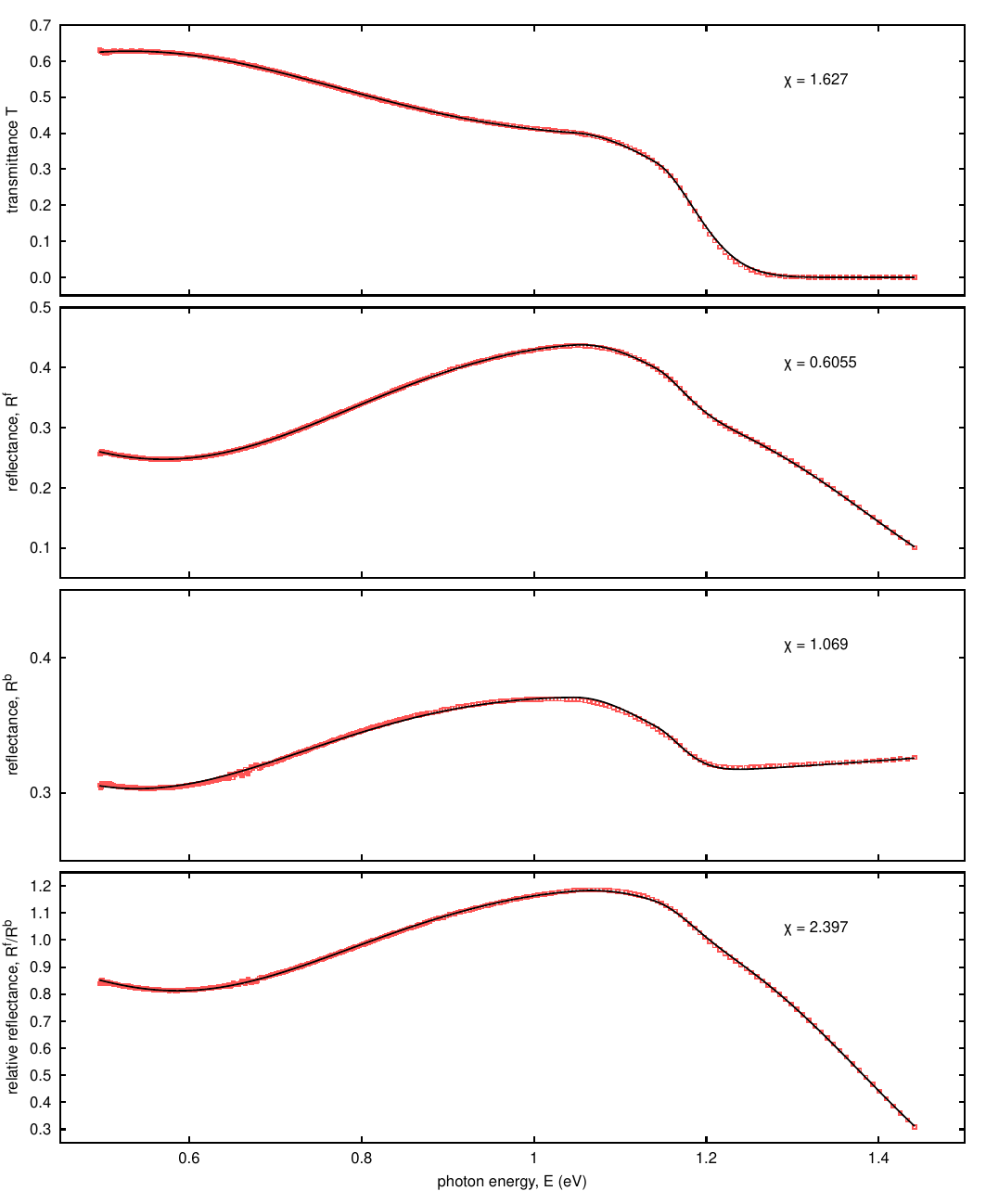}
\caption{Spectral dependencies of transmittance $T$, reflectance from front side $R^{\rm f}$, reflectance from back side $R^{\rm b}$ and relative reflectance $R^{\rm f}/R^{\rm b}$ of \SampleTThirtyFive.
Measured in near-IR region by Perkin Elmer Lambda 1050+ spectrophotometer.
} \label{fig.TR-NIR-T35}
\end{figure}

\begin{figure}[h!]
\includegraphics[width=\textwidth]{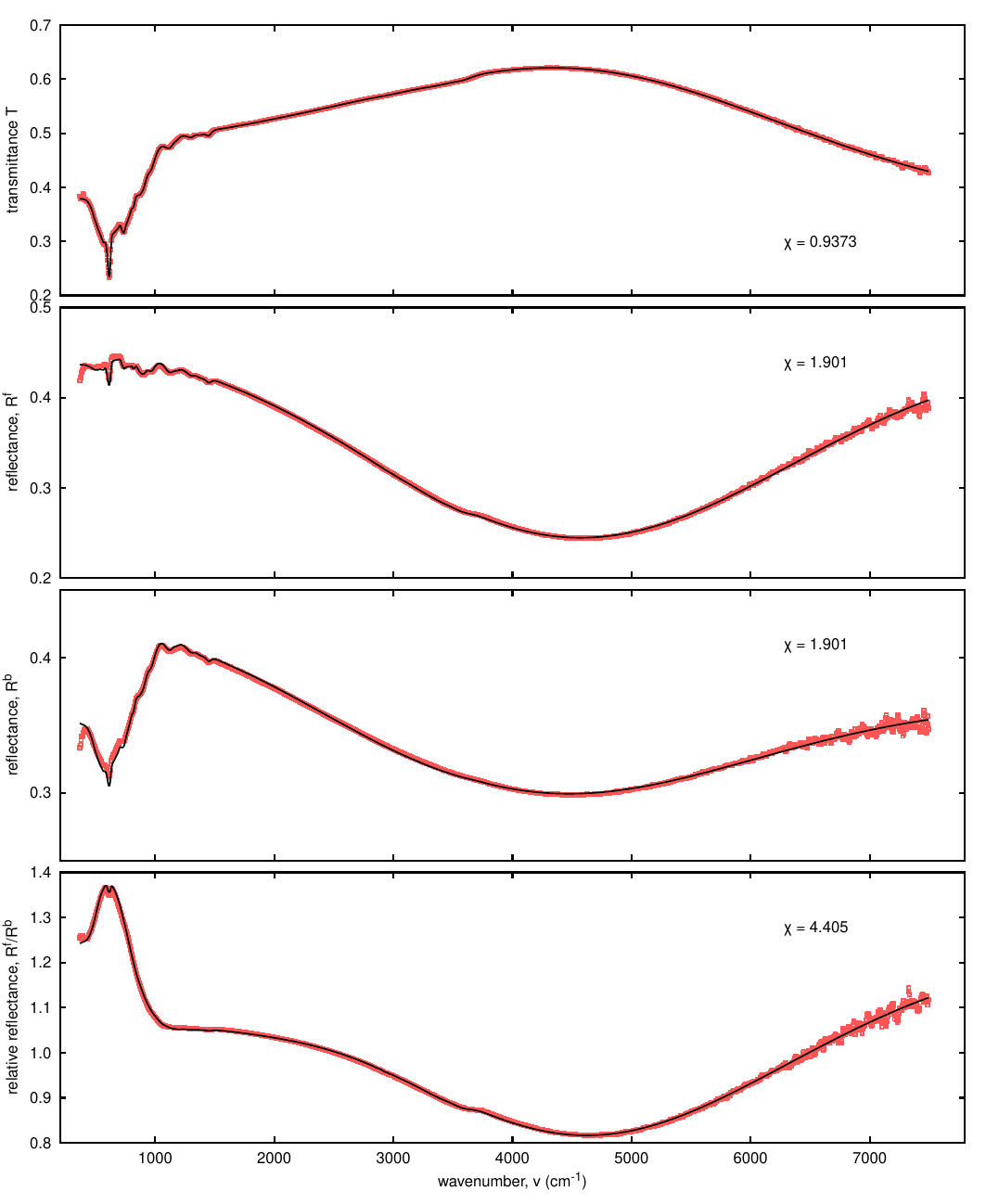}
\caption{Spectral dependencies of transmittance $T$, reflectance from front side $R^{\rm f}$, reflectance from back side $R^{\rm b}$ and relative reflectance $R^{\rm f}/R^{\rm b}$ of \SampleTThirtyFive.
Measured in mid-IR region by Bruker Vertex 80v spectrophotometer.
} \label{fig.TR-MIR-T35}
\end{figure}

\begin{figure}[h!]
\includegraphics[width=\textwidth]{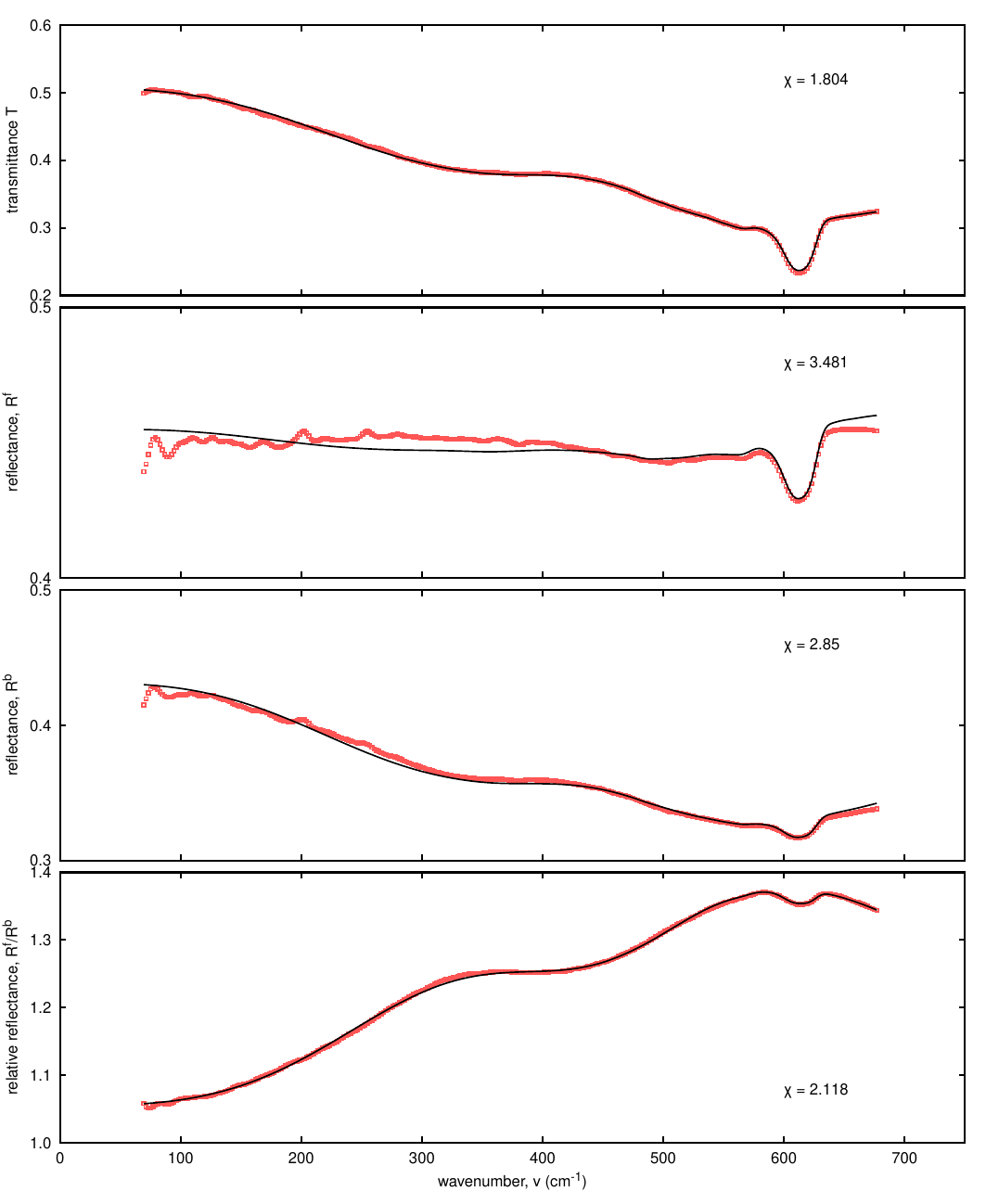}
\caption{Spectral dependencies of transmittance $T$, reflectance from front side $R^{\rm f}$, reflectance from back side $R^{\rm b}$ and relative reflectance $R^{\rm f}/R^{\rm b}$ of \SampleTThirtyFive.
Measured in far-IR region by Bruker Vertex 80v spectrophotometer.
} \label{fig.TR-FIR-T35}
\end{figure}

\begin{figure}[h!]
\includegraphics[width=\textwidth]{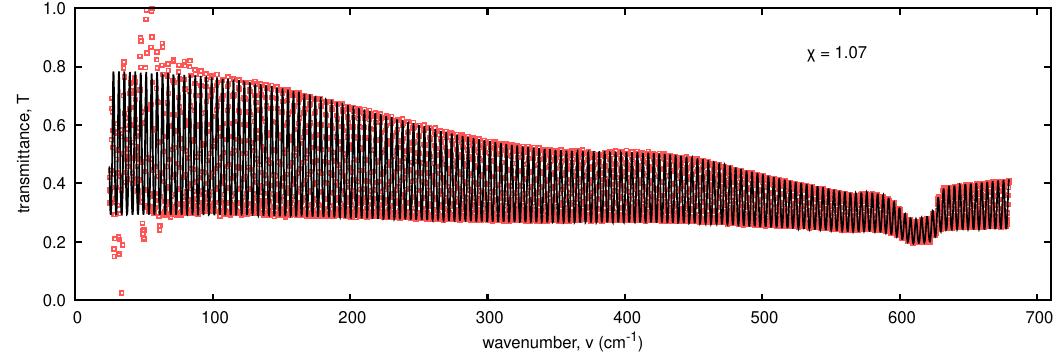}
\caption{Spectral dependencies of high-resolution transmittance $T$ of \SampleTThirtyFive.
Measured in far-IR region by Bruker Vertex 70v spectrophotometer.
} \label{fig.T-FIRHR-T35}
\end{figure}

\clearpage
\subsection{Results of optical characterization of \SampleTThirtySix}

\begin{table}[h!]
\caption{Basic parameter characterizing \SampleTThirtySix.}\label{tab.par-T36}
\centering\begin{tblr}{lc}
 \hline
 parameter & value \\
 \hline
 total thickness of the \SampleTThirtySix & $285.5$\,nm \\
 $d_{\rm f}$ thickness of the upper inhomogeneous layer & $272.9\pm0.5$\,nm \\
 $d_{\rm b}$ thickness of the bottom homogeneous layer & $12.6\pm0.2$\,nm \\
 $E_{\rm g}$ bandgap of the \AlTaO & $4.608\pm0.005$\,nm \\
 $\sigma$ rms of the heights of roughness & $2.3\pm0.2$\,nm \\
 $\tau$ autocorrelation length of roughness & $31\pm5$\,nm \\
 \hline
\end{tblr}
\end{table}

\begin{figure}[h!]
\includegraphics[width=\textwidth]{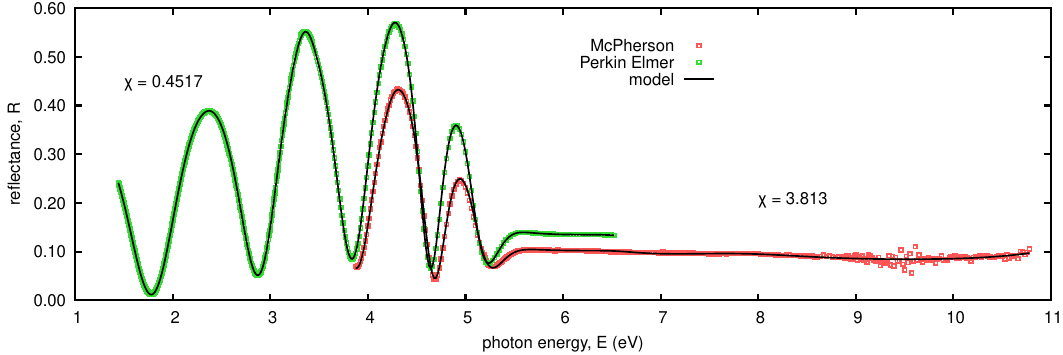}
\caption{Spectral dependencies of near-normal reflectance $R$ of \SampleTThirtySix.
} \label{fig.R-VUVV-T36}
\end{figure}

\begin{figure}[h!]
\includegraphics[width=\textwidth]{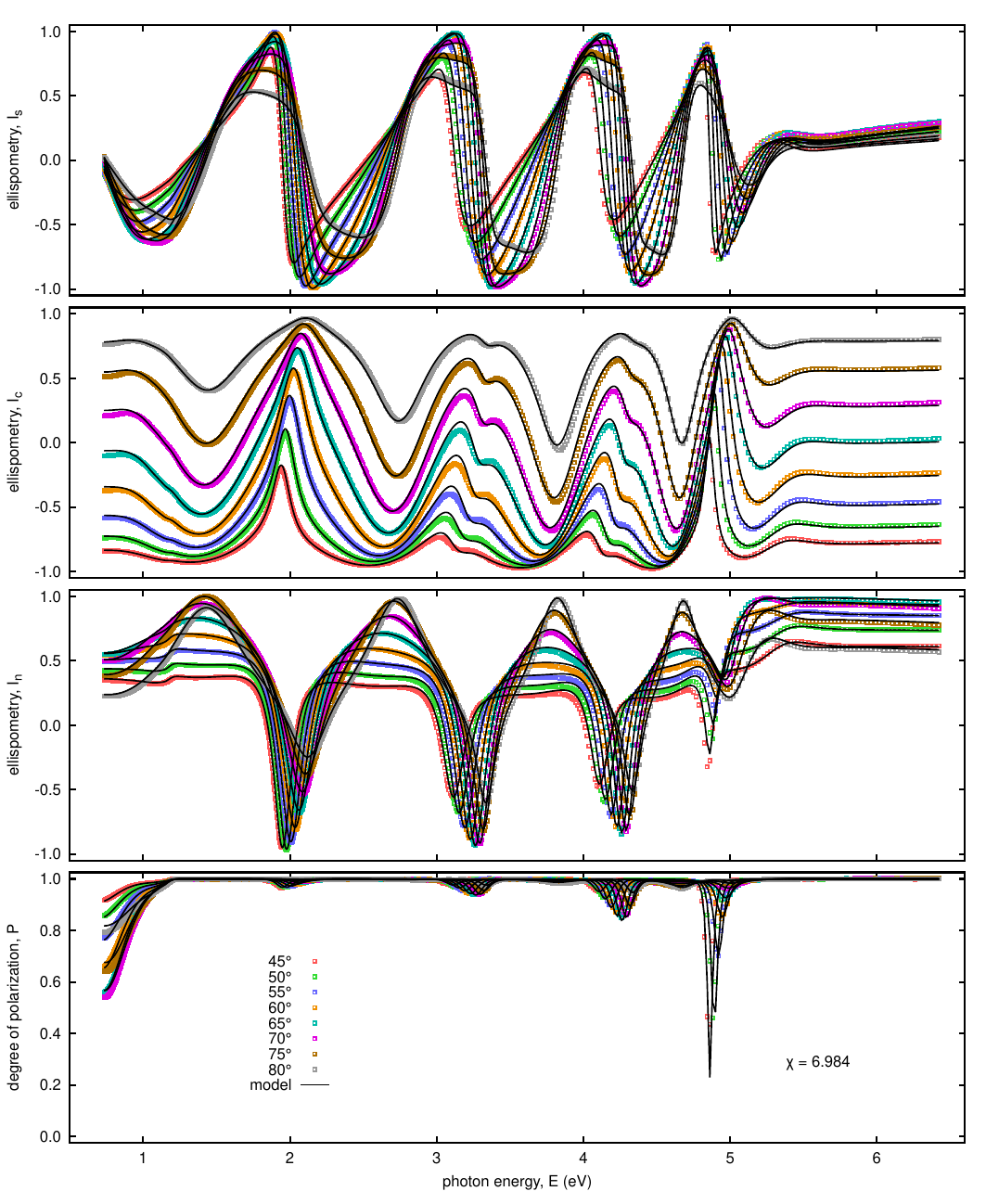}
\caption{Spectral dependencies of independent normalized Mueller matrix elements $\Ell$ in reflected light from front side of \SampleTThirtySix.
} \label{fig.E-UVNf-T36}
\end{figure}

\begin{figure}[h!]
\includegraphics[width=\textwidth]{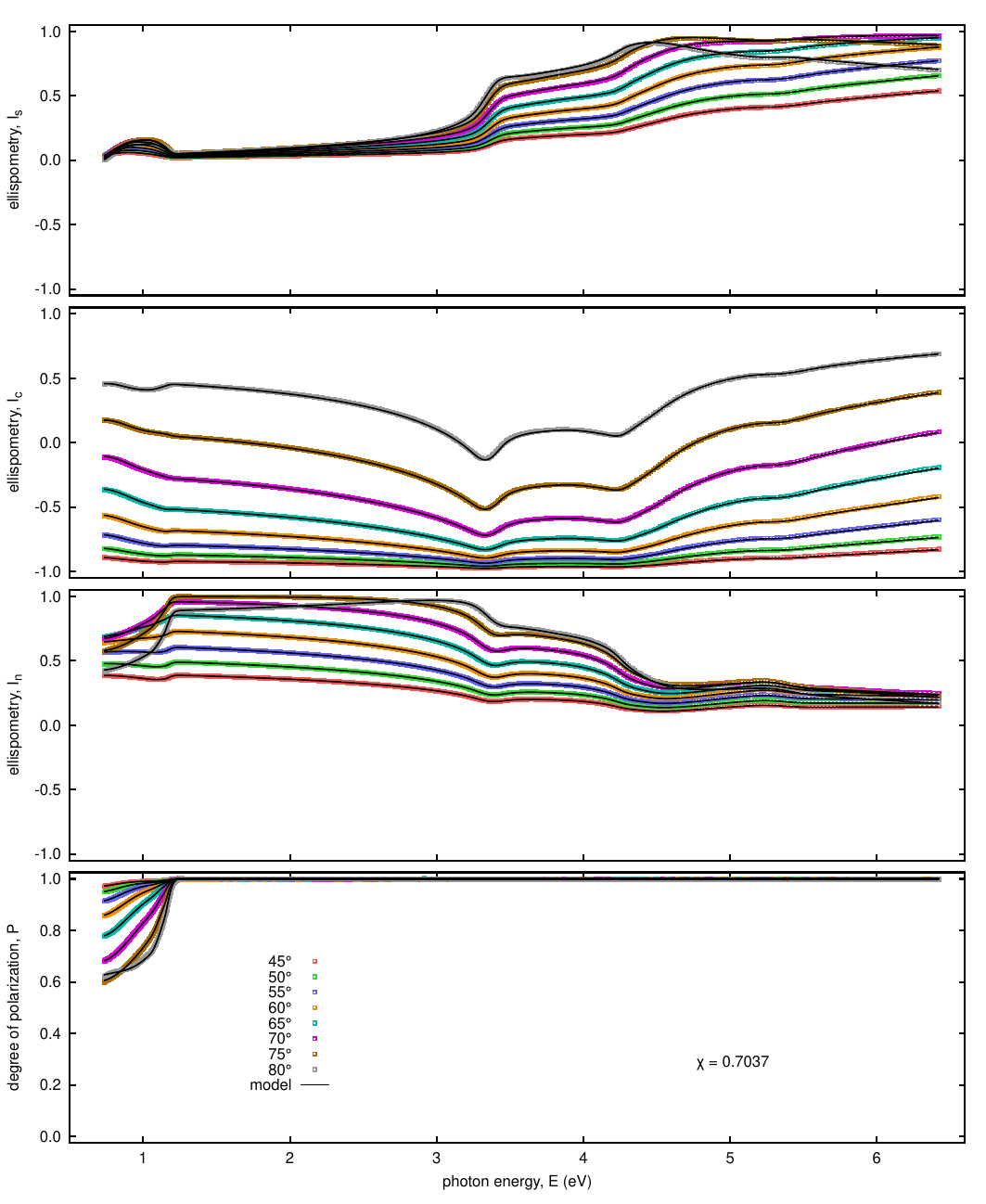}
\caption{Spectral dependencies of independent normalized Mueller matrix elements $\Ell$ in reflected light from back side of \SampleTThirtySix.
} \label{fig.E-UVNb-T36}
\end{figure}

\begin{figure}[h!]
\includegraphics[width=\textwidth]{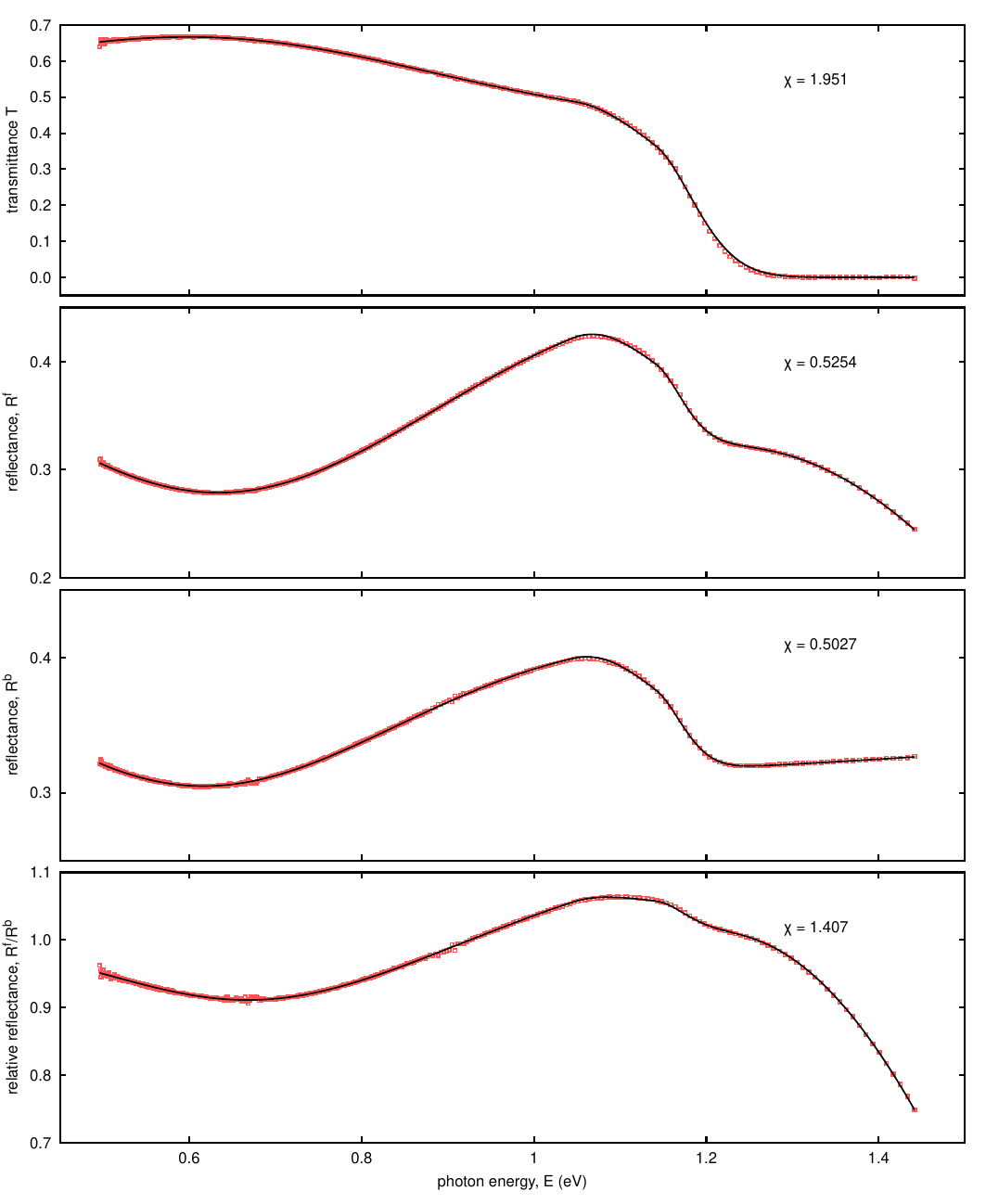}
\caption{Spectral dependencies of transmittance $T$, reflectance from front side $R^{\rm f}$, reflectance from back side $R^{\rm b}$ and relative reflectance $R^{\rm f}/R^{\rm b}$ of \SampleTThirtySix.
Measured in near-IR region by Perkin Elmer Lambda 1050+ spectrophotometer.
} \label{fig.TR-NIR-T36}
\end{figure}

\begin{figure}[h!]
\includegraphics[width=\textwidth]{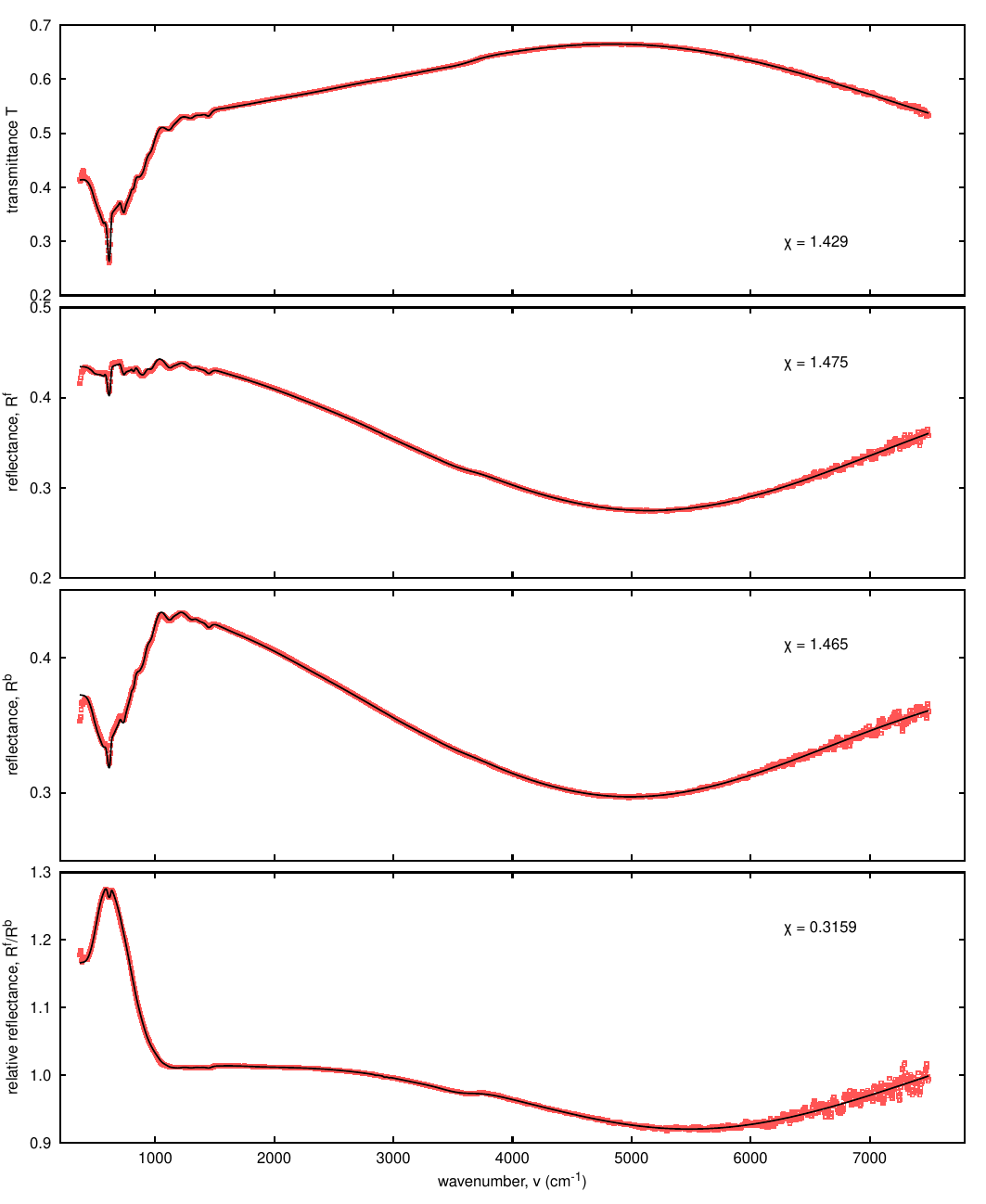}
\caption{Spectral dependencies of transmittance $T$, reflectance from front side $R^{\rm f}$, reflectance from back side $R^{\rm b}$ and relative reflectance $R^{\rm f}/R^{\rm b}$ of \SampleTThirtySix.
Measured in mid-IR region by Bruker Vertex 80v spectrophotometer.
} \label{fig.TR-MIR-T36}
\end{figure}

\begin{figure}[h!]
\includegraphics[width=\textwidth]{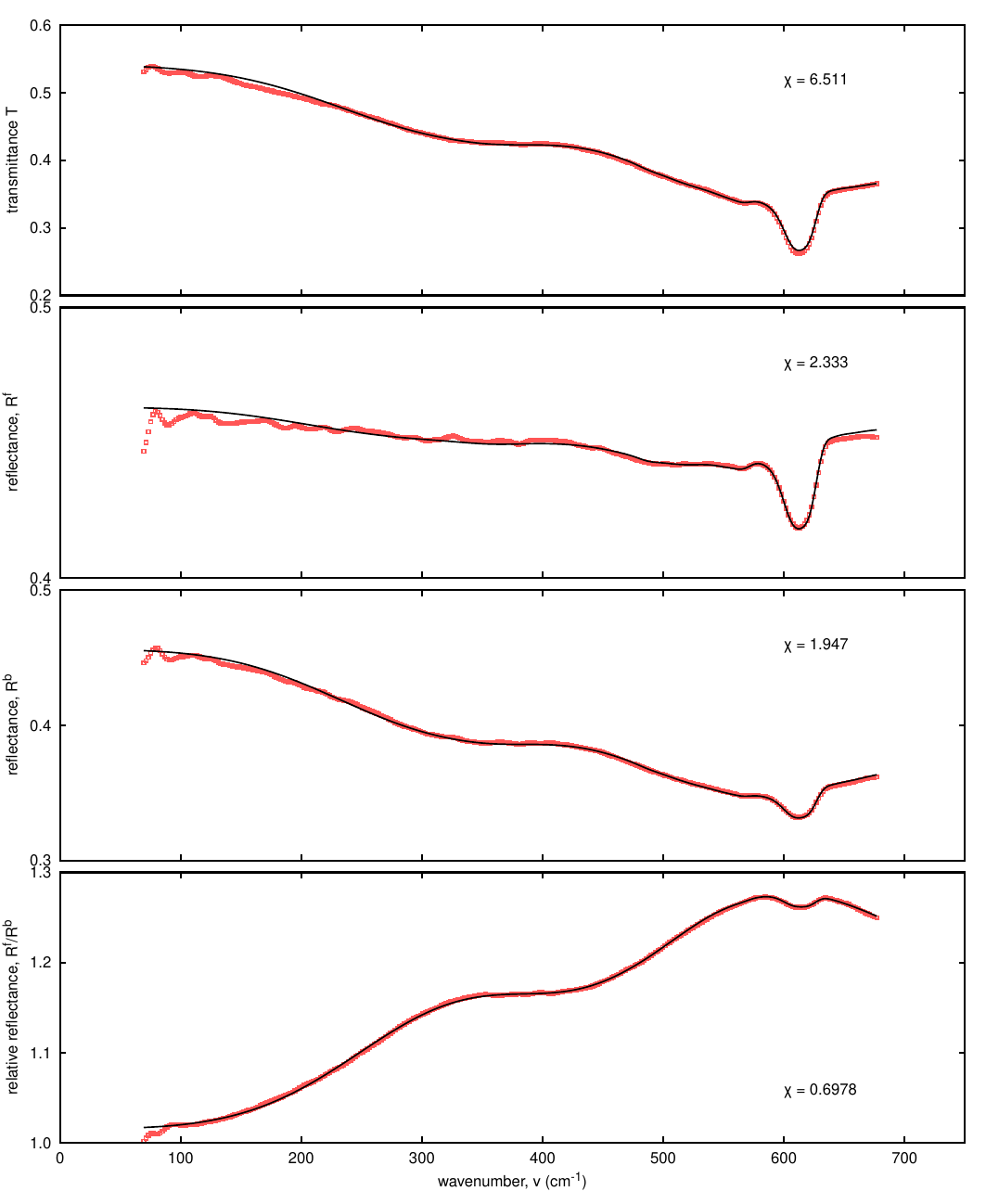}
\caption{Spectral dependencies of transmittance $T$, reflectance from front side $R^{\rm f}$, reflectance from back side $R^{\rm b}$ and relative reflectance $R^{\rm f}/R^{\rm b}$ of \SampleTThirtySix.
Measured in far-IR region by Bruker Vertex 80v spectrophotometer.
} \label{fig.TR-FIR-T36}
\end{figure}

\begin{figure}[h!]
\includegraphics[width=\textwidth]{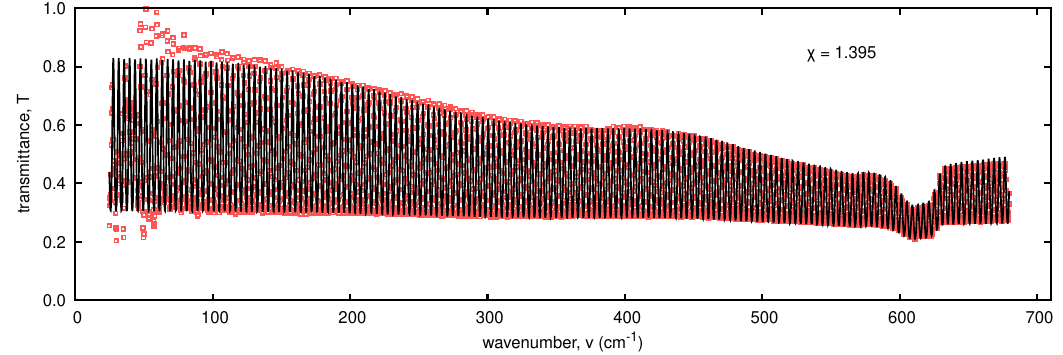}
\caption{Spectral dependencies of high-resolution transmittance $T$ of \SampleTThirtySix.
Measured in far-IR region by Bruker Vertex 70v spectrophotometer.
} \label{fig.T-FIRHR-T36}
\end{figure}

\clearpage

\subsection{Comparison of response functions of \AlTaO with previously determined response functions of \AlO and \TaO films}

\begin{figure}[h!]
\includegraphics[width=\textwidth]{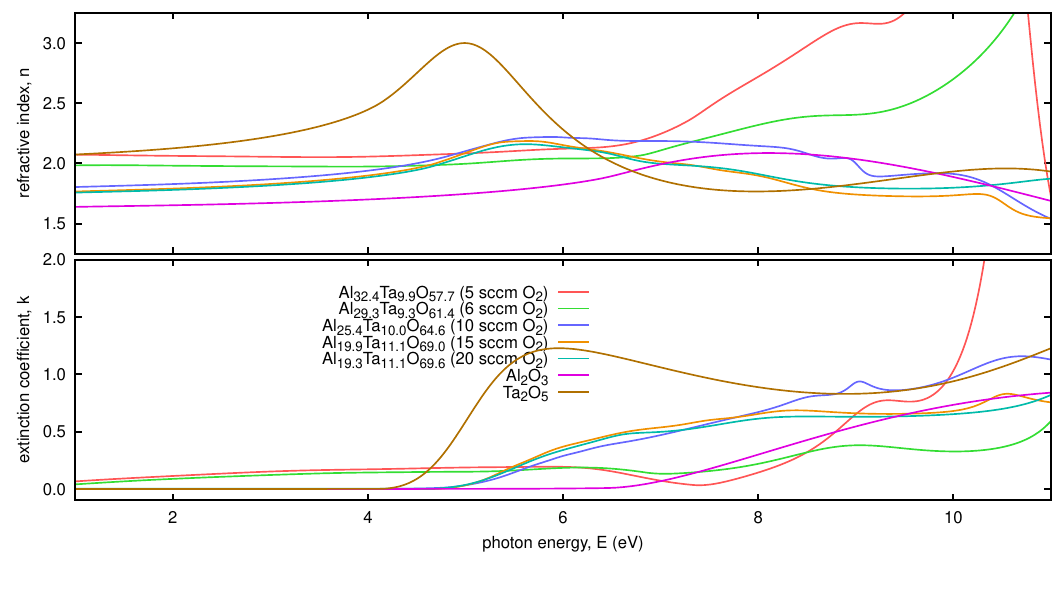}
\caption{Comparison of spectral dependencies of optical constants of \SampleTThirtyTwo, \SampleTThirtyThree, \SampleTThirtyFour, \SampleTThirtyFive, and \SampleTThirtySix with the previously determined optical constants of \AlO and \TaO films.
} \label{fig.nk-comp}
\end{figure}

\begin{figure}[h!]
\includegraphics[width=\textwidth]{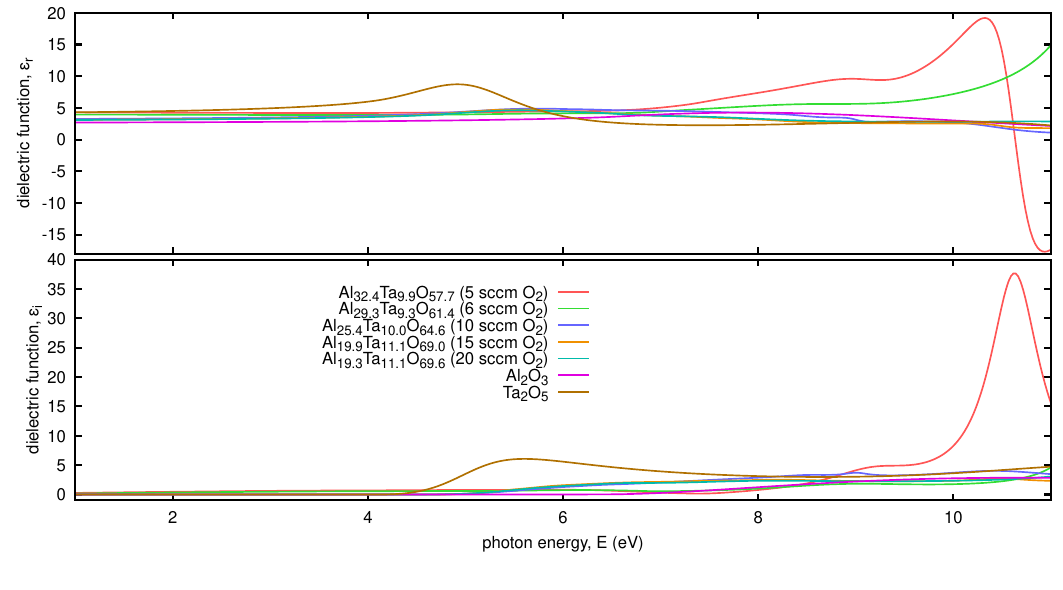}
\caption{Same data as in Fig.~\ref{fig.nk-comp} but for dielectric function $\hat\eps = (n+\iu k)^2$.
} \label{fig.eps-comp}
\end{figure}

\begin{figure}[h!]
\includegraphics[width=\textwidth]{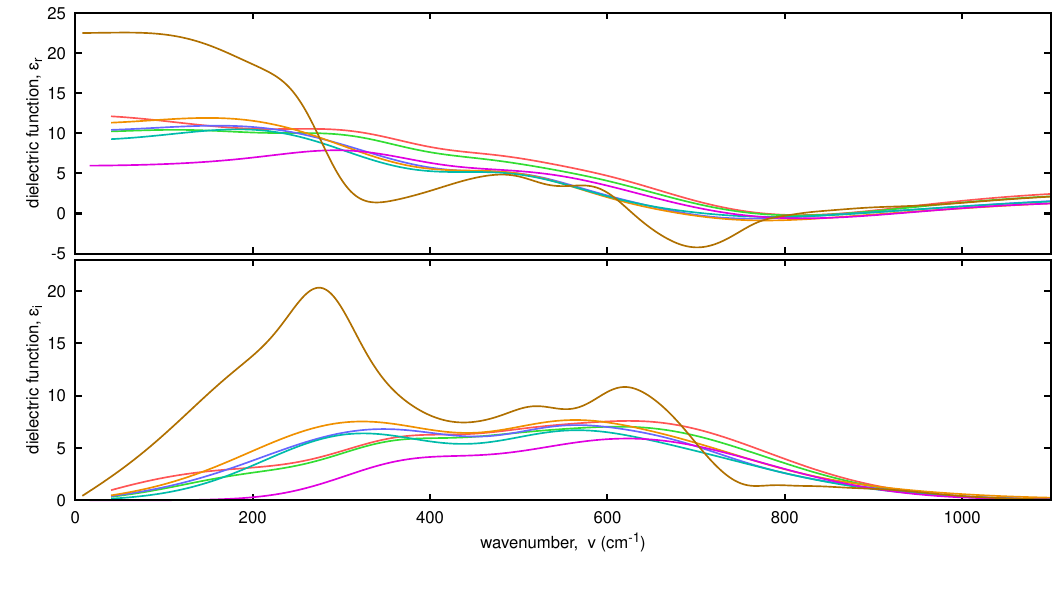}
\caption{Same data as in Fig.~\ref{fig.eps-comp} but for spectral range of phonon excitation in IR region.
} \label{fig.eps-comp-IR}
\end{figure}

\clearpage

\bibliographystyle{elsart-num}
\bibliography{references}